\documentclass[10pt]{article}
\usepackage[utf8]{inputenc}
\usepackage{multicol}
\usepackage{amsmath}
\usepackage{amssymb}
\usepackage{graphicx}
\usepackage{xcolor}
\usepackage[normalem]{ulem}
\usepackage{hyperref}
\usepackage{caption}
\usepackage{subcaption}
\hypersetup{colorlinks, urlcolor=blue, allcolors=blue}
\newcommand{\norm}[1]{\left\lVert#1\right\rVert}

\usepackage[affil-it]{authblk}
\usepackage{algorithm}
\usepackage[noend]{algpseudocode}
\usepackage[semicolon]{natbib}
\setcitestyle{square}
\usepackage{geometry}
\usepackage{graphicx}
\usepackage[outdir=Images/]{epstopdf}
\graphicspath{ {Images/} }
\usepackage{bm}
\usepackage{dirtytalk}
\usepackage[normalem]{ulem}

\usepackage{cancel}

\providecommand{\keyword}[1]{{\textit{Keywords:}} #1}

\title{A Bayesian multiscale CNN framework to predict local stress fields in structures with microscale features}
\author[1,2,*]{Vasilis Krokos}
\author[2]{Viet Bui Xuan}
\author[3]{Stéphane P. A. Bordas}
\author[4]{Philippe Young}
\author[5,1,*]{Pierre Kerfriden}

\affil[1]{School of Engineering, Cardiff University, Cardiff, UK}
\affil[2]{Synopsys Northern Europe Ltd, Exeter, UK}
\affil[3]{Institute for Computational Engineering, Faculty of Science, Technology and Communication, University of Luxembourg, Luxembourg}
\affil[4]{School of Engineering, Mathematics and Physical Sciences, University of Exeter, Exeter, UK}
\affil[5]{Centre des Matériaux, Mines ParisTech / PSL University, Evry, France}
\affil[*]{Corresponding authors: Vasilis Krokos, KrokosV@cardiff.ac.uk; Pierre Kerfriden, pierre.kerfriden@mines-paristech.fr}

\date{January 19, 2022}

\begin{document}

\maketitle

\begin{abstract}%

    Multiscale computational modelling is challenging due to the high computational cost of direct numerical simulation by finite elements. To address this issue, concurrent multiscale methods use the solution of cheaper macroscale surrogates as boundary conditions to microscale sliding windows. The microscale problems remain a numerically challenging operation both in terms of implementation and cost. In this work we propose to replace the local microscale solution by an Encoder-Decoder Convolutional Neural Network that will generate fine-scale stress corrections to coarse predictions around unresolved microscale features, without prior parametrisation of local microscale problems. We deploy a Bayesian approach providing credible intervals to evaluate the uncertainty of the predictions, which is then used to investigate the merits of a selective learning framework. We will demonstrate the capability of the approach to predict equivalent stress fields in porous structures using linearised and finite strain elasticity theories.
    
\end{abstract}

\keyword{Multiscale stress analysis, Convolutional neural network, Surrogate modelling, Bayesian machine learning}


\section{Introduction}
    
    Multi-scale structural analyses are prominent in mechanical and bio-mechanical engineering (e.g. composite materials such as carbon-reinforced polymers or concrete, porous materials such as bones). Full Finite Element Analysis (FEA) for stress prediction is usually prohibitively expensive for those structures, as the finite element mesh needs to be very dense to capture the effect of the fine scale features. Therefore, a common approach is to split the problem into a macroscale mechanical problem, and local microscale computations. The macroscale problem diffuses the overall stress field in the entire structure without fully resolving the material, while the local microscale computations are needed to correct the macroscale fields and characterise the constitutive law to be used at the macroscale.
    
    Multiscale computational modelling can be approached in two ways. Homogenisation, be it through the principles of micromechanics \citep{zohdiwrigger} or asymptotic expansions \citep{sanchezpalencia74}, performs all microscale computations over a representative volume element (RVE), assuming that the macroscale displacement gradients do not vary over the material sample. When the scales cannot be separated, scientists resort to domain decomposition-based approach. The results of homogenisation are applied to the boundary of regions of interest for concurrent microscale corrections to be performed \citep{Ghosh_2004, Oden_2006, Kerfriden_Allix_2009, Multiscale_Zhu, Paladin_Kerfriden_2016}. These approaches are computationally more expensive and practically more intrusive than methods based on RVEs. However, their deployment is necessary when predicting the microscale response to fast macroscale gradients, for instance due to sharp macroscale geometrical features. The proposed method belongs to this latter class of methods.
    
    Multiscale computational modelling may be coupled to offline/online acceleration methods such as model order reduction (MOR) techniques \citep{BARRAULT2004667, Ryckelynck_RedBasis, goury_amsallem_bordas_liu_kerfriden_2016} and meta-models \citep{PolynomialChaos1991, Kriging2009}. The idea is to realise many expensive computations in advance, subject to parameter variations, approximate the family of generated solutions using statistical regression, and use the statistical model online to produce solutions inexpensively. Meta-models directly produce an analytical mapping between parameters and solutions, whilst the more advanced MOR techniques require online conditioning of the a generative statistical model (PCA for instance) to partial satisfaction of the PDE system, making the latter approach intrusive. Both approaches share the fundamental requirement for an appropriate adhoc parametrisation of the PDE system. In addition to being cumbersome, such parametrisation are, in the case of meta-models based on polynomial chaos expansion or Kriging, further restricted to be of low-dimension to circumvent the curse of dimensionality \citep{Constantine2013}.
    
    In this paper, we propose to develop a meta-modelling approach to inexpensively generate microscale mechanical corrections given the result of coarse scale simulations. The meta-model will not require any knowledge of the PDE system to produce the fine-scale corrections. Moreover, we wish the meta-model to produce results for a variety of microstructures, characterised by parameters of random fields and exemplified by realisations of these fields. Our developments and examples are dedicated to 2D porous media with random distributions of circular and elliptical voids. Parametrising such distributions with small number of parameters is notoriously difficult. To circumvent this fundamental difficulty, we propose to convert the field of material properties into an image and use a Convolutional Neural Network (CNN) as surrogate model, guided by the fact that the CNNs used in computer vision perform statistical tasks extremely well in parameter space of very large dimensions (i.e. the number of pixels in an image). Our approach is not a priori limited to any particular statistics of the geometric features but in practice we will show that we have to stay close to the ones represented in the training set. Of course, the price that needs to be paid for these properties will be the necessity to generate thousands of examples of fine-scale simulations in order to achieve reasonable performance. Other works like the ones of \citep{BESSA2017633, Hengyang2019} opt for a dataset created by reduced order models to reduce the computational cost of performing the necessary FE simulations.
    
    \begin{figure}[htb]
        \includegraphics[width=\linewidth]{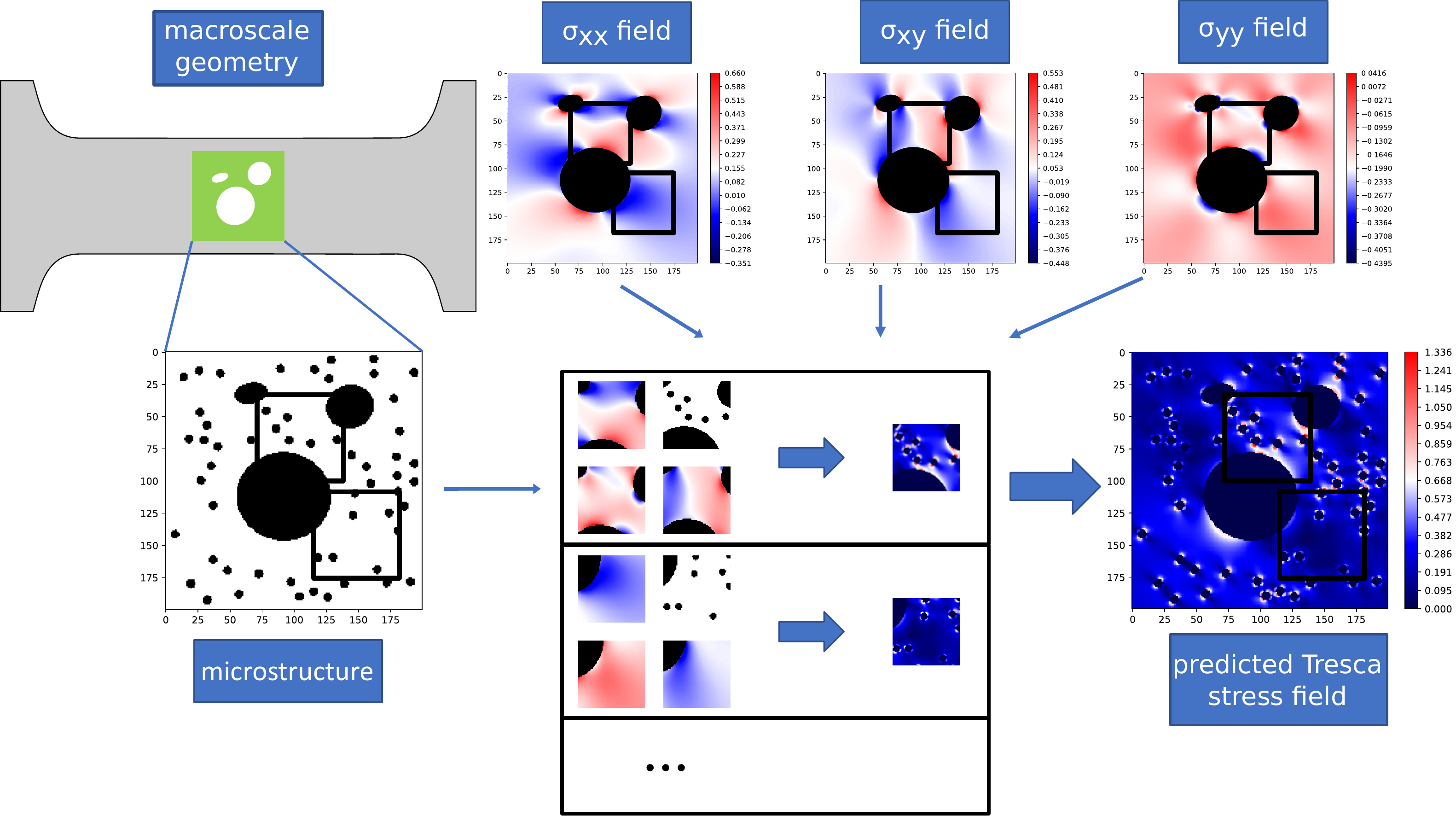}
        \caption{The multifidelity deep learning approach that is pursued in this article to predict microscale stress fields given macroscale stress fields. The inputs of the CNN are the macro scale stress tensor components, computed by macroscale FEA and converted into images, and a binary representation of the micro scale geometry within the sliding window (called "patch" in the article). The CNN predicts the microscale stress field in the window, i.e. it corrects the macro stress field based on the knowledge of the local microstructure. Entire microscale solution may be reconstructed by assembling such patch-level predictions. The methodology does not require any scale separation, but it relies on the Saint-Venant's principle.}
        \centering
        \label{fig:Intro}
    \end{figure}

    The input of the CNN can be the full macro scale stress fields, not only hand-crafted averages such as the ones used in homogenisation theory \citep{Homo}. Consequently, we can assume scale separability and apply arbitrary boundary conditions at infinity. Additionally, real medical or industrial data are hard to find and often expensive, so we aim to train our CNN on simpler, artificial, datasets and find a way to transfer our knowledge to real cases. To achieve that we are going to train our CNN using only patches of the geometry, so the CNN will be completely agnostic to the overall structure and will learn to identify the effect of microscale features on the macro stress field. An overview of the method can be seen in [Fig \ref{fig:Intro}].
    \par 
  
    Although Neural Networks (NNs) are used in a wide variety of applications, they usually do not incorporate uncertainty information in their predictions. As a consequence, the effect of deviating from the snapshot is not measured properly, as opposed to methodologies based on Gaussian Processes, which are more confidently used by mechanical engineers (e.g. Kriging), geostatisticians and practictioners of uncertainty quantification in general. Moreover, the availability of uncertainty measures in the predictions is useful to develop methods of active learning based on greedy selection of data \citep{JonesBayesianOptimisation, MadayPateraReducedBasis}. For these two reasons, our CNN meta-model will be Bayesian, thereby incorporating uncertainty estimates by construction, in a similar fashion to Gaussian processes.
    
    Bayesian Neural Networks are not very popular, owing to the unavailability of tractable posterior distributions, the multimodality of posterior distributions, and the difficulty of interpreting and calibrating the prior hyperparameters. Recently, significant developments have allowed researchers to partially tackle the tractability issue. Modern Bayesian Neural Networks are based on variational inference, whereby the parameters of families of possible posterior distributions are optimised using (stochastic) gradient descent in order to minimise the KL divergence between the surrogate distribution and the true, untractable posterior distribution \citep{Bishop}. Successful variants of these ideas include variational Bernoulli dropout \citep{gal2016dropout}, variational Gaussian dropout \citep{kharitonov2018variational} and Bayes by Backpropagation \citep{blundell2015weight}, these different variants being closely related to one another. We will make use of the latter method, which assumes independence and gaussianity of the posterior distribution of the weights of the network, making use of the reparametrisation trick to perform backpropagation over weight means and variances. 
    
    Based on the efficient BNN architecture described above, we will investigate the use of uncertainty-driven selective learning. To this end, we will only add to the training set microscale examples for which the error estimate is large. We will show that such approaches are beneficial in terms of training time. Eventually, microscale examples could be \textit{designed}, i.e. optimised to maximise the network uncertainty, but this more advanced conceptual idea is not investigated in the present contribution.

    Deep learning is being actively investigated worldwide for solving challenging issues in computational  mechanics and bio mechanics. One of the earliest works of using deep learning models as surrogates for FEA is from \citep{DLAorta2018} who developed an image-image deep learning framework to predict the aortic wall stress distribution where the mechanical behaviour in the FEA model was described by a fibre-reinforced hyperelastic material model. Other NNs with fully connected layers have then been used for stress predictions for non-linear materials but simple beam structures as shown by \citep{RoewerDespres2018TOWARDSFS, meister2018fast}. Later, \citep{MENDIZABAL2020101569} used a CNN for the prediction of nonlinear soft tissue deformation on more complicated structures such as a liver but without any kind of microscale features. Moreover, \citep{ResCNN} deployed a CNN model for stress prediction on simple structures with geometric features but not on multi-scale problems as the size of these features was comparable to the size of the structure.
    Also, \citep{Nie2020} used a GAN to analyze mechanical stress distributions on a set of structures encoded as high and low resolution images. A variety of loading and boundary conditions has been used and some of them resembled the effect of isolated microstructural features. Recently, \citep{sun2020predicting} based on the architecture of \citep{ResCNN} created an Encoder-Decoder CNN for the prediction of stress field on Fiber-reinforced Polymers but their samples come from a single specimen and with a single FE simulation implying low generalisation ability both in terms of different structures and loading/boundary conditions. Additionally, they predict only the $z$ component of the stress tensor and they report a value of about 70\% in their evaluation metric. Lastly, \citep{Cracks2021} used a Convolutional Aided Bidirectional Long Short-term Memory Network to predict the sequence of maximum internal stress until material failure. 
    \par
    
    Our paper is organised as follows. In section [\ref{Governing Equations}] we present the reference multiscale mechanical model that we aim to solve online and we introduce useful definitions for our methodology. In section [\ref{CNN section}] we discuss the architecture and the input and output of the CNN. Finally, in section [\ref{Numerical Examples}] we apply our method to different problems and show the results. 

\section{Methods and Governing Equations} \label{Governing Equations}

    In this section we will discuss the reference multiscale mechanical model that we aim to solve online using the trained CNN and we will also introduce definitions and notations that will be necessary for us to explain our methodology. 


\subsection{Problem statement: hyperelasticity}
    
        We consider a 2D body occupying domain $\Omega_0 \in \mathbb{R}^2$ with a boundary $\partial \Omega_0 = S_0$. The body is subjected to prescribed displacements $U_D$ on its boundary $\partial \Omega_{u,0}$ and prescribed tractions $T_D$ on the complementary boundary $\partial \Omega_{T,0}$ = $\partial \Omega \backslash  \partial \Omega_{u,0}$. We consider displacement $ \textbf{u} : \Omega_{0} \rightarrow \mathbb{R}^2$ and the associated deformed configuration $\Omega = \{ \hat{X} / \exists\ X \in \Omega_0, \hat{X} = X + \textbf{u}(X) \}$. The boundary value problem of hyperelasticty consists in finding $\textbf{u}(X) = \text{arg } \displaystyle{\min_{\textbf{u}^{*}} E_p(\textbf{u}^{*}) }$ 
        where the potential energy is defined as
        
        \begin{equation}
            E_p(u) = \int_{\Omega_0} W(\pmb{E}) \,d\Omega_0 - 
                     \int_{\Omega_0} \pmb{f_0} \textbf{u} \,d\Omega_0 +
                     \int_{\partial \Omega_{T,0}} \pmb{T_0} \textbf{u} \, dS_0
            \label{eq:PDE}
        \end{equation}
        We consider a linear Saint-Venant–Kirchhoff material model defined by its strain energy density
        \begin{equation}
            W(\pmb{E}) = \frac{1}{2} \lambda [\textrm{tr}(\pmb{E})]^2 + \mu\textrm{tr}(\pmb{E}^2)
            \label{eq:strain-energy}
        \end{equation}
        where the Green-Lagrange strain tensor $\pmb{E}$ is defined by
        \begin{equation}
            \pmb{E} = \frac{1}{2} (\pmb{F}^{T} \pmb{F} - I)
            \label{eq:Green-Lagrange strain tensor}
        \end{equation}
        with the definition of the deformation tensor 
        \begin{equation}
            \pmb{F} = \frac{\partial \hat{X} }{ \partial X}
        \end{equation}
    In the equations above, $\lambda$ and $\mu$ are the Lamé elasticity parameters and $I$ denotes the identity tensor.

        The Cauchy stress tensor $\pmb{\sigma}$ may be calculated as follows:
        \begin{equation}
            \pmb{\sigma} = \frac{1}{J} \pmb{F} \frac{\partial W}{\partial \pmb{E}} \pmb{F}^{T}
            \label{eq:nonlinear-Cauchy-stress}
        \end{equation}

        where the jacobian of the deformation tensor reads as
        \begin{equation}
            J = \textrm{det}(\pmb{F})
        \end{equation}
    
    The prescribed volume force $\pmb{f_0}$ and prescribed surface tractions $\pmb{T_0}$
         expressed in the reference configuration may be expressed as a function of their physical counterparts $\pmb{f}$ and $\pmb{T}$ as
                 \begin{equation}
            \pmb{f_0} = J\pmb{f}
        \end{equation}
        \begin{equation}
            \pmb{T_0} = J \norm{\pmb{F}^{-T} \pmb{n_0}} \pmb{T}
        \end{equation}
        where  $\pmb{n_0}$ is the unit normal vector in the reference configuration.
    
    for the special case of linear elasticity the Green-Lagrange strain tensor is replaced by the linearised strain tensor $\pmb{\epsilon}$:
    
        \begin{equation}
            \pmb{E} \approx \pmb{\epsilon} =  \frac{1}{2} ( \nabla \textbf{u} + (\nabla \textbf{u})^\top ) 
            \label{eq:linear-strain-energy}
        \end{equation}
        
        \begin{equation}
            \pmb{\sigma} = \frac{\partial W(\pmb{\epsilon})}{\partial \pmb{\pmb{\epsilon}}}
            \label{eq:linear-Cauchy-stress}
        \end{equation}
        
    \subsection{Equivalent stress}
    
        We are interested in predicting the stress field and more specifically an equivalent stress field indicating potential crack initiation sites. A possible choice is the Tresca stress which be calculated from the stress tensor as follows. The stress tensor can be rotated 
                \begin{equation}
                    \sigma' = Q \cdot \sigma \cdot Q^\top
                    \label{eq:transform}
                \end{equation}
        using rotation matrix
                        \begin{equation}
                    Q =  
                        \begin{bmatrix}
                            \cos(\theta) & -\sin(\theta) \\
                            \sin(\theta) & \cos(\theta)
                        \end{bmatrix} 
                    \label{eq:rot}
                \end{equation}
        The components of the stress tensor obtained after rotation are as follows
                 \begin{subequations}
                    \begin{align} 
                        \sigma'\!_{xx} &= \sigma_{xx} \cos^2 \theta + \sigma_{yy} \, \sin^2 \theta + 2 \, \tau_{xy} \, \sin \theta \cos \theta \label{eq:s'_x} \\
                        \sigma'\!_{yy} &= \sigma_{xx} \sin^2 \theta + \sigma_{yy} \, \cos^2 \theta - 2 \, \tau_{xy} \, \sin \theta \cos \theta \label{eq:s'_y} \\
                        \tau'\!_{xy}   &= (\sigma_{yy} - \sigma_{xx}) \sin \theta \cos \theta + \tau_{xy} (\cos^2 \theta - \sin^2 \theta) \label{eq:s'_xy}
                    \end{align}
                \end{subequations}       
        From [Eq.  \ref{eq:s'_xy}] we can find that there must be an angle $\theta_p$ such that the shear stress after rotation is zero:
                        \begin{equation}
                    \tan (2 \theta_P) = { \frac {2 \tau_{xy}} {\sigma_{xx} - \sigma_{yy}} }
                     \label{eq:theta_p}
                \end{equation}
        After inserting $\theta_p$ into [Eq.  \ref{eq:s'_x}, \ref{eq:s'_y}], the 2 principal stress components can be obtained: 
            \begin{equation}
                    \sigma_{\text{max}}, \sigma_{\text{min}} = {\frac{\sigma_{xx} + \sigma_{yy}}{2}} \pm \sqrt{ \left( \frac{\sigma_{xx} - \sigma_{yy}}{2}\right)^{2} + \tau_{xy}^2 }
                    \label{eq:principal_stresses}
                \end{equation}
        The Tresca stress $\sigma_T$ is equal to $\sigma_T = {\frac{1}{2}}$ ($\sigma_{\text{max}}$-$\sigma_{\text{min}})$. 

    \subsection{Finite element solver}        
    
        We use a standard P1 finite element method to discretise the hyperelasticity problem in space. We use unstructured triangular meshes that conform to the boundary of the domain. To help the Newton algorithm reach regions of quadratic convergence, the load is applied in steps of increasing amplitude. 
        
        The finite element fields are converted into image by standard interpolation from meshes to cartesian grids.

    \subsection{Global-local framework}
    
    We assume that macroscale mechanical quantities can be computed by solving an homogenised model over the computational domain, using a relatively coarse finite element mesh whose typical length-scale is unrelated to the microscale structure. We restrict ourselves to homogenised models obtained by solely modifying the constitutive relation of the material.
    
    All the examples presented in this paper are dedicated to hyperelasticity in a porous medium made of an homogeneous matrix with a random distribution of spherical voids. At the coarse-scale level, the voids are ignored and the fine-scale constitutive law is used as an homogeneous material model for the entire structure, without further adjustment of the elasticity coefficients. We could have used various homogenisation schemes to obtain macroscopically accurate homogenised coefficients, but the approach followed in this paper does not require the use of such a finely tuned homogenisation model.
    
    The macroscale mechanical fields being solved for using an inexpensive finite element solver, we now assume that it is possible to relocalise the solution field over an arbitrary window $B \subset \Omega_0$, which may intersect arbitrarily with boundary $\partial \Omega_0$. More precisely, we surmise that there exists a function 

    \begin{equation}
        \begin{array}{rccccl}
        \mathcal{F} : \qquad & \mathcal{L}_2(B)^3  & \times & \mathcal{M}(B) & \rightarrow & \mathcal{L}_2(\hat{B}) \\ 
        & 
        \begin{pmatrix}
        \displaystyle \sigma^M_{xx} \\
        \displaystyle \sigma^M_{yy} \\
        \displaystyle \tau^M_{xy} \\
        \end{pmatrix}
        & \times & g & \mapsto & \sigma_T
        \end{array}
    \end{equation}
    
\begin{figure}[h]
            \begin{center}
                \includegraphics[scale=1]{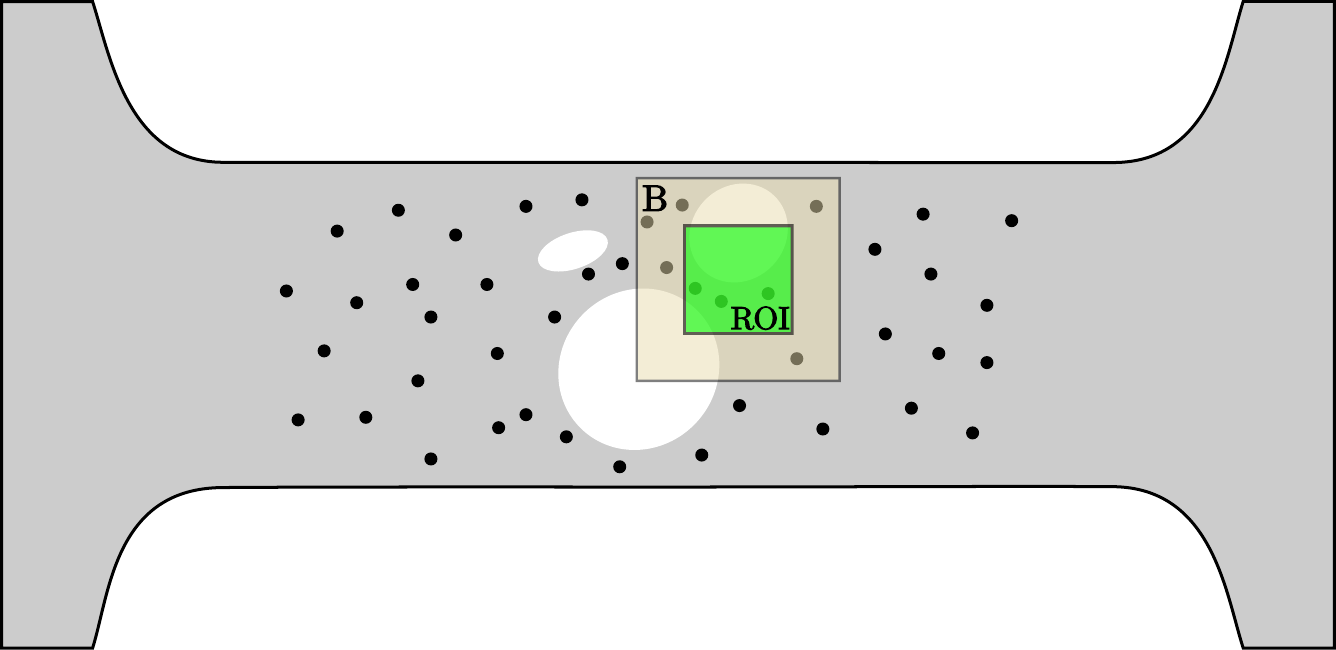}
                \caption{Porous material. Three elliptical macroscale features are represented in the specimen. Black areas correspond to microscale pores. In brown, we see a subregion $\textrm{B}$ (patch) where the macro stress field and the microscale geometrical information are available and in green the Region of interest (RoI) where the micro Tresca is to be predicted.}
                \label{fig:Overview}
            \end{center}
        \end{figure}

    In the previous expression, $B$ is the window over which relocalisation is performed and $\hat{B} \subset B$ is the region of interest (ROI), which is strictly contained in $B$ (see [Fig \ref{fig:Overview}]). Field $g : B \rightarrow \mathbb{R}^{n_m}$ is an indicator of the material phase present at point $X \in B$. In our case of porous medium, $g(X)$ is binary, 0 indicating void, 1 indicating the elastic material. Finally, superscript $M$ indicates that the field have been computed using the homogenised surrogate. With these notations at hand, the previous expression indicates that knowing the macroscopic stress field over the window, and the geometry of the structure and the precise position and nature of the material at the microscale, $\mathcal{F}$ will produce a map microscale Tresca stress field in the ROI that is accurate, i.e. close to that produced by the direct numerical simulation over the entire computational domain.
    
    Essentially, the role of the CNN developed further below is to learn function $\mathcal{F}$. The fact that such a function exists is backed up by the following qualitative observations, but will ultimately be proved numerically.
    \begin{itemize}
        \item knowing the exact stress field over the boundary of the window, an elasticity problem with Neumann boundary conditions may be solved over the window to provide the exact microscale stress field in $\hat{B}$. Hence, the stress field provided as input of $\mathcal{F}$ may carry all the information that is needed to relocalise the stress field exactly over $B$. Of course, in practice, this information will be degraded by the fact that it is approximated by an homogenised model solved using a coarse finite element solver.
        \item an approximate stress field being available on the boundary of $B$, and providing that standard self-equilibrium conditions are met, the Neumann boundary value problem may produce poor relocalised stresses in a band $B \backslash \hat{B}$ located in the vincinity of the boundary of the window (in virtue of Saint-Venant's principle). This is the reason why our predictions will be made in $\hat{B}$ and not in $B$, the boundary stresses of which cannot be expected to be produced with sufficient accuracy.
    \end{itemize}
    
    Finally, we note that $\mathcal{F}$ is a function to be learned by examples, which is why we do not expect the choice of the homogenised model to have a significant impact on the quality of the result. The Neural Network will be given sufficient amount of macro/micro stress pairs to compensate for systematic macroscale inaccuracies.
    
\section{CNN} \label{CNN section}

    \subsection{Input-Output}

\begin{figure}[h]
            \includegraphics[width=\linewidth]{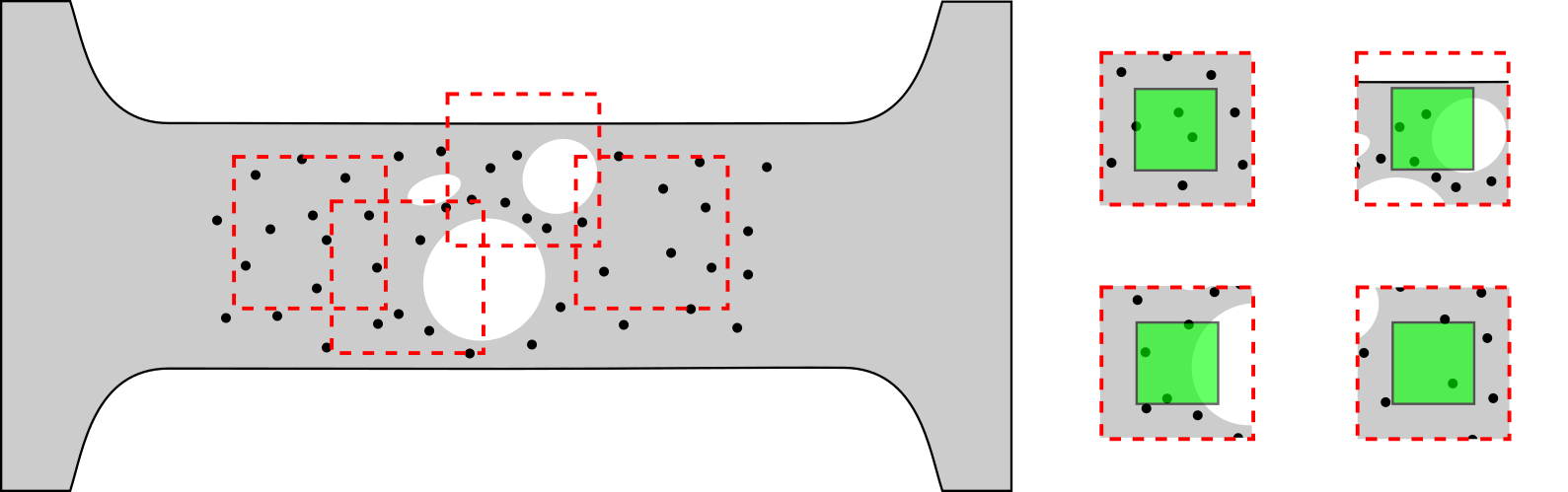}
            \caption{On the left the original structure and 4 patches that correspond to the red squares. On the right the extracted patches that will be fed to the Neural Network.}
            \centering
            \label{fig:Patches}
        \end{figure}

\begin{figure}[h]
            \centering
            \includegraphics[width=0.7\linewidth]{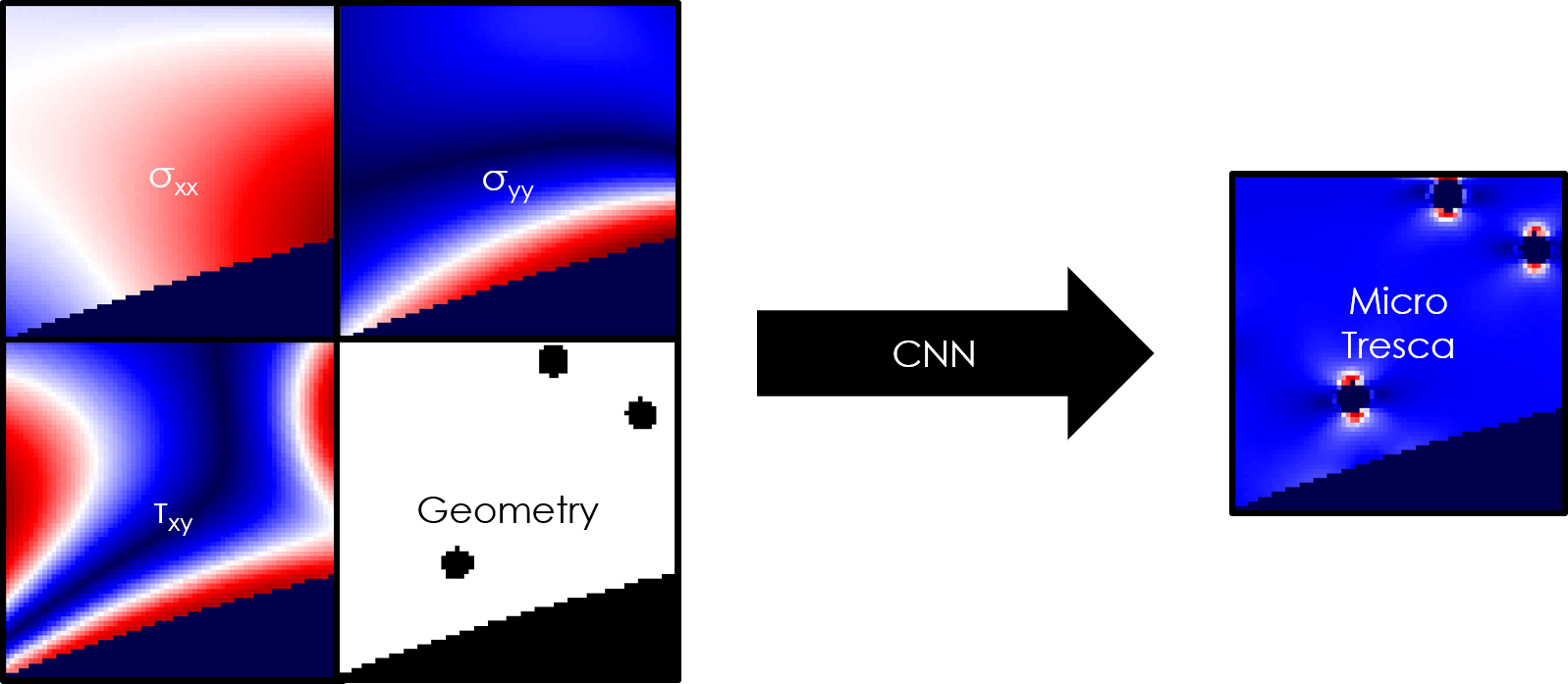}
            \caption{On the left the input of the CNN (the three components of the macroscopic stress fields, converted into images, plus the binary image corresponding to the indicator function of the microstructure) and on the right the output, which is the microscale Tresca stress}
            \label{fig:InOut}
        \end{figure}

        As already discussed the input of the CNN are patches extracted from the structure [Fig \ref{fig:Patches}]. The input of the CNN is a 3D array of size $[N_x \times N_y \times N_{C}]$ where: $N_x$ and $N_y$ are the size of the input image along the $x$ and $y$ direction respectively and $N_{C}$ is the number of channels of every data point. Each data point has 4 channels namely $\sigma_{xx}$, $\sigma_{yy}$, $\tau_{xy}$ and $Geometry$ corresponding to the $xx$, $yy$, $xy$ component of the macro stress tensor and a binary image of the geometry respectively. The output of the model is an $[N_x \times N_y]$ image corresponding to the micro Tresca stress. Note that we are only interested in the ROI of the patch so all the statistics during training and inference are calculated there. Because we want to identify the effect of micro scale features on the macro scale stress we will scale the output with a number that reflects the intensity of the macro stress field. This number is the sum of the absolute principal stresses of the macro stress tensor $|\sigma_{\text{max}}| + |\sigma_{\text{min}}|$ from [Eq. \ref{eq:principal_stresses}]. The micro stress in areas away from micro scale features should be the same as the macro scale stress because these features only have a local effect. This suggests that the output should be constant away from the micro scale features and change rapidly very close to them. That is clearly visible in [Fig \ref{fig:InOut}]. 
        \par
        
    \subsection{Layers}
        
        Although Deep Neural Networks have been successfully used to model very complex tasks, their training is challenging for a number of reasons. In this work we will use two layers that will allow us to efficiently train our DNN namely Batch Normalization (BN) and Residual Blocks.
        
        \subsubsection{Batch Normalization}
        
            In DNNs the distribution of each layer’s inputs changes during training, as the parameters of the previous layers change, this phenomenon is known as internal covariate shift. This slows down the training by requiring lower learning rates and careful parameter initialization \citep{ioffe2015batch}. BN aims at reaching a stable distribution of activation values throughout training. To achieve that, BN normalizes the output of a previous activation layer by subtracting the batch mean and dividing by the batch standard deviation. After the normalization, BN tries to scale and shift the normalized output by adding two trainable parameters to each layer. \citep{ioffe2015batch, santurkar2019does}. Additionally, BN makes the optimization landscape significantly smoother. This smoothness induces a more predictive and stable behaviour of the gradients, allowing for faster training \citep{santurkar2019does}. Lastly, there is a large consensus that the BN should be used before dropout and the activation function \citep{ioffe2015batch, li2018understanding, gabrin2020dp}
 
        \subsubsection{Residual Blocks}
        
\begin{figure}[h]
                \includegraphics[width=\linewidth]{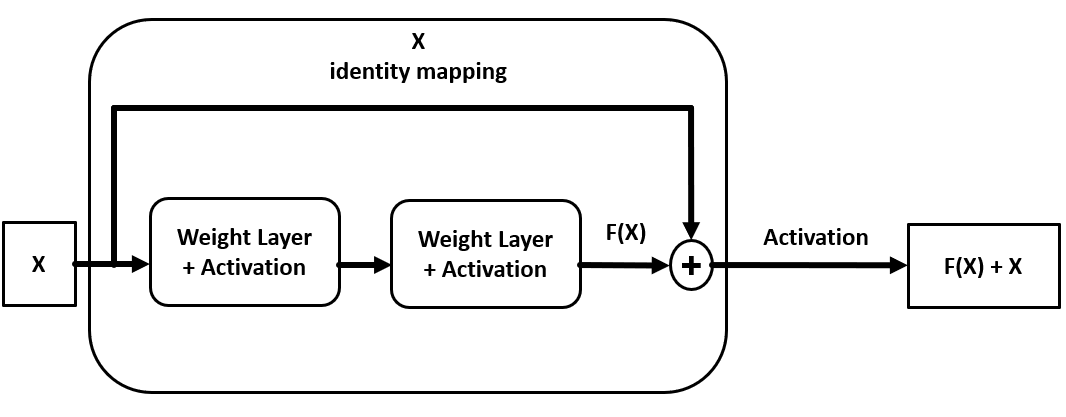}
                \caption{Structure of a generic residual block with input $X$ and output $F(X)+X$.}
                \centering
                \label{fig:GenericResBlock}
            \end{figure}

           Another problem with DNNs is the vanishing gradients causing the NN to not be able to learn simple functions like the identity function between input and output \citep{chapter-gradient-flow-2001, sussillo2015random}. The current way to train DNNs is through residual blocks \citep{he2015deep, kim2016deeplyrecursive, zagoruyko2017wide, lim2017enhanced}. With residual blocks the NN itself can choose its depth by skipping the training of a few layers using skip connections. As we can see from [Fig  \ref{fig:GenericResBlock}] even if the NN chooses to ignore some layers ($F(X)=0$) it will learn to map the input of the block to the output of the block. In this case the expression of the output would be simplified to: $F(X) + X = 0 + X = X$. This way we can use a large number of residual blocks and the network will simply ignore the ones it does not need. The name residual comes from the fact that the network tries to learn the residual, $F(X)$, or in other words the difference between the true output, $F(X)+X$, and the input, $X$.
        
        \subsubsection{Squeeze and Excitation block}
        
\begin{figure}[h]
                \includegraphics[width=\linewidth]{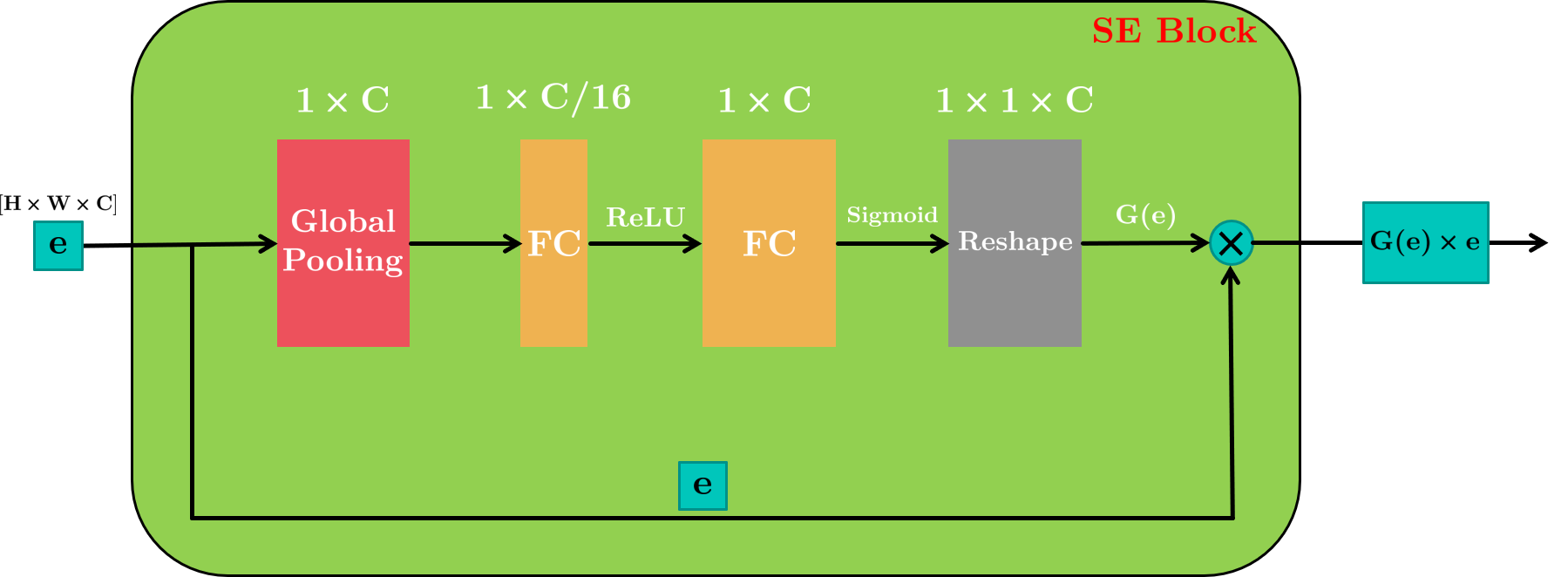}
                \caption{Structure of a generic SE block with input $e$ and output $G(e) \times e$. FC stands for Fully Connected layer}
                \centering
                \label{fig:SEBlock}
            \end{figure}

            In the residual blocks of our CNN we are using another kind of block: the Squeeze and Excitation block, SE. The SE can adaptively recalibrate channel-wise feature responses by explicitly modelling interdependencies between channels resulting in improved generalization across datasets and improvement in the performance \citep{SE2, Li_2018_ECCV, hu2019squeezeandexcitation}. The input of the SE has $C$ channels, height $H$ and width  $W$, $[H \times W \times C]$. The input decreases in size using a global-averaging pooling layer resulting in a linear array of size $[1 \times C]$.  After that, two fully connected layers downsample and then upsample the linear array. Firstly the linear array is downsampled by a factor of 16, $[1 \times C/16]$, as this is indicated to result in optimum performance \citep{hu2019squeezeandexcitation}, then a ReLU activation function is applied before upsampling again using a factor of 16 $[C/16 \cdot 16 = 1 \times C]$ and in the end a Sigmoid activation function is applied. Lastly, the linear array is reshaped to size $[1 \times 1 \times C]$ and multiplied with the input of the SE block [Fig \ref{fig:SEBlock}].
    
        \subsection{Architecture}
        
\begin{figure}[h]
                \includegraphics[width=\linewidth]{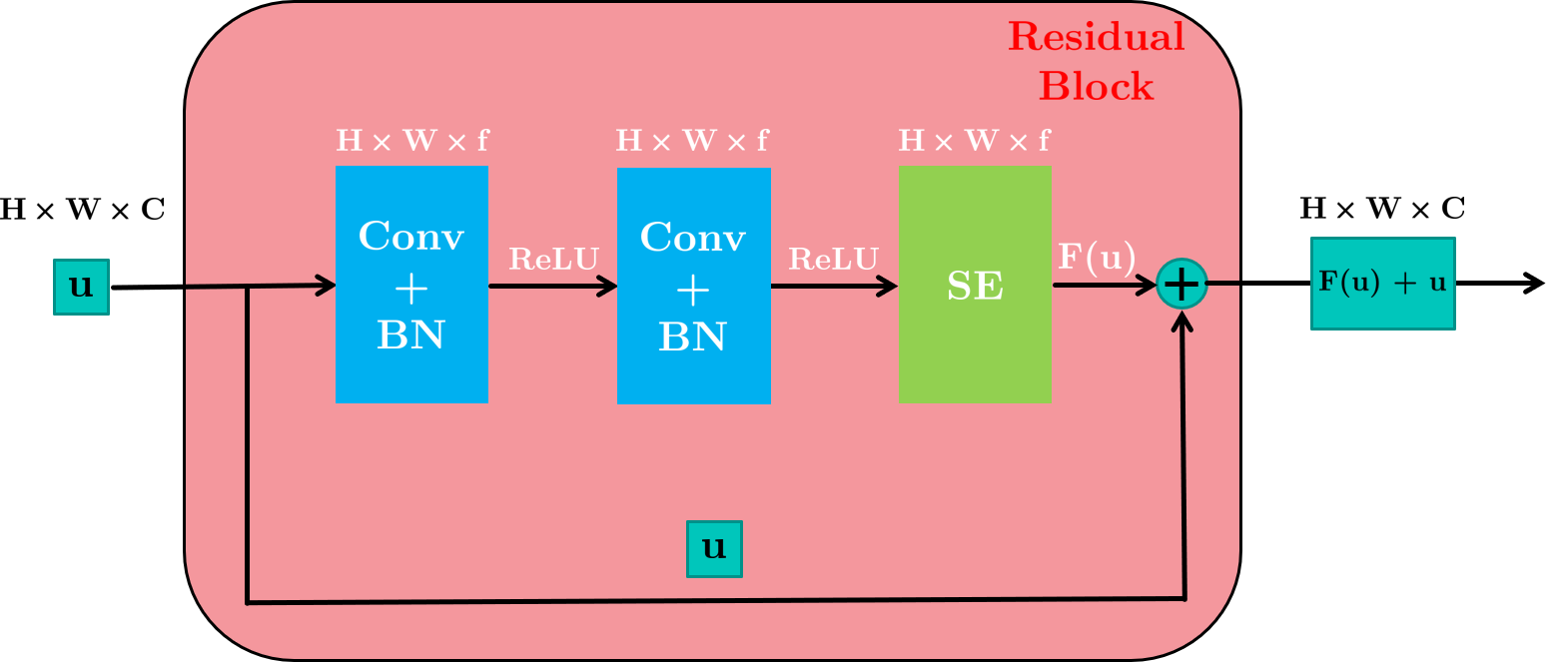}
                \caption{Structure of the residual block we are using with input $u$ and output $F(u)+u$.}
                \centering
                \label{fig:ResBlock}
            \end{figure}

            The residual blocks we will use in this work consist of two convolution layers, followed by a BN layer and a ReLU activation function each, and a SE block in the end [Fig \ref{fig:ResBlock}]. The input and output of this block has exactly the same size as we choose the number of filters for the convolution layers to be the same as the number of filters at the input of the residual block. 
            
\begin{figure}[h]
                \includegraphics[width=\linewidth]{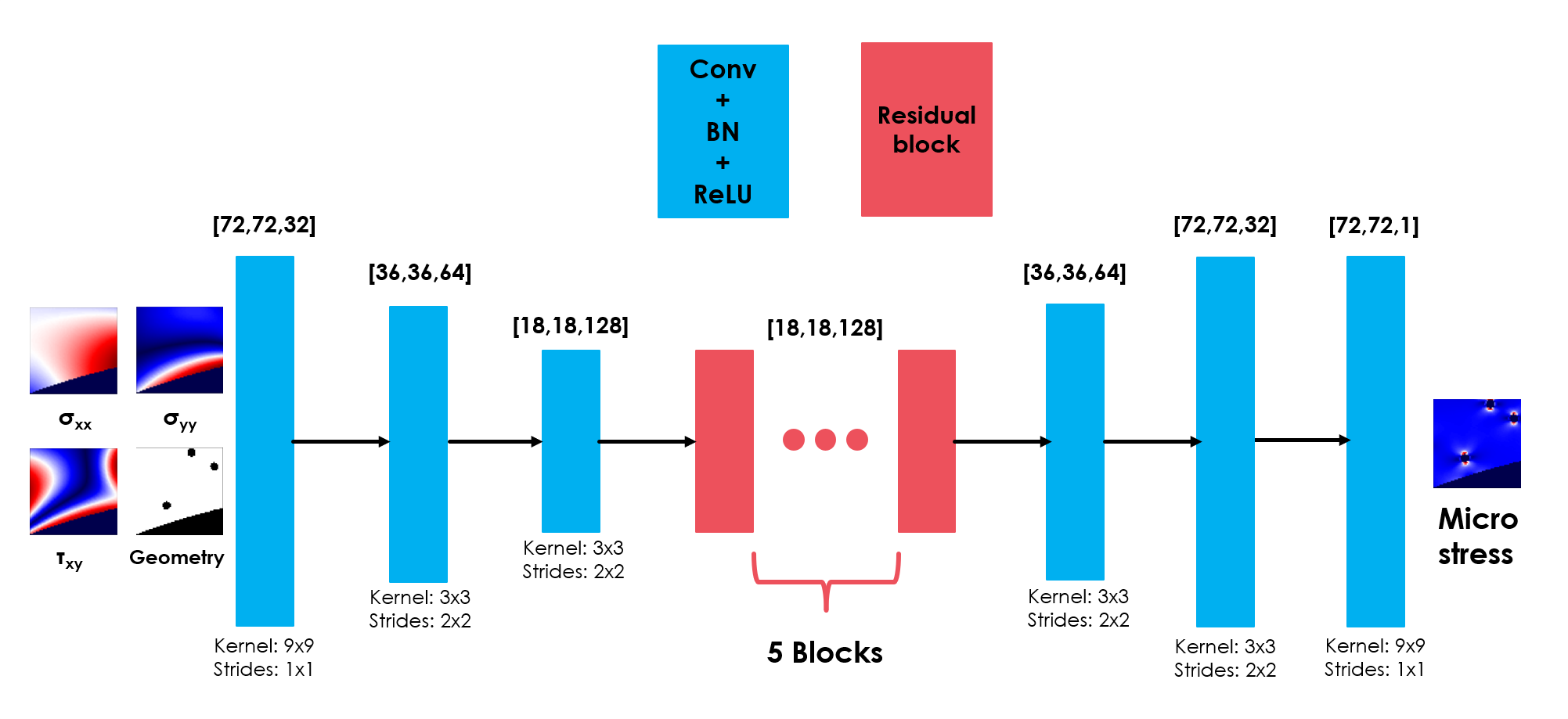}
                \caption{Structure of the CNN. $\sigma_{xx}$, $\sigma_{yy}$ and $\tau_{xy}$ are the stress components on the $x$, $y$ and $xy$ direction respectively.}
                \centering
                \label{fig:CNN}
            \end{figure}

            The architecture of the network is inspired by the “StressNet”, proposed by \citep{ResCNN}. Three convolution layers with increasing number of filters will downsample the input, after that five residual blocks are applied to the resulting array before using 3 deconvolution layers with a decreasing number of filters to upsample to the original dimension but with 1 channel instead of 4 [Fig \ref{fig:CNN}].
            
\clearpage 

    \subsection{Bayes By Backprop}
        
\begin{figure}[h]
            \includegraphics[width=0.7\linewidth]{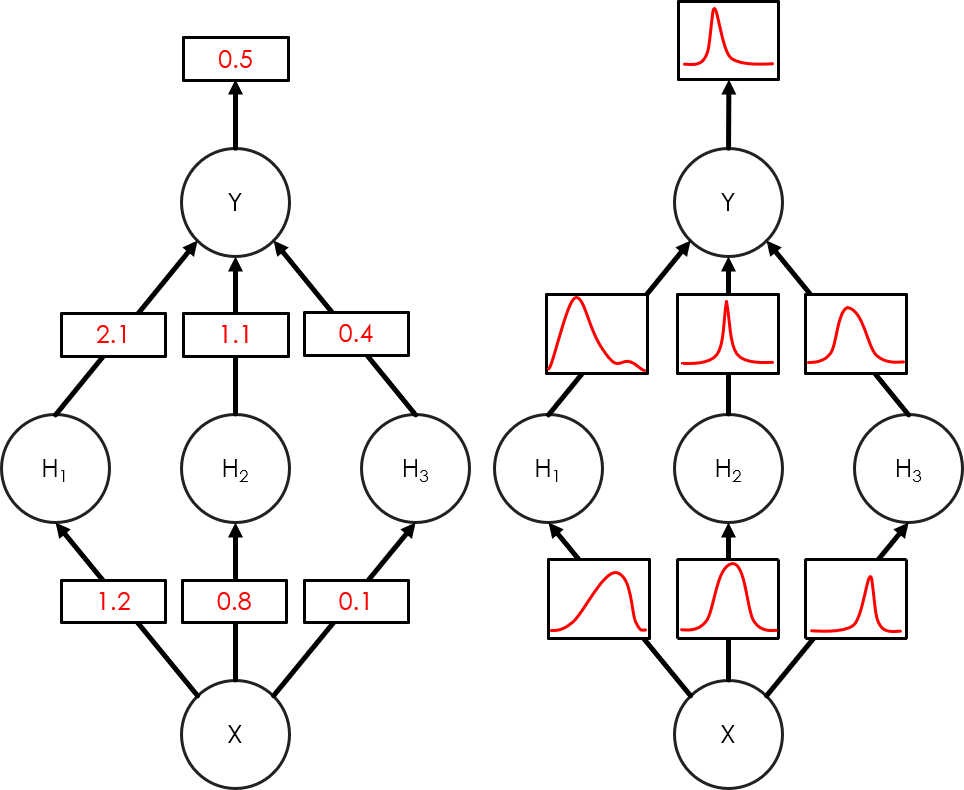}
            \centering
            \caption{On the left we have a sketch of a plain Neural Network with constant weights and on the right a Bayesian Neural Network where the weights are replaced by distributions. In both sketches the biases have been omitted for simplicity.}
            \centering
            \label{fig:BNN_sketch}
        \end{figure}

        In our effort to include uncertainty information into our prediction we will deploy a Bayesian framework, as described by \citep{blundell2015weight}, to introduce uncertainty in the weights of the network. To achieve that we will replace the constant weights of a deterministic neural network with a distribution over each weight as seen in [Fig \ref{fig:BNN_sketch}]. The output of this probabilistic model, for an input $x \in \mathbb{R}^n$ and a set of possible outputs $y \in \mathbb{R}$, will be a probability distribution over all possible outputs. The distribution of the weights before observing the data is called prior distribution, $P(\omega)$, and it incorporates our prior beliefs for the weights. The goal is to calculate the posterior, the distribution of weights after observing the data, because during training and of course inference the weights of the network are sampled from the posterior. The goal of Bayesian inference is to calculate the posterior distribution of the weights given the training data, $P(w|D)$. Unfortunately, the posterior is intractable for NNs but can be approximated by a variational distribution $q(w|\theta)$ \citep{Hinton1993, NIPS2011_4329}, parameterised by $\theta$. Variational learning finds the parameters $\theta^{opt}$ that minimise the Kullback-Leibler (KL) divergence between the approximate posterior and the true Bayesian posterior. This KL divergence between the approximate posterior and the true Bayesian posterior is the loss function and is defined as follows [Eq. \ref{eq:Loss}]:
        
        
        \begin{equation} \label{eq:Loss}
            \begin{split}
                \theta^{opt} &= \text{arg } \displaystyle{\min_{\theta} \text{KL}[ q(w|\theta) || P(w|D) ] } \\
                             &= \text{arg } \displaystyle{\min_{\theta} \int_{}^{} q(w|\theta) \text{log} \frac{ q(w|\theta) }{ P(w)P(D|w) } \,dw } \\
                             &= \text{arg } \displaystyle{\min_{\theta} [\text{KL}[ q(w|\theta) || P(w) ] } - \mathbb{E}_{q(w|\theta)}[ \text{log} P(D|w)]]
            \end{split}
        \end{equation}
        
        The first term of the loss is the KL divergence between the approximate posterior and the prior. It is obvious that the prior is introducing a regularization effect because the KL divergence penalizes complexity by forcing the approximate posterior to be close to the prior. The second part is the negative log likelihood. This is a data  dependent term and it forces the network to fit the data.
        
        Here we consider the approximate posterior to be a fully factorised Gaussian \citep{NIPS2011_4329}. During one forward pass we sample the weights from the posterior but during back propagation we would have to define the gradient of the loss with respect to this sampling procedure, which is of course not possible. Instead we use the reparameterization trick \citep{kingma2014autoencoding}. This procedure is well described in \citep{blundell2015weight}. Instead of having a parameter-free operation (sampling) we can obtain the weights of the posterior by sampling a unit Gaussian shifting it by a mean $\mu$ and scaling it by a standard deviation $\sigma$. This standard deviation is parameterised as $\sigma = \text{log} (1+ \text{exp}(\rho))$ and thus it is always positive. So the weights are sampled according to the following scheme: $w = \mu + \text{log} (1+ \text{exp}(\rho))$, and the variational parameters to be optimised are $\theta = (\mu, \rho)$.
        \par
        
    \subsection{Adapting the Bayes by Backprop method to our case}
        
        The architecture of the BNN remains the same as the deterministic case but we replace all the convolutional and dense layers with the respective Bayesian layers. We use a Gaussian as predictive distribution,  $P(y|x, w)$, corresponding to a squared loss \citep{blundell2015weight}. We choose to have Gaussian distributions as priors which corresponds to L2 regularization \citep{blundell2015weight}. There is a single prior distribution for each layer, so all the weights in a layer share exactly the same $\mu$ and $\rho$. During backpropagation we optimise the prior by considering the gradients of loss not only with respect to the posterior but also the prior parameters. This allows us to change the prior hyperparameters of our model by using the available data, this process is known as empirical Bayes. In the results section we will demonstrate the advantages that empirical Bayes offers to our model. The mean prediction and the variance for the Bayesian CNN are calculated by passing the same input through the network multiple times.

    \subsection{Selective Learning}
        
        A selective learning process, in a supervised learning framework, assumes that a large pool of unlabelled data is available while there is a very expensive function that labels this data. The aim of selective learning is to identify which of the unlabelled data contain useful information so that only these are labelled. To achieve that, an acquisition function needs to be formulated to identify the useful data. \citep{Tsymbalov_2018} suggest that the uncertainty extracted from a Bayesian Neural Network is a sensible acquisition function for this task. This is also intuitively a sensible conclusion because high uncertainty in the prediction of the network means that the input is far away from the training data distribution.
        \par
        
        In our case the pool of unlabelled data is a large number of coarse scale FE simulations. The expensive function that labels these data is the fine scale FE simulations that take into account the micro features of the geometry. The aim of selective learning in our case is to identify the stress cases and/or microstructural patterns that are sufficiently different from the ones that are already in the labeled set and perform the fine scale FE simulations only to these cases. This is very important for our application because the cases that will be identified by the SL framework are exactly the cases that contain unknown interactions that our network would have predicted with a large error. For the acquisition function we need an uncertainty metric. Here we chose the max variance present in each prediction.
        
\section{Numerical Examples} \label{Numerical Examples}

    In this section we will present results for linear and nonlinear models both for deterministic and probabilistic CNNs. We will compare the CNN prediction with the FE prediction in the ROI level where we will be able to compare the 2 stress distributions but also we will compare the max values in all the ROIs in the test set. Additionally, we will demonstrate the use of selective learning and we will show that it can lead to a 50\% reduction in the labelled data requirement.
    
    \subsection{Linear elasticity} \label{linear elastic}
    
        \subsubsection{Training dataset}
    
\begin{figure}[h]
                \begin{center}
                    \includegraphics[width=0.5\linewidth]{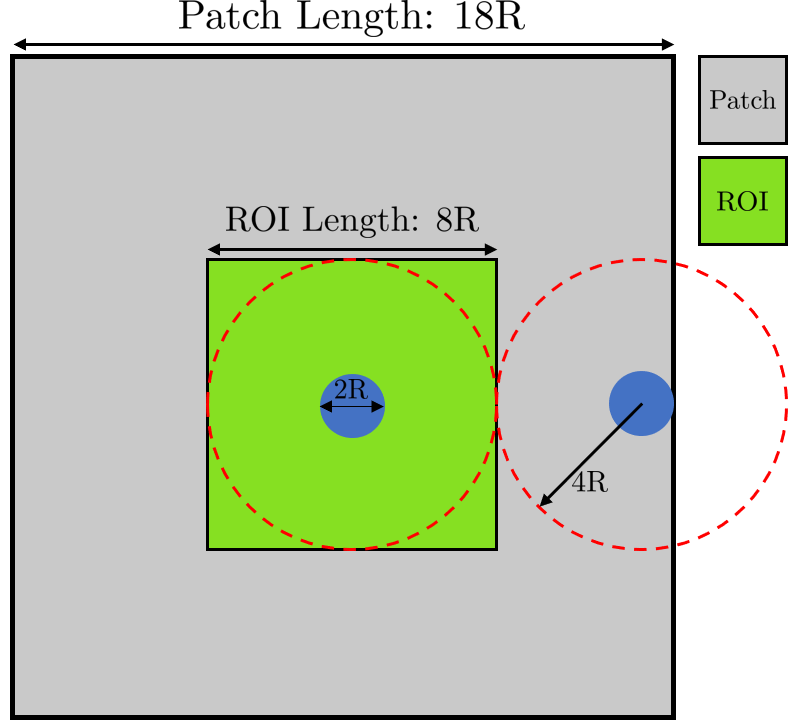}
                    \centering
                    \caption{A sketch of the patch. In grey we see the patch, in green the region of interest and in blue we can see disks of radius equal to that of the microscopic pores.}
                    \centering
                    \label{fig:ROI_4}
                \end{center}
            \end{figure}

            For the purpose of training our model we have assumed a distribution of elliptical pores as macroscale features. We consider all the microscale features as disks with the same radius, $R$. The Young's modulus and the Poisson ratio of the structure are 1 and 0.3 respectively. We consider the linear elastic case and we will discuss the non-linear elasticity in section [\ref{Non Linear Section}]. We assume that for a distance larger than 4 radii from the center of the microscale features the micro effect on the global stress field is negligible, for instance in the case of an infinite plate under uni-axial loading the max stress at $r=4R$ is 1.04 times the macro stress \citep{SIFs}. Based on that the micro feature length is $2R$, and the interaction length is equal to $3R$. Given those 2 parameters we conclude that the patch length should be $18R$ and the ROI should be a $[8R \; \times \; 8R]$ window in the middle of the patch as shown in [Fig  \ref{fig:ROI_4}]. We chose $R = 4$ so the input is of size $[72 \times 72 \times 4]$, the output is of size $[72 \times 72]$ and the ROI is of size $[32 \times 32]$.
    
\begin{figure}[h]
                \centering
                \begin{subfigure}{.5\textwidth}
                  \centering
                  \includegraphics[width=\linewidth]{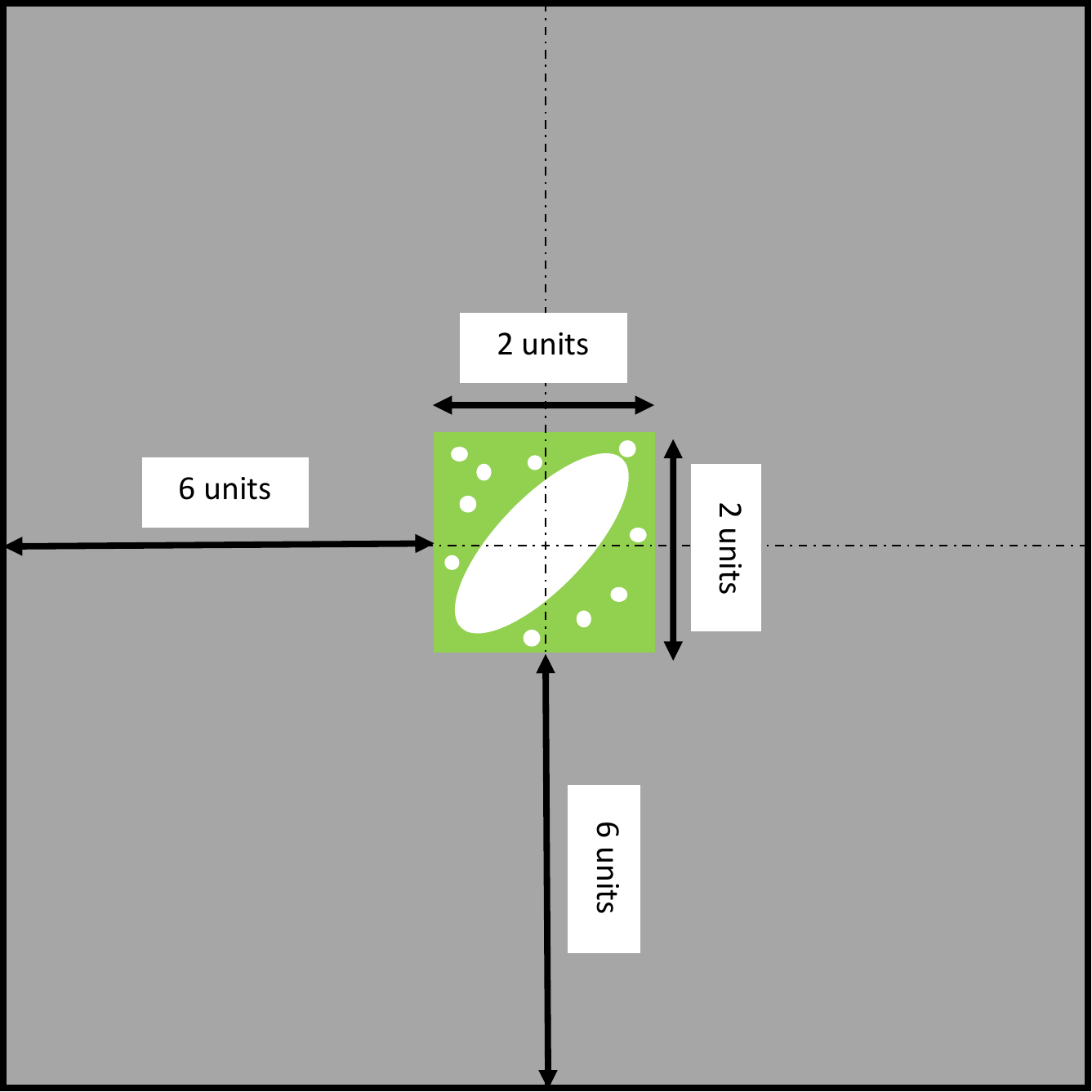}
                  \caption{}
                  \label{fig:buffer:a}
                \end{subfigure}%
                \hfill
                \begin{subfigure}{.37\textwidth}
                  \centering
                  \includegraphics[width=\linewidth]{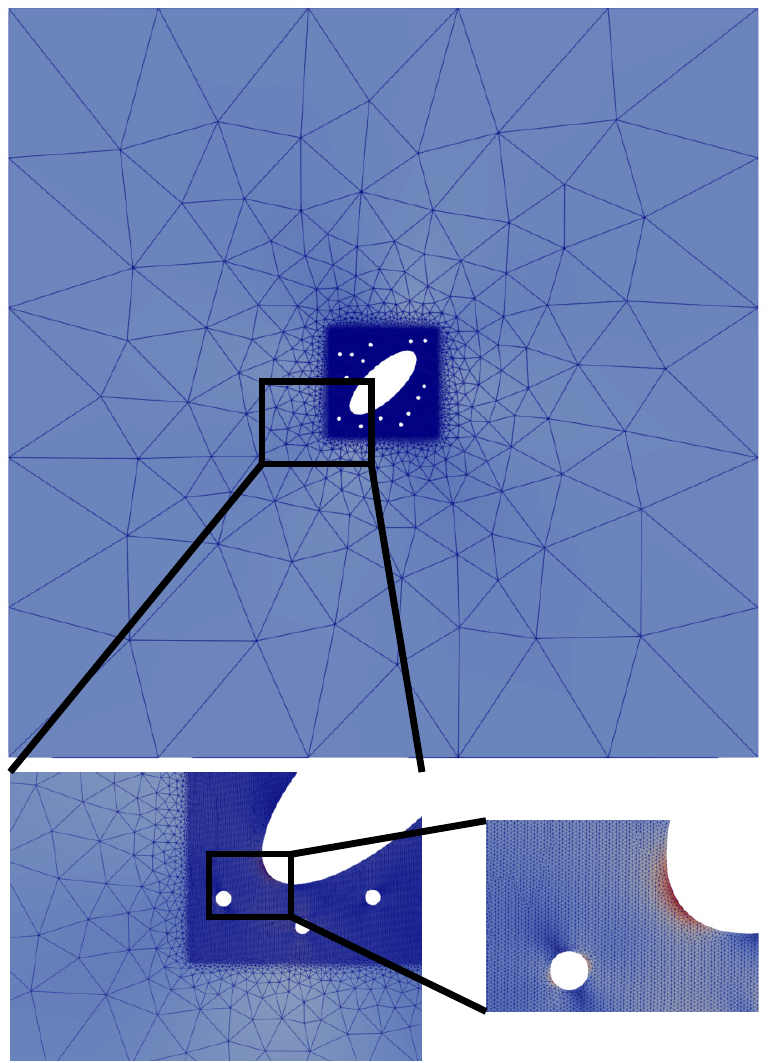}
                  \caption{}
                  \label{fig:buffer:b}
                \end{subfigure}
                \caption{On the left (a), a sketch of the buffer zone in grey and on the right (b) an example where the mesh and buffer area are visible.}
                \label{fig:buffer}
            \end{figure}

            In this example we want to study the interaction between elliptical macroscale features and spherical microscale pores in an infinite domain. In order to achieve this, the boundary conditions are applied to a buffer area where the mesh is much coarser, as can be seen in [Fig  \ref{fig:buffer}]. The buffer area allows us to apply boundary conditions without introducing boundary effects on the fine mesh area. Additionally, because the mesh in the buffer area is very coarse the computational cost remains practically the same. We apply displacement as boundary conditions [Eq.  \ref{eq:BCs}].
            
            \begin{align}
                u &= 
                \begin{bmatrix}
                    E_{xx} & E_{xy} \\
                    E_{xy} & E_{yy}
                \end{bmatrix}
                (X - X_0)^\top
            \label{eq:BCs}
            \end{align}
            
            where $E_{xx}$ is the far field displacement along the $x$ direction, $E_{xy}$ is the far field displacement along the $xy$ direction, $E_{yy}$ is the far field displacement along the $y$ direction, $X$ is the position of a point in $\mathbb{R}^2$ and $X_0$ is the initial position of the center of the body in $\mathbb{R}^2$.
    
        \subsubsection{Scaling}
        
            Differences in the scales across input variables may increase the difficulty to model the problem, for example increased difficulty for the optimizer to converge to a local minimum or unstable behaviour of the network, thus a standard practice is to pre-process the input data usually with a simple linear rescaling \citep{Bishop}. In our case we will scale the data not only to improve the model but also to restrict the space we have to explore. The space that we have to cover is infinite because the input can take any real value. Fortunately, Tresca stress scales linearly with the components of the stress tensor. From [Eq. \ref{eq:principal_stresses}] it is trivial to show that if we replace $\sigma_{xx}, \; \tau_{xy}, \; \sigma_{yy}$ to $k \cdot \sigma_{xx}, \; k \cdot \tau_{xy}, \; k \cdot \sigma_{yy}$ where $k$ is some scaling factor, then $\sigma_{\text{max}}' = k \cdot \sigma_{\text{max}}$ and $\sigma_{\text{min}}' = k \cdot \sigma_{\text{min}}$ where $\sigma_{\text{max}}'$ and $\sigma_{\text{min}}'$ are the new principal stresses after scaling the input and thus $\text{Tresca}'$ = ${\frac{1}{2}}(\sigma_{\text{max}}'-\sigma_{\text{min}}')$ = ${\frac{1}{2}}(k \sigma_{\text{max}}- k\sigma_{\text{min}})$ = $k \cdot \text{Tresca}$ where $\text{Tresca}'$ is the Tresca stress after scaling. Additionally, because we model linear elastic problems scaling the load terms with a scaling factor $k$ will result in a local and global stress field multiplied by $k$ as well. Here $k^{-1}$ is the maximum stress value present in all the 3 stress components over the patch. This scaling of the input values to the range $[0, 1]$ allows us to make predictions on input data of any possible scale. We just have to calculate $k$, multiply the input by it to transfer it to the desired scale and then multiply the output with $k^{-1}$ to get the true output.

        \subsubsection{Numerical Example with 1 Ellipse} \label{Initial Dataset}
    
\begin{figure}[!h]
                \begin{center}
                    \includegraphics[width=0.5\linewidth]{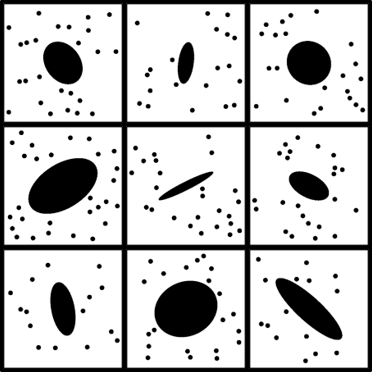}
                    \caption{Nine examples from the "1 ellipse" dataset}
                    \centering
                    \label{fig:Geom_0}
                \end{center}
            \end{figure}

            Firstly we created an initial dataset with very simple examples  [Fig  \ref{fig:Geom_0}]. A single ellipse in the middle playing the role of the macroscale feature, creating a diverse macroscopic stress field. Also, a few micro features are randomly positioned around the ellipse, accounting for the micro scale features that will affect the macro stress field. All the micro features have a circular shape and the same radius, $R = 4$ units. From 500 examples, generated in 43 hours on an Intel\textsuperscript{\textregistered} Core\textsuperscript{\texttrademark} i7-6820HQ CPU, we extracted 33,000 patches, 5,000 of which where used as a validation set. No rotations were applied to this dataset as data augmentation technique. The patches were extracted such that the union of all the ROIs is equal to the entire domain $\Omega$. Specifically, the ROI of the first patch is aligned with the top right corner of the domain and the rest of the patches are created using a sliding window and a stride equal to half the length of the ROI in each dimension. The patches that do not contain any micro or macro scale features are discarded. 
            \par
            
            Experiments on this dataset showed very positive results. Training with 28,000 patches and validating on 5,000 unseen patches resulted in a validation accuracy of 96\% when training with the Adam optimizer for 600 epochs, which required about 6 hours on a NVIDIA T4 GPU. The loss function that we used is the mean squared error applied only in the ROI as already discussed. The concept of accuracy in a regression task with images needs to be discussed. The process followed to define accuracy is summarised in [Algorithm \ref{alg:acc}] and it is described with more detail as follows. 
            
            \begin{enumerate}
            
              \item We take the max of each prediction in the ROI, this is because we are primarily interested in the max values as these values will indicate if the material will fail or not
              
              \item We define an error metric between the real max value, $y_{\text{FE}}$, and the max value in our prediction, $y_{\text{NN}}$. Here we use the relative error defined as:   $ relative\_error = |y_{\text{NN}} - y_{\text{FE}}| / y_{\text{FE}}$
              
              
              \item We set a threshold for the acceptable error. In this case we will use 10\%. To sum up, 96\% validation accuracy means that in the validation set 96\% of the max values in the ROI were predicted with a relative error less than 10\%
              
            \end{enumerate}
            
            \begin{algorithm}
            \caption{Calculate accuracy}\label{alg:acc}
            \begin{algorithmic}[1]
            \State $N = \textbf{length}(datapoints)$
            \State $accepted = \textbf{zeros}(N)$
            \For{$point \textbf{ in } datapoints$}\
                \State $y_{NN} = \textbf{max}(prediction[point])$\
                \State $y_{FE} = \textbf{max}(ground\_truth[point])$\
                \State $error$ = $|y_{NN}-y_{FE}|/y_{FE}$\
                \If {$error \leq threshold$}\
                    \State $accepted[point] = 1$
                \EndIf
            \EndFor\label{accendfor}
            \State $accuracy$ = $\textbf{sum}(accepted)/N$
            \State \textbf{return} $accuracy$
            \end{algorithmic}
            \end{algorithm}
            
\begin{figure}[!h]
                \begin{center}
                    \includegraphics[width=0.7\linewidth]{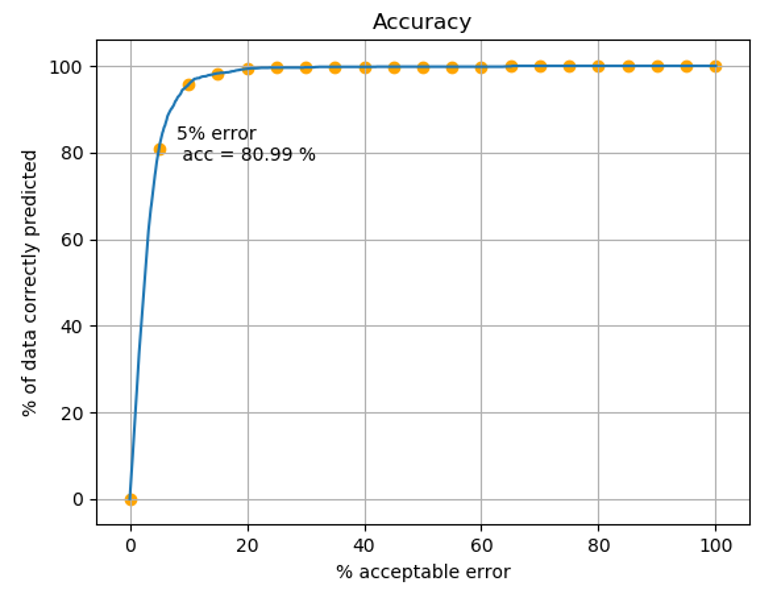}
                    \caption{Accuracy as function of the threshold. Here, accuracy is defined as the percentage of patches in the dataset for which the relative error between the max NN prediction and the max FE result in the ROI is less than a predefined threshold.}
                    \centering
                    \label{fig:thrs}
                \end{center}
            \end{figure}

            This 10\% threshold is arbitrarily chosen and it should be more application specific because different applications have different error requirements. We have constructed a diagram that shows the accuracy as a function of the threshold [Fig \ref{fig:thrs}]. 
            
\begin{figure}[h]
                \centering
                \begin{subfigure}{.5\textwidth}
                  \centering
                  \includegraphics[width=\linewidth]{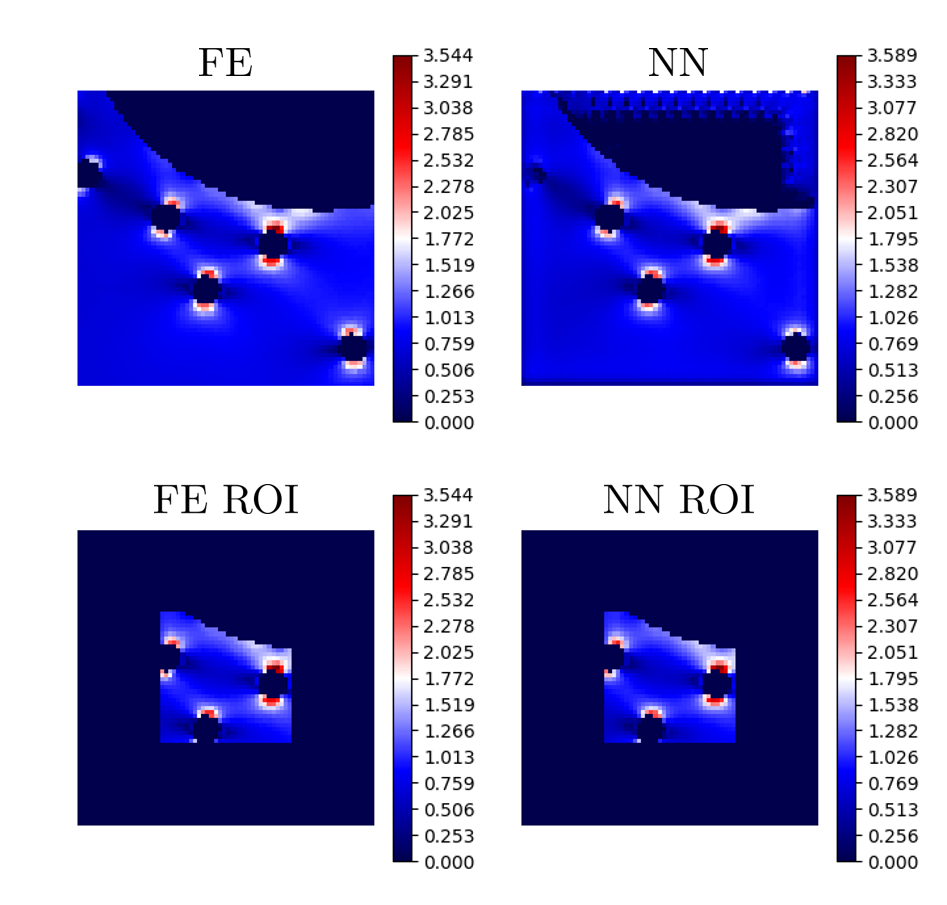}
                  \caption{}
                  \label{fig:Exmp_0:a}
                \end{subfigure}%
                \begin{subfigure}{.5\textwidth}
                  \centering
                  \includegraphics[width=\linewidth]{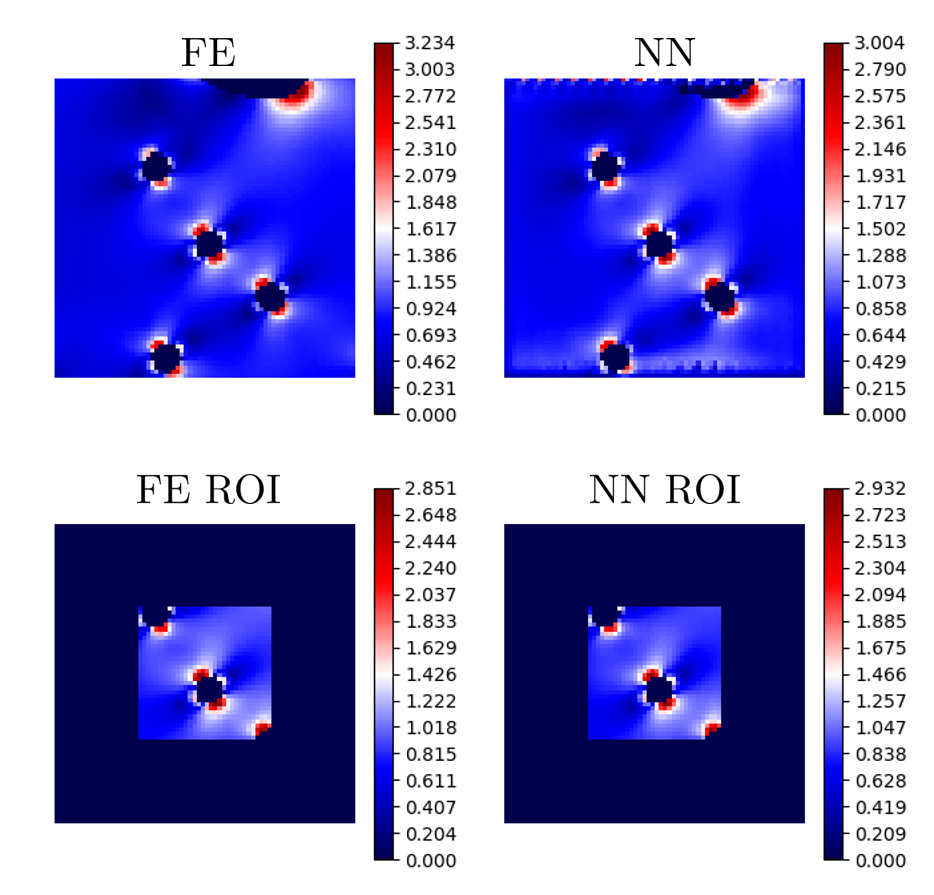}
                  \caption{}
                  \label{fig:Exmp_0:b}
                \end{subfigure}
                \caption{ Evaluation of the Neural Network performance on 2 patches. On the top left of each example we see the scaled Tresca stress field computed by FEA and converted into an image for the whole patch and on the top right the NN prediction for the whole patch. On the second row we see the same but for the ROI.}
                \label{fig:Exmp_0}
            \end{figure}

\begin{figure}[!h]
                \begin{center}
                    \includegraphics[width=.95\linewidth]{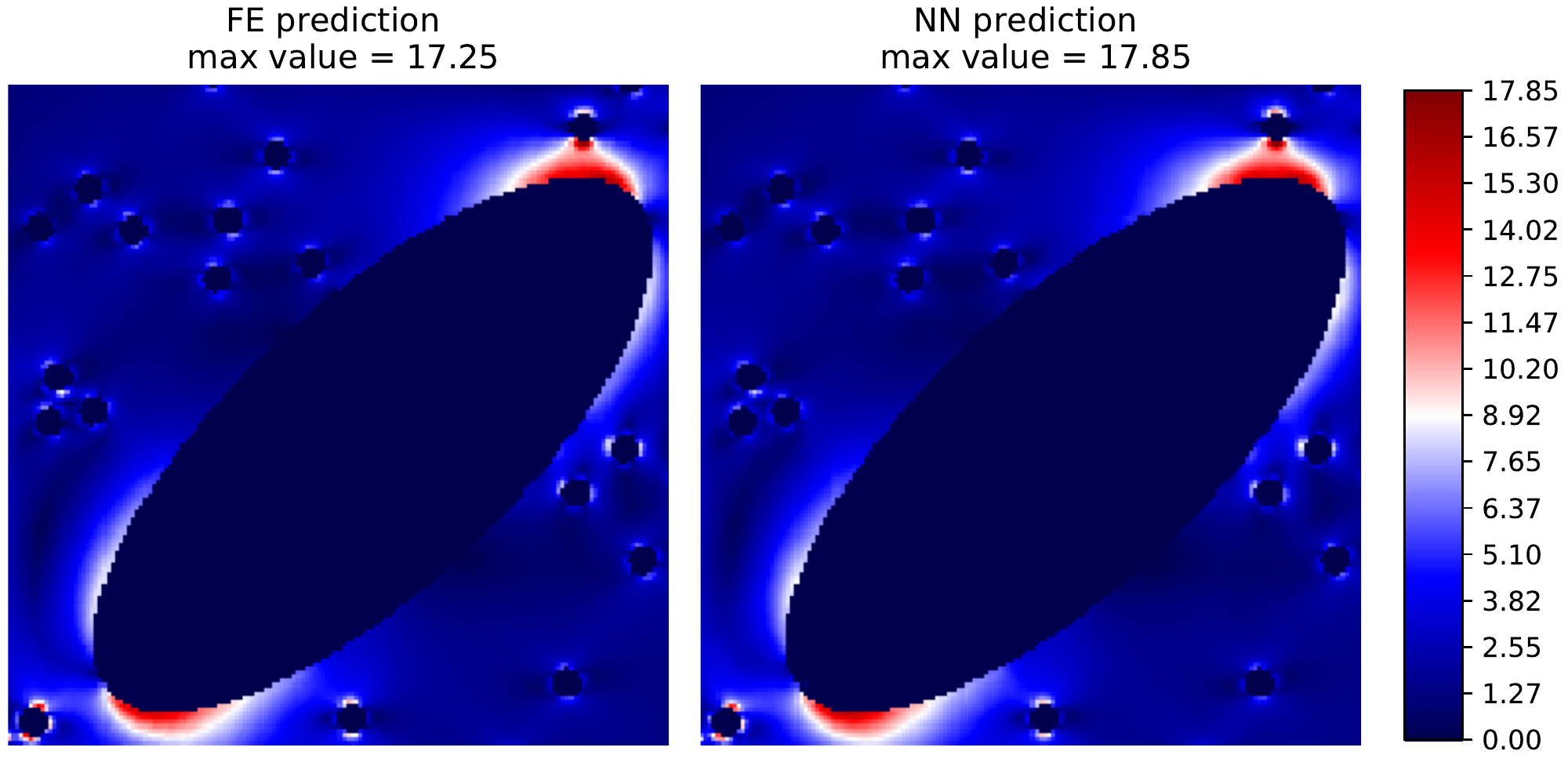}
                    \caption{Comparison between the Tresca stress field computed by FEA and converted into an image, on the left, and an image reconstructed using the NN predictions on the patch level, on the right.}
                    \centering
                    \label{fig:Exmp_whole}
                \end{center}
            \end{figure}

\begin{figure}[!h]
                \begin{center}
                    \includegraphics[width=\linewidth]{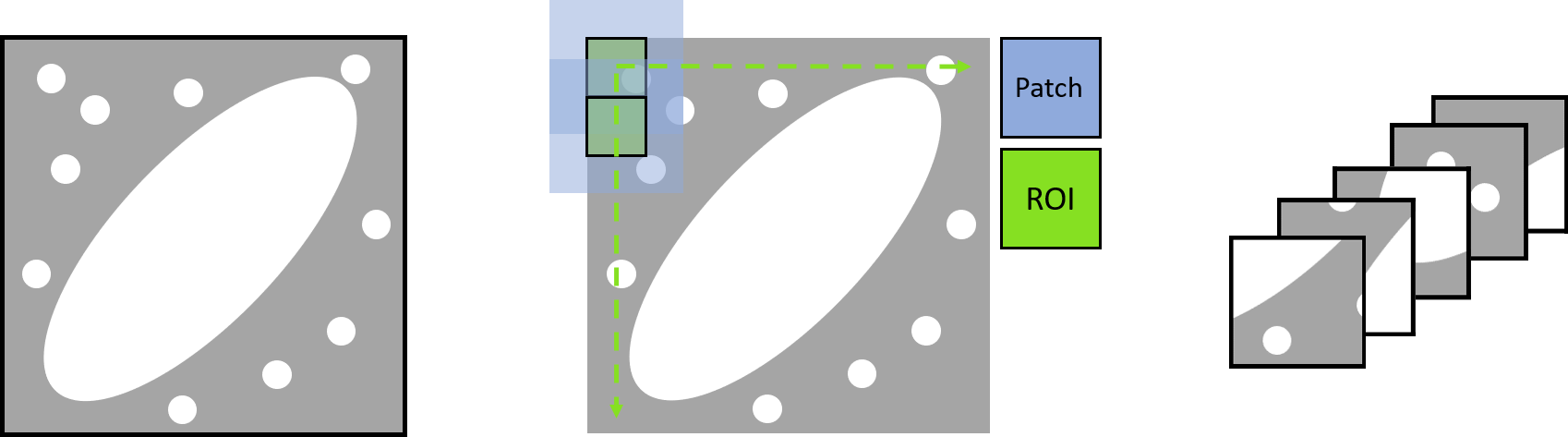}
                    \caption{Patch generation for full image prediction. The corner of the ROI is aligned to the corner of the image and then a sliding window of size equal to the size of the ROI is used.}
                    \centering
                    \label{fig:patch_full}
                \end{center}
            \end{figure}

            We present results from 2 random patches [Fig \ref{fig:Exmp_0}] and then a result on the whole structure [Fig \ref{fig:Exmp_whole}]. For the whole image the true Tresca is displayed and not the scaled version of it. The prediction is made again on the patch level but then the original image is reconstructed. This is possible if we align one corner of the ROI with a corner of the image and use a sliding window equal to the size of the ROI as can be seen in [Fig \ref{fig:patch_full}]. We can see that in all cases the CNN was able to accurately reconstruct the full micro stress field but it was also able to predict the max values with a very small error. More specifically it is clear that away from the micro scale features the micro scale field is constant. We can also see that very close to the micro scale features we have a very steep rise of the micro Tresca stress. The micro stress field is accurately predicted even in complicated cases where more than one micro features are interacting or micro features and macro scale features are interacting.
            
\begin{figure}[h]
                \centering
                \begin{subfigure}{.5\textwidth}
                  \centering
                  \includegraphics[width=\linewidth]{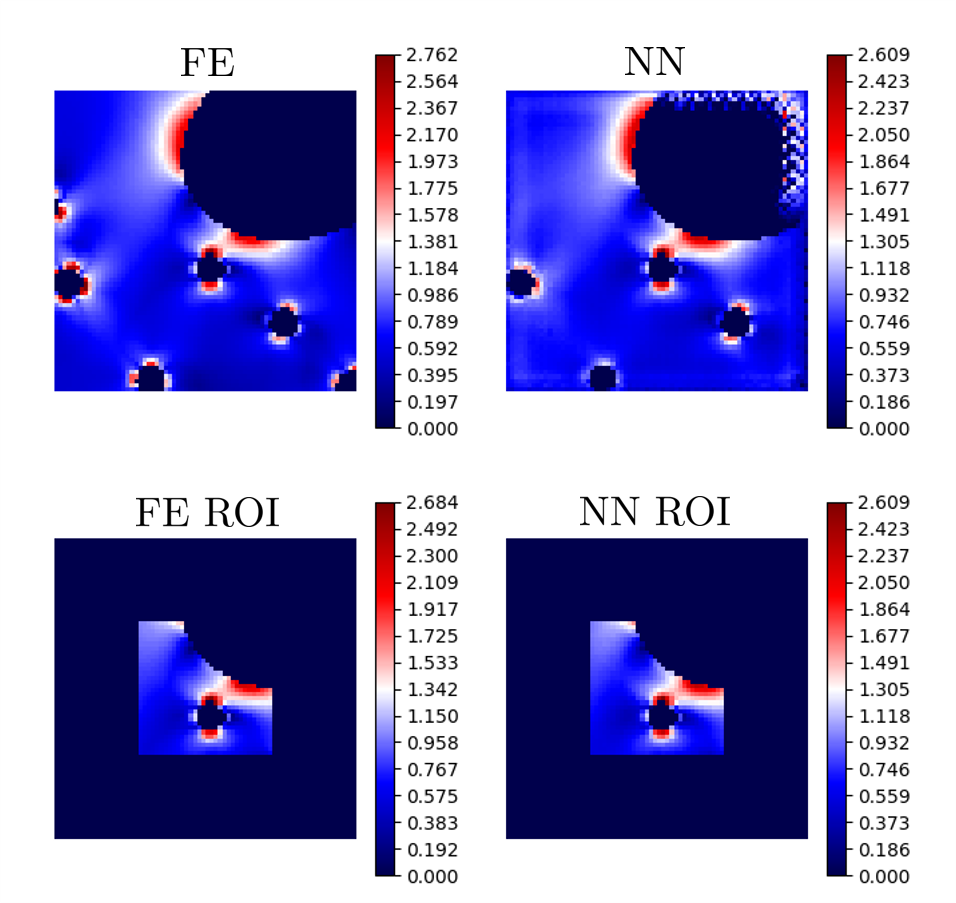}
                  \caption{}
                  \label{fig:BigSmall:a}
                \end{subfigure}%
                \begin{subfigure}{.5\textwidth}
                  \centering
                  \includegraphics[width=\linewidth]{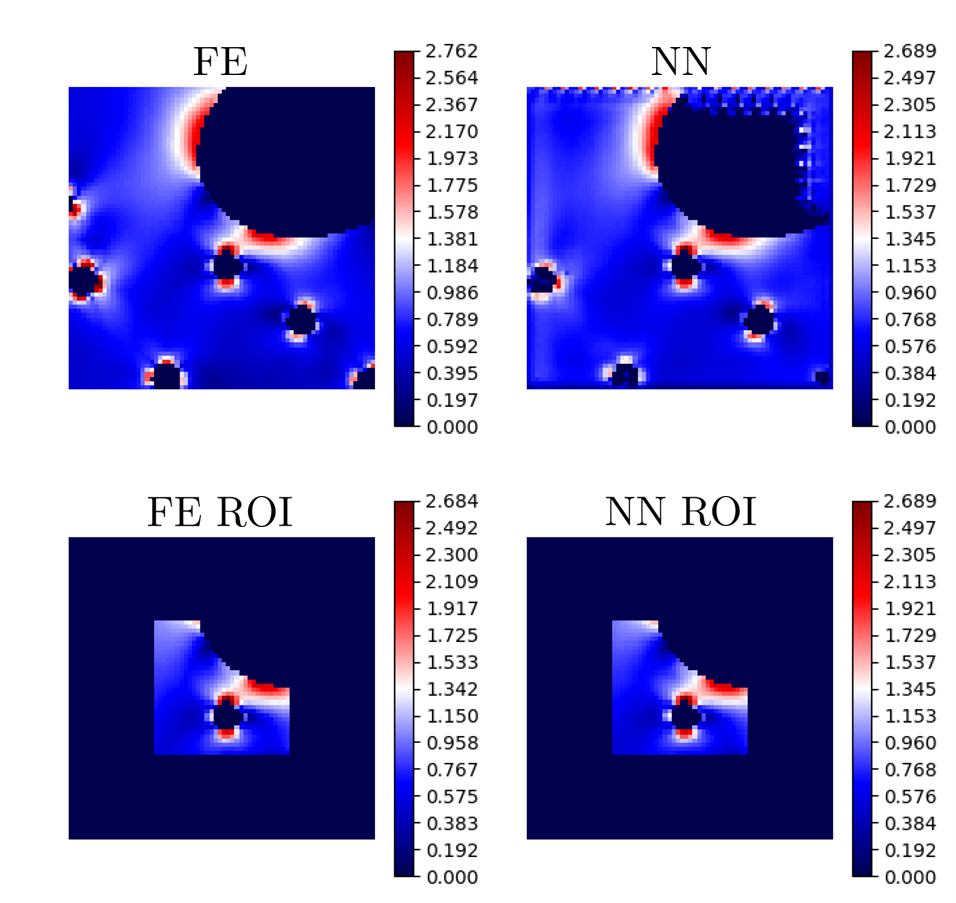}
                  \caption{}
                  \label{fig:BigSmall:b}
                \end{subfigure}
                \caption{A prediction for the same input for 2 identical NNs trained with 10,000 patches (a) and 28,000 (b). The two ROI predictions look very similar but we can observe that the maximum predicted value from the NN that was trained with the larger dataset is closer to the real value compared to the maximum predicted value from the NN that was trained with the smaller dataset.}
                \label{fig:BigSmall}
            \end{figure}

            Lastly, we will investigate the effect of training with less data on the CNN accuracy. We randomly chose 10,000 patches, almost 30\% of the available data, and train the NN with exactly the same settings, an example can be found in [Fig \ref{fig:BigSmall}]. We noticed that for the 10\% threshold the accuracy is 6\% higher for the large dataset even though we used almost 3 times as much data. This may look like a small increase but it means that the mispredicted cases climbed from 4\% to 10\%. In reality most of the data that we rejected when we created the smaller training dataset (10,000 patches) were very similar to the ones accepted in the smaller dataset. A small portion of them contained new interactions that our network would have learned from. These cases are exactly the cases we are interested in because they contain complex examples that create strong interactions and sharp increase in the micro stress field. In section [\ref{Selective Learning}] we will demonstrate a Selective Learning framework that will allow us to identify these cases and train our network only on them. This way we keep the computational cost to the minimum while preserving the same level of accuracy.
    
        \subsubsection{Numerical Example with 3 Ellipses} \label{Advanced Dataset}
        
           Even though the CNN we trained seems to work well for the data it was trained on we do not expect the same level of accuracy as we depart from this dataset, although the method is fully non-parametric and the trained NN can make prediction for any unseen micro and macro geometries. Specifically, we would expect a decrease in accuracy in the following cases.
           
           \begin{enumerate}
            
              \item Spatially fast varying macro stress field, generated by macroscale features not present in the training dataset
              
              \item Microscale features not present in the training dataset, for instance non circular holes 
              
              \item Patterns of microscale features not present in the training dataset, for instance different distribution of circular holes
              
            \end{enumerate}

\begin{figure}[h]
                \begin{center}
                    \includegraphics[width=0.5\linewidth]{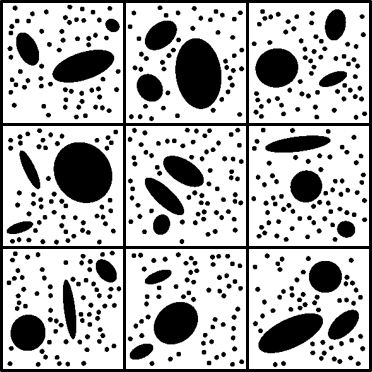}
                    \caption{Nine random examples from the "3 ellipses" dataset}
                    \centering
                    \label{fig:Dataset_2}
                \end{center}
            \end{figure}

           To tackle this problem we created a new, more interesting, family of data with the expectation that this would add more complexity [Fig \ref{fig:Dataset_2}]. At first, we used the old CNN to make predictions on the new dataset. We observed that the accuracy dropped from 96\% to 72\%. This implies two things. Firstly, the drop in accuracy means that the new dataset contains information that the network had never seen before or was unable to learn from (due to the sparsity of the examples), thus training in this dataset will help the CNN to generalise better. Lastly, the concept of making the knowledge transferable seems to be working as we were able to make reasonable, but not perfect, predictions on a new family of data. This suggests that we managed to learn interactions between micro scale features and the macro stress field and not just the structures themselves.
        
            Training a CNN with the new dataset proved to be more challenging. By using 23,000 patches as training set (almost as many as with the original case) and 5,000 patches as a validation set we obtained, with the same settings, a validation accuracy of 74\% in contrast to the 96\% in the first case. We believe that this happens not only because more micro scale features are present in each case but also because the 3 ellipses are creating a much more complicated macro stress field. From experiments we found out that, as more and more new patches are added, the accuracy tends to increase slower and slower. This happens because the new patches added tend to contain less and less new information. 
            
            When we tried to use this CNN to make predictions on the 1 Ellipse dataset the accuracy slightly improved, compared to the CNN trained with the 1 Ellipse dataset, from 96\% to 96.7\%. This small increase was expected since this CNN is trained with a dataset that contains all the necessary information to make predictions on the 1 Ellipse dataset and even more information that may or may not be useful. The very small increase in accuracy implies that the miss predicted cases are underrepresented in both datasets.
            
            A common technique used to improve the performance of CNNs is data augmentation. Common data augmentation techniques for image data are shifting, flipping, rotating and zooming. Here we use rotation as data augmentation technique. We rotate mechanically the stress tensor [Eq. \ref{eq:transform}] and “physically” the images. 

\begin{figure}[h]
                \begin{center}
                    \includegraphics[scale=0.5]{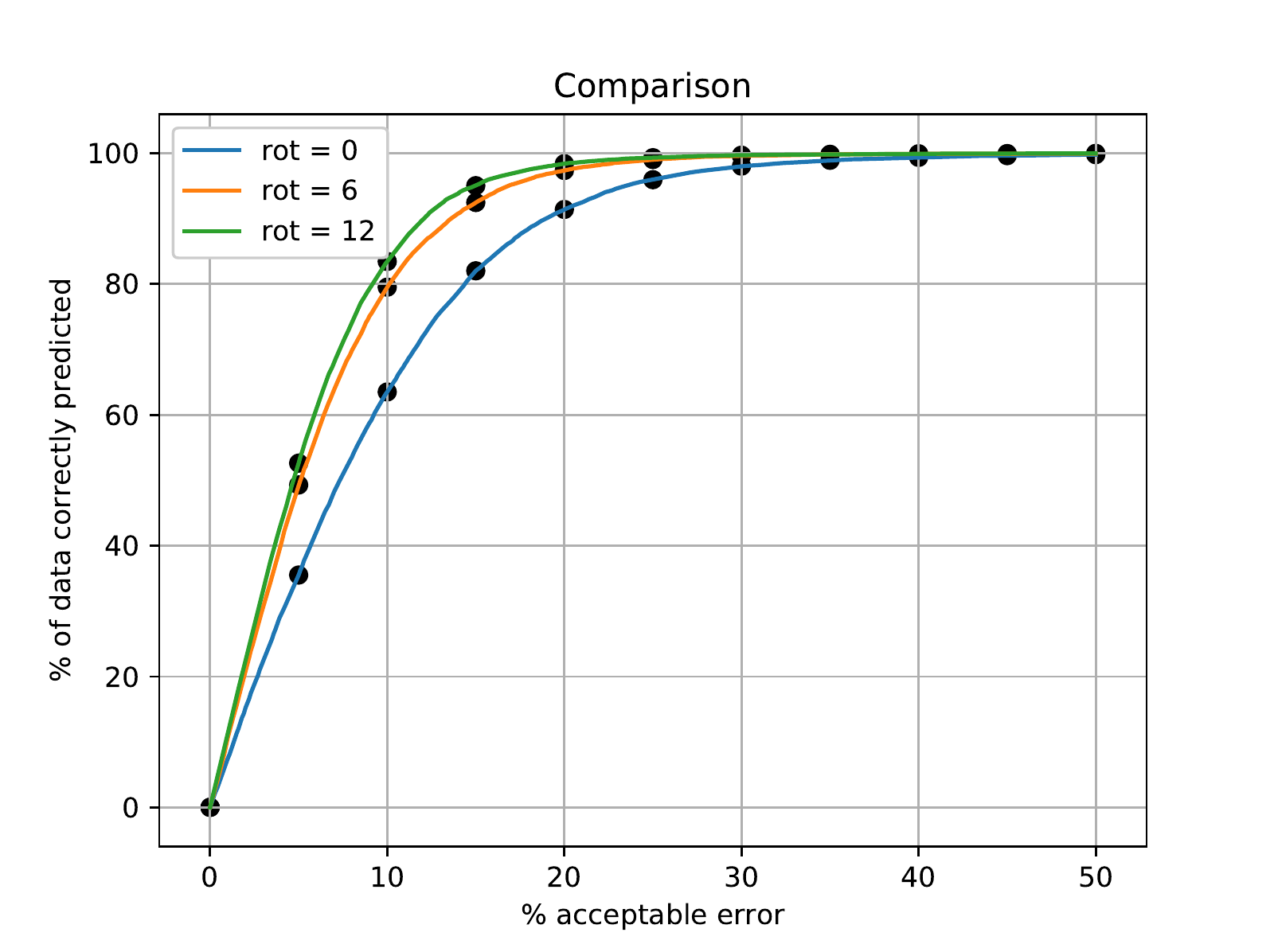}
                    \caption{Comparison between 3 CNNs trained with the same settings but different datasets. Blue line corresponds to the original dataset with no rotation, orange line to a dataset with 6 rotations and finally the green line to a dataset with 12 rotations. We can observe that the accuracy increases as the number of rotations increases.}
                    \centering
                    \label{fig:comp_1_6_12}
                \end{center}
            \end{figure}

\begin{figure}[h]
                \begin{center}
                    \includegraphics[width=\linewidth]{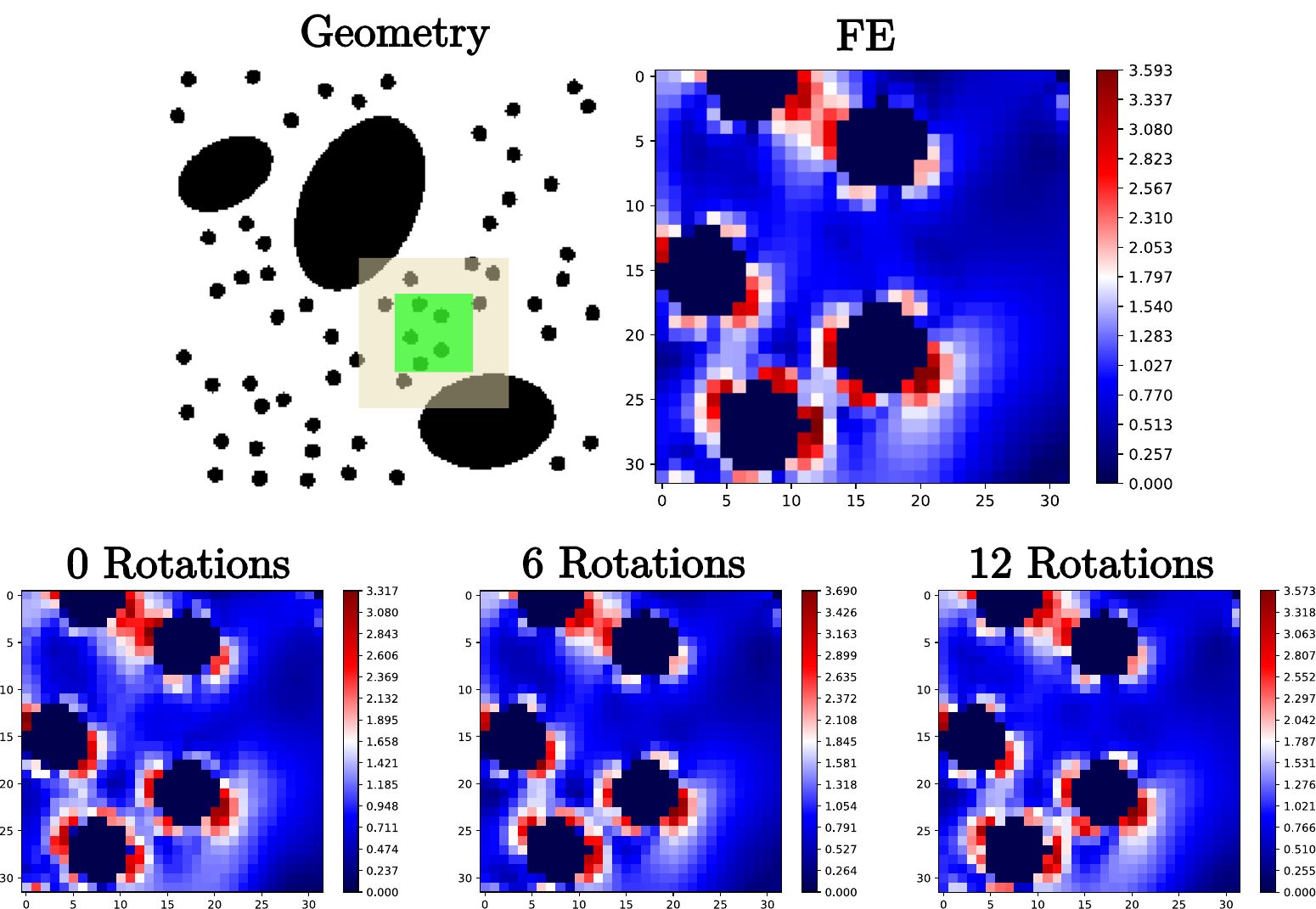}
                    \caption{Top left corner the structure, the patch (with light brown) and the ROI (with green) for the prediction. Top right corner the scaled Tresca stress field in the ROI computed by FEA and converted into an image. Bottom from left to right, prediction in the ROI from a NN trained with a dataset with 0, 6 and 12 rotations respectively. We observe that even though those 3 images look quantitatively very similar the predicted maximum value approaches the one calculated by the FE simulation as the number of rotations increases.}
                    \centering
                    \label{fig:comp_1_6_12_image}
                \end{center}
            \end{figure}

\begin{figure}[ht] 
              \begin{subfigure}[b]{0.5\linewidth}
                \centering
                \includegraphics[width=0.75\linewidth]{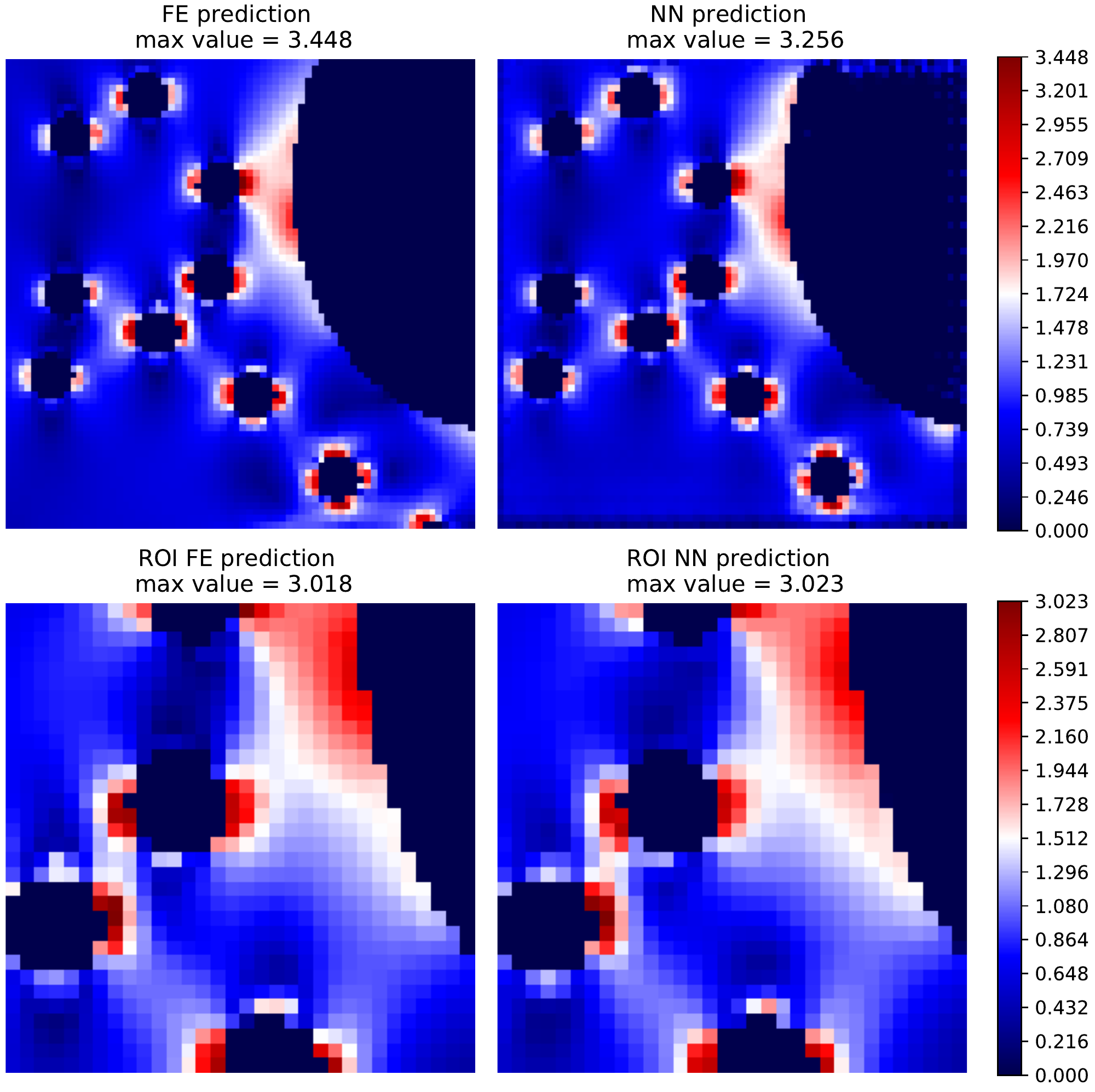} 
                \caption{} 
                \label{fig:4Exmpls:a} 
                \vspace{4ex}
              \end{subfigure}
              \begin{subfigure}[b]{0.5\linewidth}
                \centering
                \includegraphics[width=0.75\linewidth]{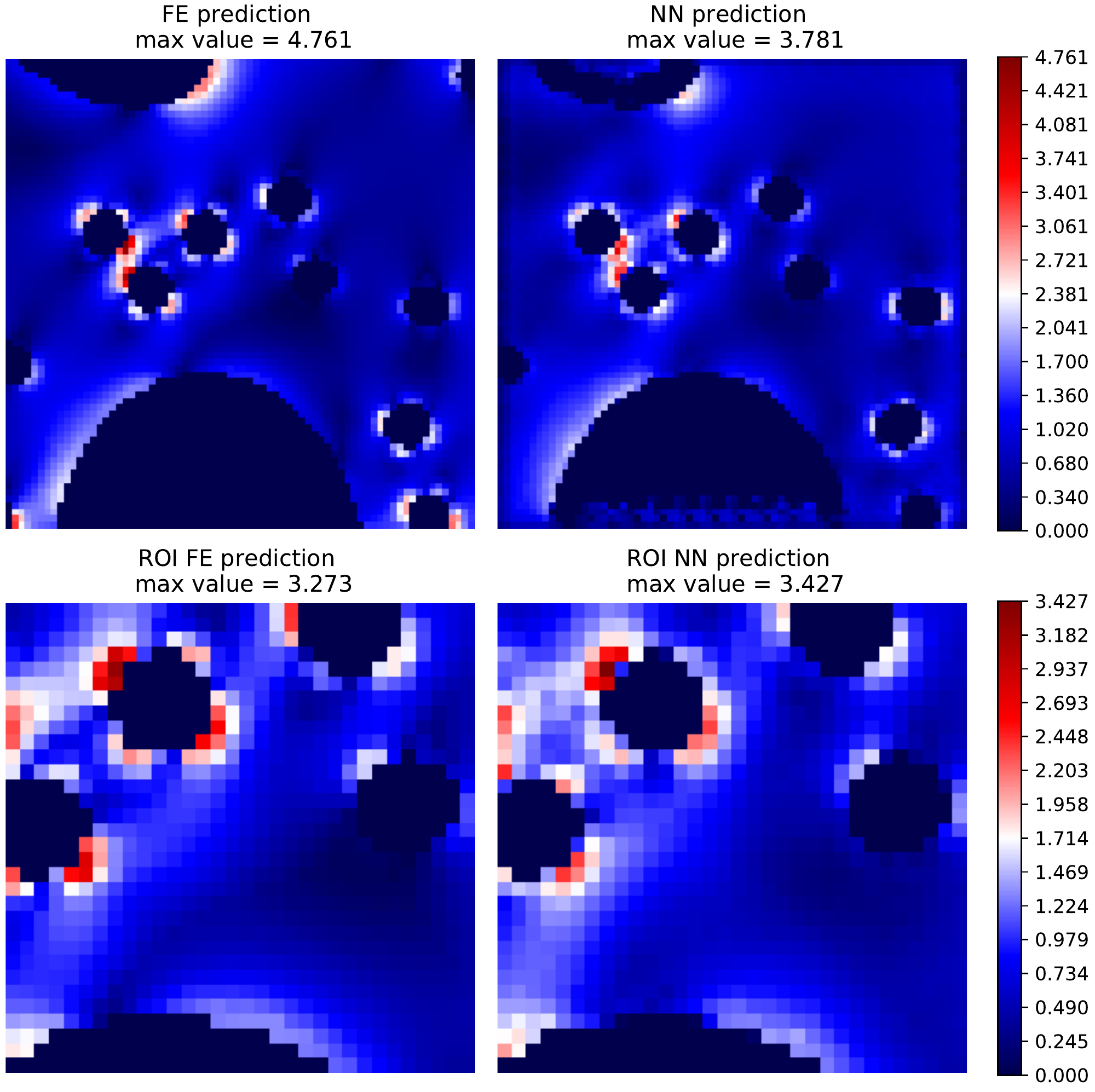} 
                \caption{} 
                \label{fig:4Exmpls:b} 
                \vspace{4ex}
              \end{subfigure} 
              \begin{subfigure}[b]{0.5\linewidth}
                \centering
                \includegraphics[width=0.75\linewidth]{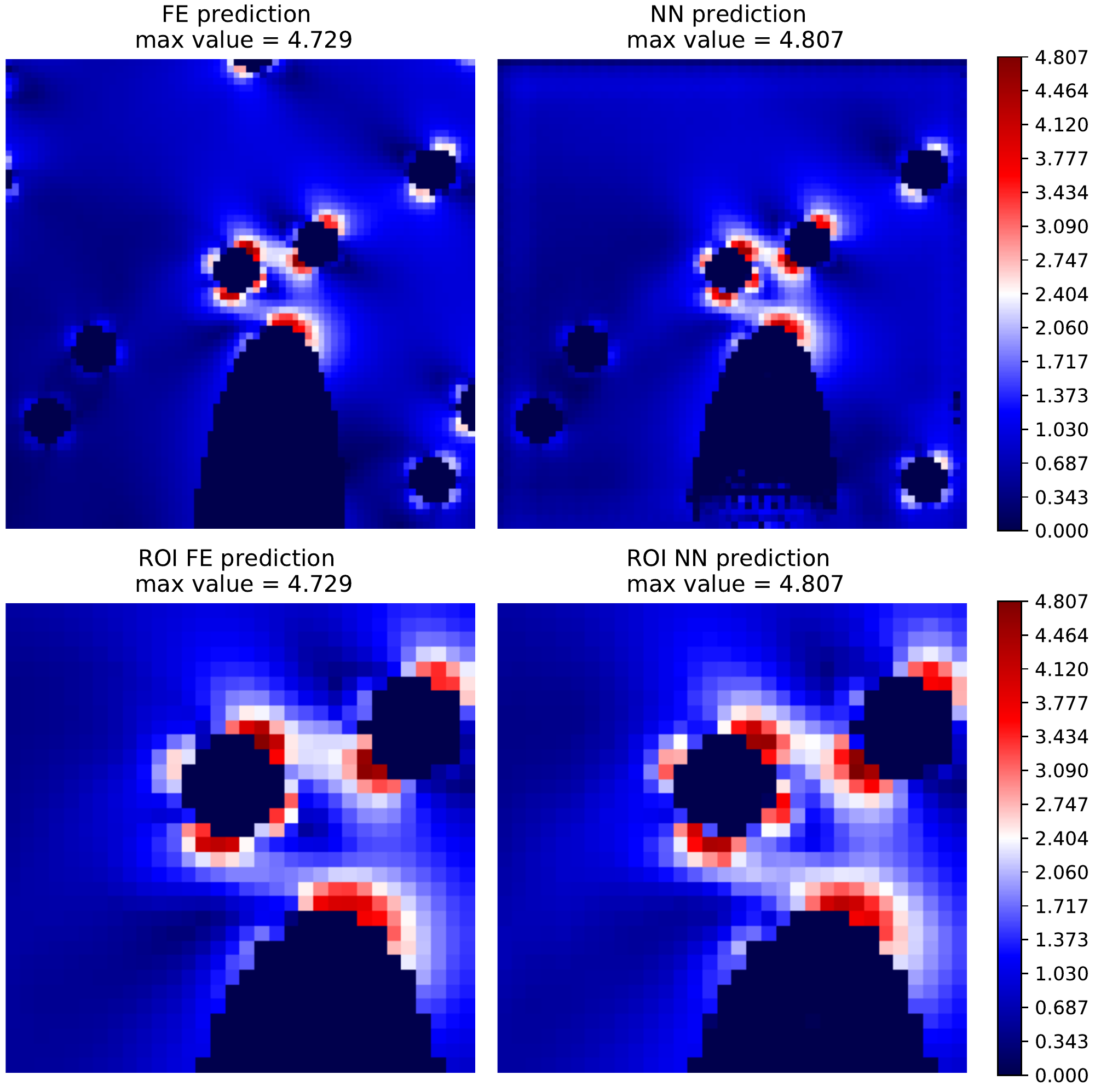}
                \caption{} 
                \label{fig:4Exmpls:c} 
              \end{subfigure}
              \begin{subfigure}[b]{0.5\linewidth}
                \centering
                \includegraphics[width=0.75\linewidth]{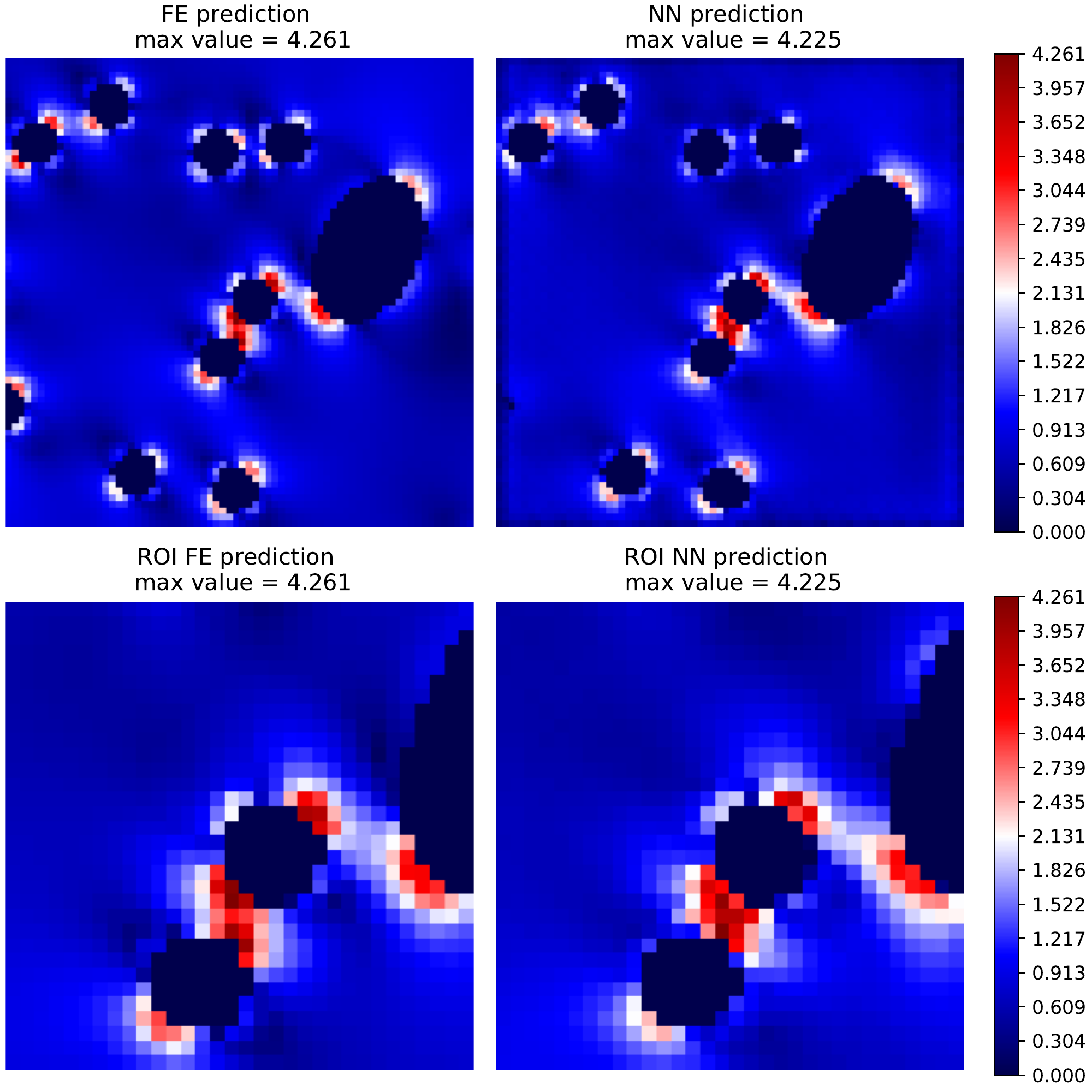}
                \caption{} 
                \label{fig:4Exmpls:d} 
              \end{subfigure} 
              
              \caption{Evaluation of the performance of the CNN trained on the 6 rotation dataset on 4 patches. In each of the 4 images, on the first row we can see the scaled Tresca stress field for the entire patch computed by FEA and converted into an image on the left and the NN prediction for the entire patch on the right. On the second row we can see the scaled Tresca stress field for the ROI computed by FEA and converted into an image on the left and the NN prediction for the ROI on the right. In all four cases we have strong interactions between micro scale features and in images (a), (c) and (d) we have strong interactions between micro and macro scale features. We can see that in all the cases the prediction in the ROI is qualitatively very similar to the FE simulation but also that the prediction for the maximum value in the ROI is very close to maximum value in the ROI calculated by FE simulations.}
              \label{fig:4Exmpls} 
            \end{figure}

            We started from an initial training set of 5,000 patches ($\approx$ 1/4 of the full set) and we rotated the dataset 6 and 12 times. After training with the same settings for all the cases, a validation accuracy of 62\%, 80\% and 82\% was achieved for the 0, 6, 12 rotations dataset respectively for the 10\% threshold [Fig \ref{fig:comp_1_6_12}]. Firstly, this means that we managed to outperform by 8\% the full dataset and secondly, we realized that rotating from 6 to 12 times didn't add a significant amount of new information even though the data are doubled. Once more, that was the motivation to start working with Selective Learning. We can see an example of a prediction with all 3 CNNs on the same input [Fig \ref{fig:comp_1_6_12_image}], where the prediction improves with the number of rotations. We can also see a prediction of the CNN trained with 6 rotations on 4 random patches [Fig \ref{fig:4Exmpls}].
          
            Moreover, we compare 2 CNNs with and without SE blocks. The 2 CNNs were trained with 27,000 training examples and validated on 3,000 validation examples. The CNN with the SE block reported accuracy of 78.98\% while the CNN without the SE Block 68.39\%. This clearly shows that adding the SE block in the Residual Blocks of the CNN substantially improves the performance.

\begin{figure}[h]
                \begin{center}
                    \includegraphics[width=\linewidth]{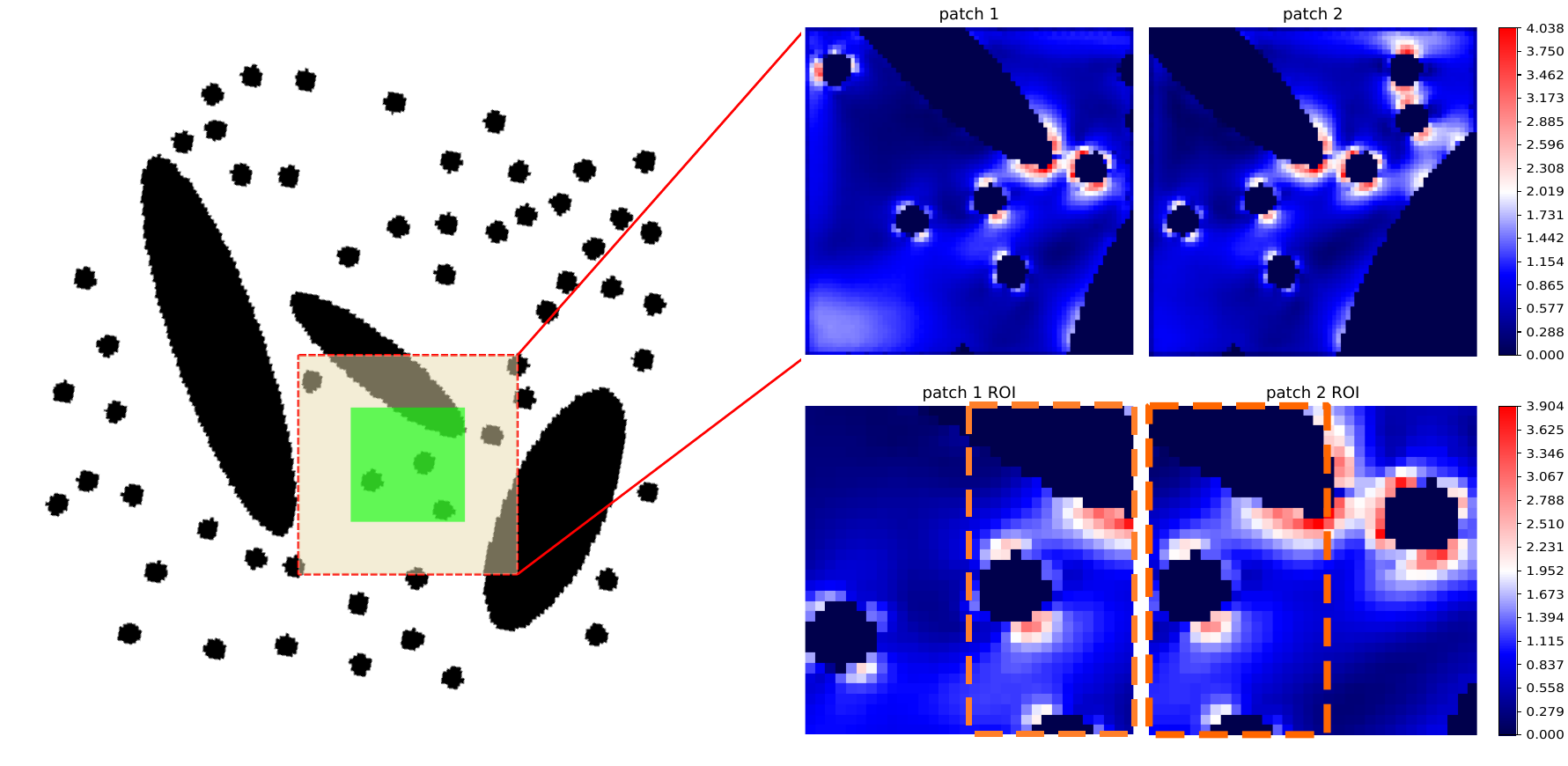}
                    \caption{Prediction of the CNN in two patches with overlapping ROIs. On the left we see the geometry where we solve the FE problem on and we can also see the patch and the ROI. For clarity we only show 1 of the 2 patches. The second patch will be created by sliding to the right a sliding window with size half the size of the ROI. On the right we see 4 plots, the 2 top plots are the CNN prediction on the entire patch and the 2 bottom plots the CNN prediction in the ROI. The orange discontinues boxes correspond to the common area of the 2 ROIs. The CNN prediction in both boxes is the same.}
                    \centering
                    \label{fig:overlapping}
                \end{center}
            \end{figure}

            Additionally, we investigate the smoothness of the solution. We compare the CNN prediction in 2 ROIs that share a common area. In [Fig \ref{fig:overlapping}] we have highlighted with orange discontinues boxes the CNN prediction in the common area of the 2 ROIs and it is very clear that the prediction is the same in both of them.

            Lastly, we perform a cross-validation study to confirm that the accuracy is not dependent on our choice of test dataset. Specifically, we used a dataset of 30,000 patches and we divided it in 5 subsamples of size 6,000 patches. We run 5 tests. Each time we used 1 of these 5 subsamples as a validation set and the rest as training test. The mean accuracy for the validation set is 0.7813 (78.13\%) and the standard deviation is 0.0174. The small value of the standard deviation implies that the CNN is stable and gives consistent results independent of the choice of test set, as long as the test set is big enough.

        \subsubsection{Numerical Example using a Bayesian Neural Network}
    
            Until now we have used a deterministic neural network for the predictions. In this section we will present results corresponding to the use of the Bayesian NN. We trained the BNN with the same 5,000 patches as in section [\ref{Advanced Dataset}] for 600 epochs and validated on 10,000 patches. That requires 2.1 times more computational time compared to the deterministic case. The accuracy of the prediction is 72\% for the 10\% threshold compared to 62\% for the deterministic case. In the Bayesian CNN case the accuracy is calculated using the MAP solution.
            \par
            
\begin{figure}[h]
                \centering
                \begin{subfigure}{.5\textwidth}
                  \centering
                  \includegraphics[width=\linewidth]{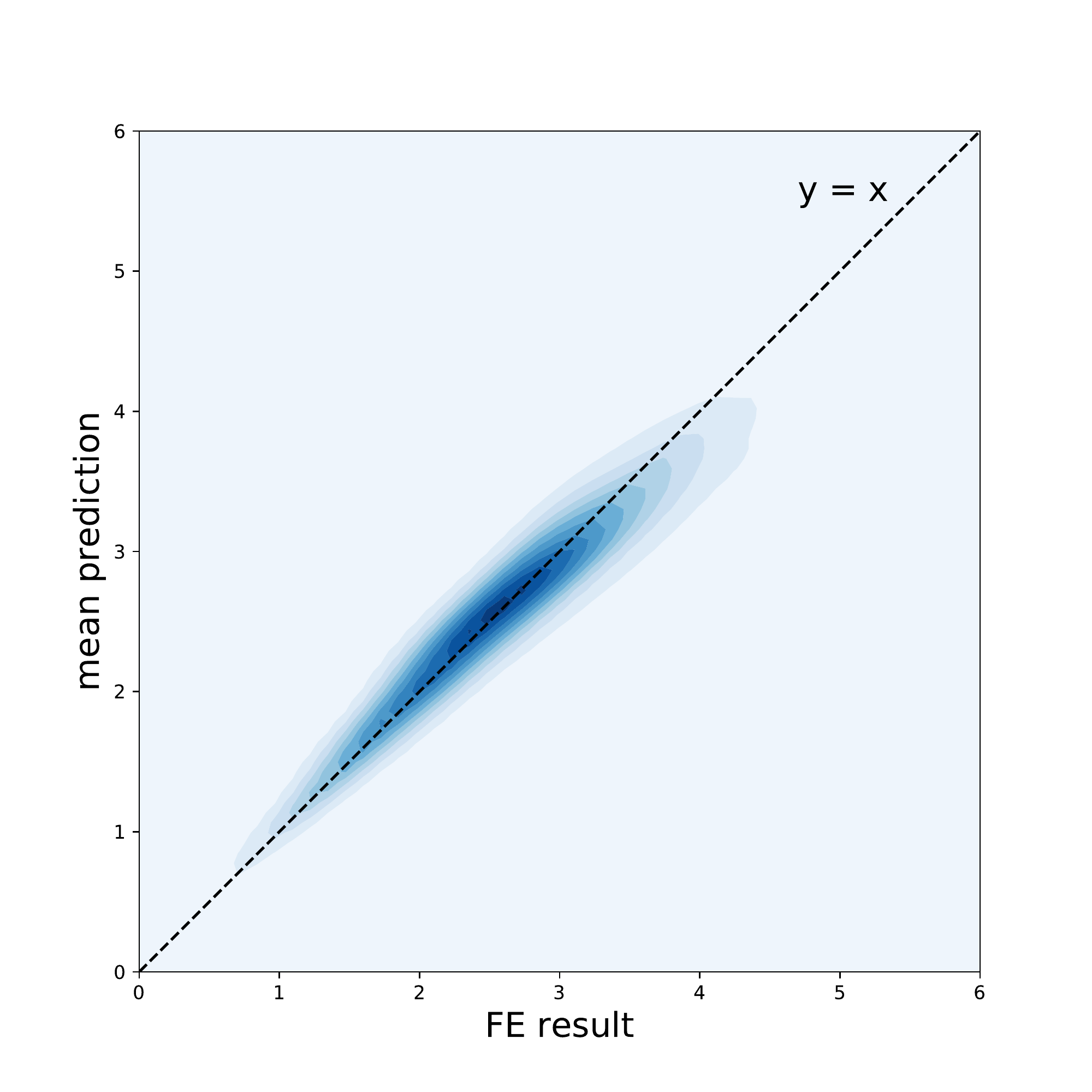}
                  \caption{}
                  \label{fig:BNN_res_0:a}
                \end{subfigure}%
                \begin{subfigure}{.5\textwidth}
                  \centering
                  \includegraphics[width=\linewidth]{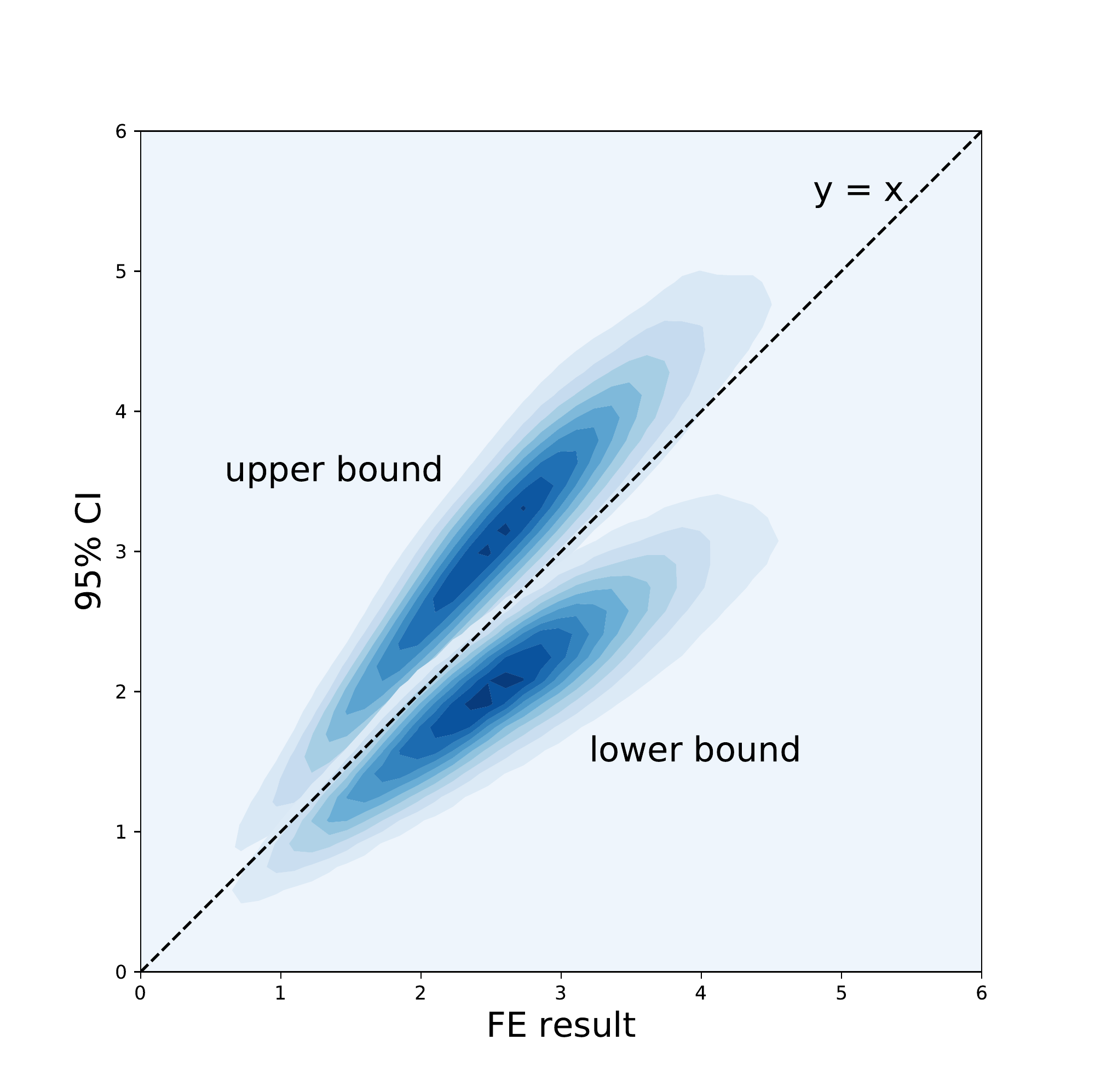}
                  \caption{}
                  \label{fig:BNN_res_0:b}
                \end{subfigure}
                \caption{In these 2 figures we see point densities where darker colors correspond to higher point density. On the left (a), a diagram showing the relationship between BNNs' mean prediction and FE results for the maximum value in the ROI. We can clearly see that most of the points are on the diagonal. On the right (b), a diagram showing the upper and lower 95\% CIs for the prediction. We can observe that for most of the points the diagonal is between the upper and lower 95\% CIs.}
                \label{fig:BNN_res_0}
            \end{figure}

\begin{figure}[h]
                \centering
                \begin{subfigure}{.33\textwidth}
                  \centering
                  \includegraphics[width=\textwidth]{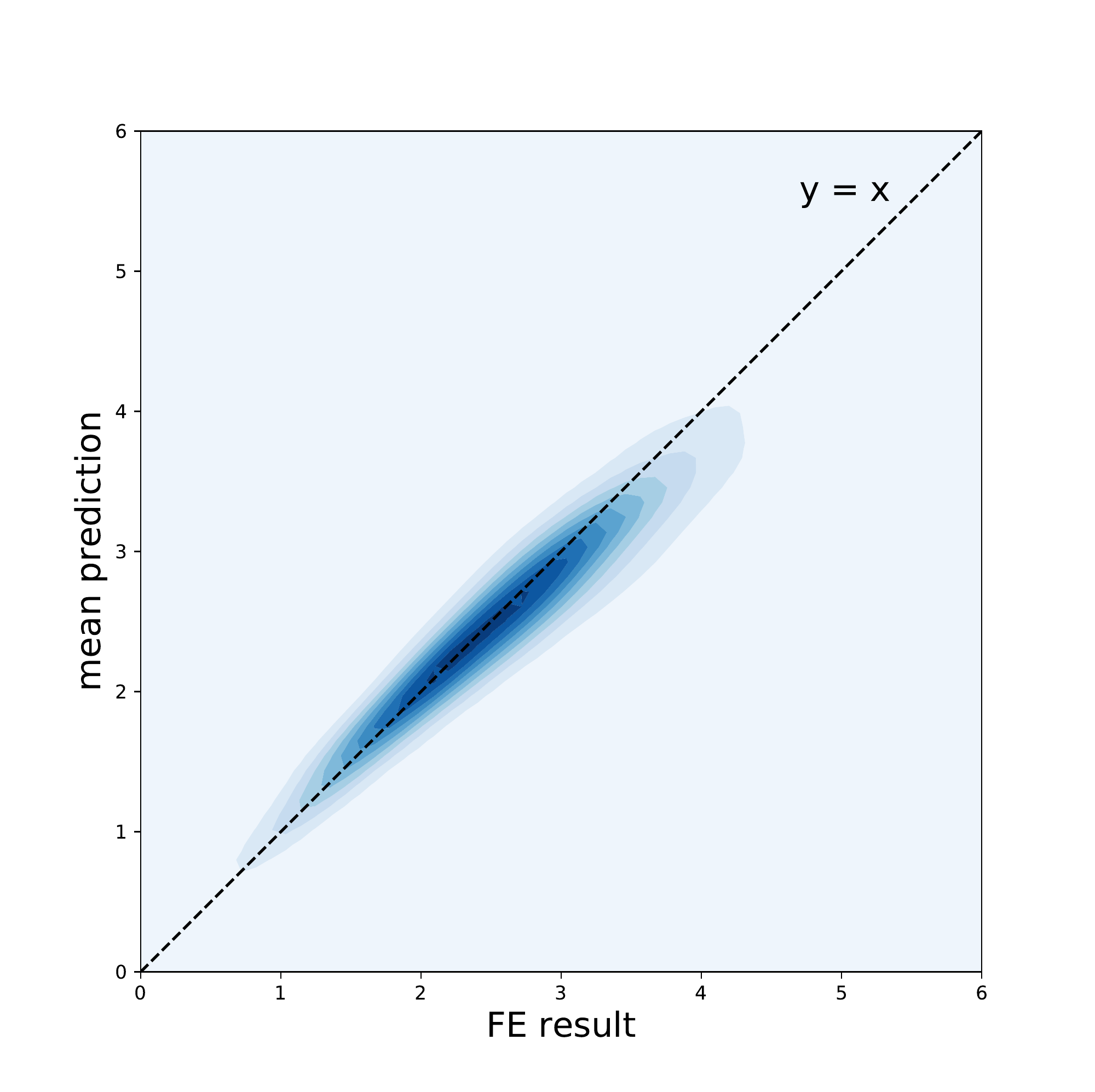}
                  \caption{Mean prediction}
                  \label{fig:BNN_error:a}
                \end{subfigure}%
                \begin{subfigure}{.33\textwidth}
                  \centering
                  \includegraphics[width=\textwidth]{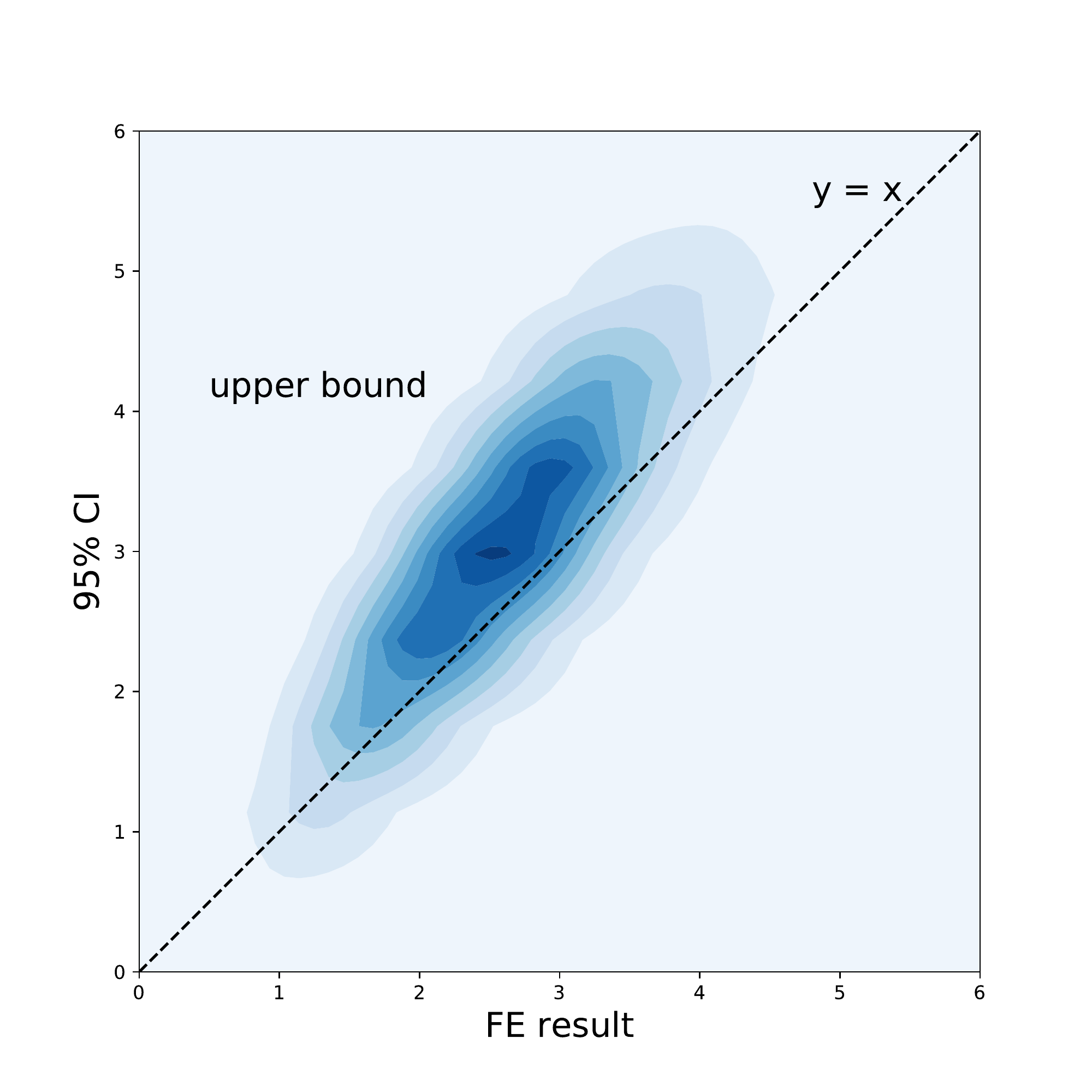}
                  \caption{Upper 95\% CI}
                  \label{fig:BNN_error:b}
                \end{subfigure}
                \begin{subfigure}{.33\textwidth}
                  \centering
                  \includegraphics[width=\textwidth]{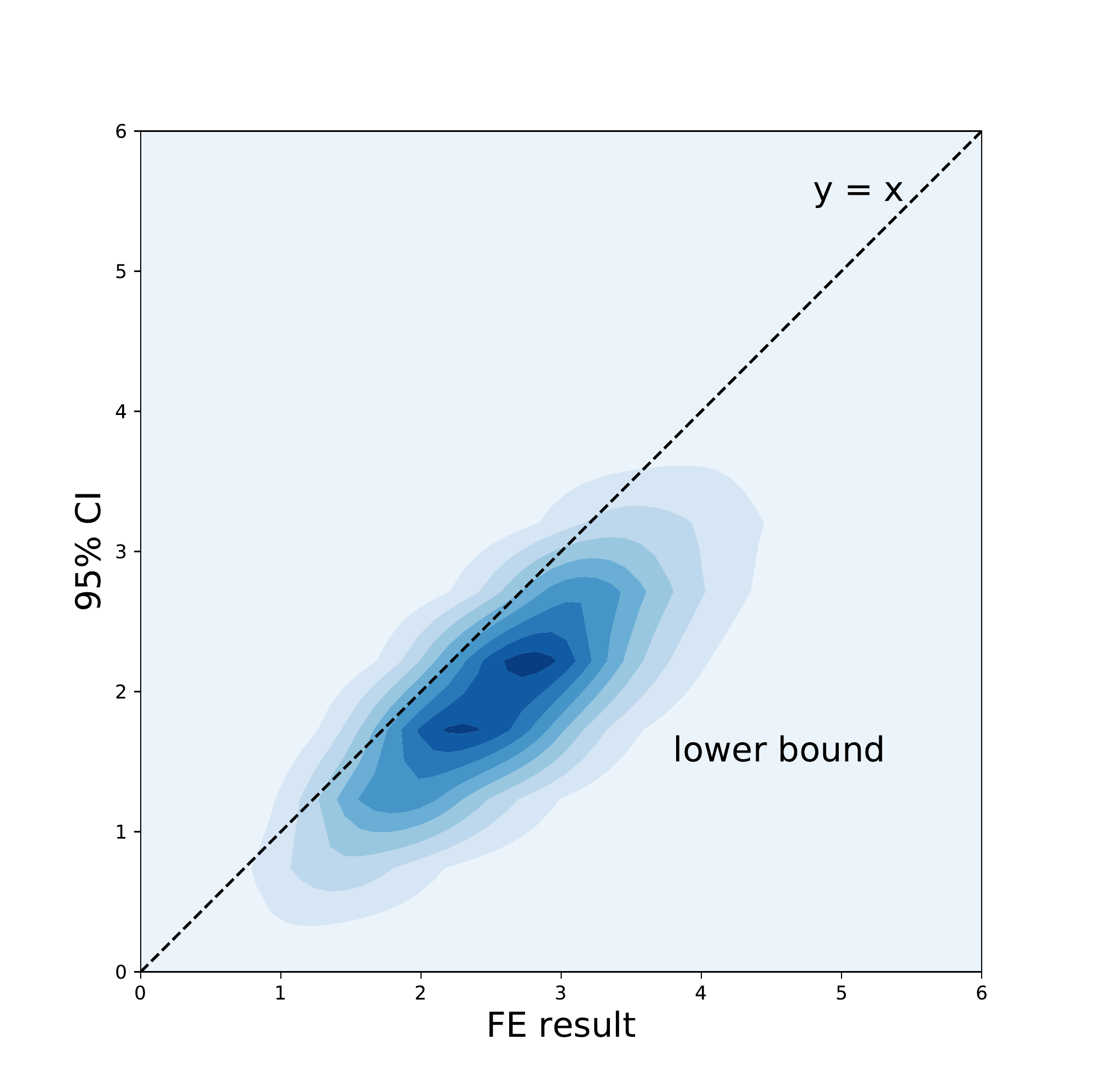}
                  \caption{Lower 95\% CI}
                  \label{fig:BNN_error:c}
                \end{subfigure}
                \caption{Three diagrams, depicting point densities where darker colors correspond to higher point density, corresponding to a BNN where the prior was not optimised during training. The prior distributions are Gaussian initialised as: $N(0, 1)$. First diagram (a) is the BNNs' mean prediction against the FE results for the maximun value in the ROI. Most of the points are on the diagonal so the NN was able to provide good mean estimations. The next two diagrams correspond to the upper 95\% CI (b) and the lower 95\% CI (c). Ideally the point densities shouldn't intersect with the diagonal. The high percentage of points below the diagonal for (b) and above the diagonal for (c) indicates that the network wasn't able to successfully quantify the uncertainty.}
                \label{fig:BNN_error}
            \end{figure}

            The mean and the variance of the BNN prediction are calculated by drawing the weights of the network from the posterior distribution 100 times and performing inference for every input. The results for a BNN where the prior was optimised during training can be found in [Fig \ref{fig:BNN_res_0}]. We can see from the first image, [Fig \ref{fig:BNN_res_0:a}], that the mean prediction is very close to the real value. We can also observe that for higher values we get higher absolute error. This is expected because those cases are represented to a lesser extent in the dataset. In the second image, [Fig \ref{fig:BNN_res_0:b}], we can observe that the diagonal is almost always, and specifically for 92\% of the patches, between the upper and lower 95\% Credible Intervals (CIs) implying that the true solution is bounded by the 95\% CIs for 92\% of the patches. 
            Lastly, we can observe a BNN where the prior parameters were not optimised during training. From [Fig \ref{fig:BNN_error:a}] we can see that the mean prediction is very good, a slight decrease of 2\% is observed in the accuracy compared to the optimised prior BNN. Nevertheless, from [Fig \ref{fig:BNN_error:b}] and [Fig \ref{fig:BNN_error:c}] we can see that the uncertainty fails to explain the error as there are many cases where the diagonal is either above the upper 95\% CI or below the lower 95\% CI. Specifically, the true value is bounded by the 95\% CIs in 82\% of the cases, a decrease of 10\% compared to the optimised prior BNN.
            \par

\begin{figure}[ht] 
              \begin{subfigure}[b]{0.5\linewidth}
                \centering
                \includegraphics[width=0.69\linewidth]{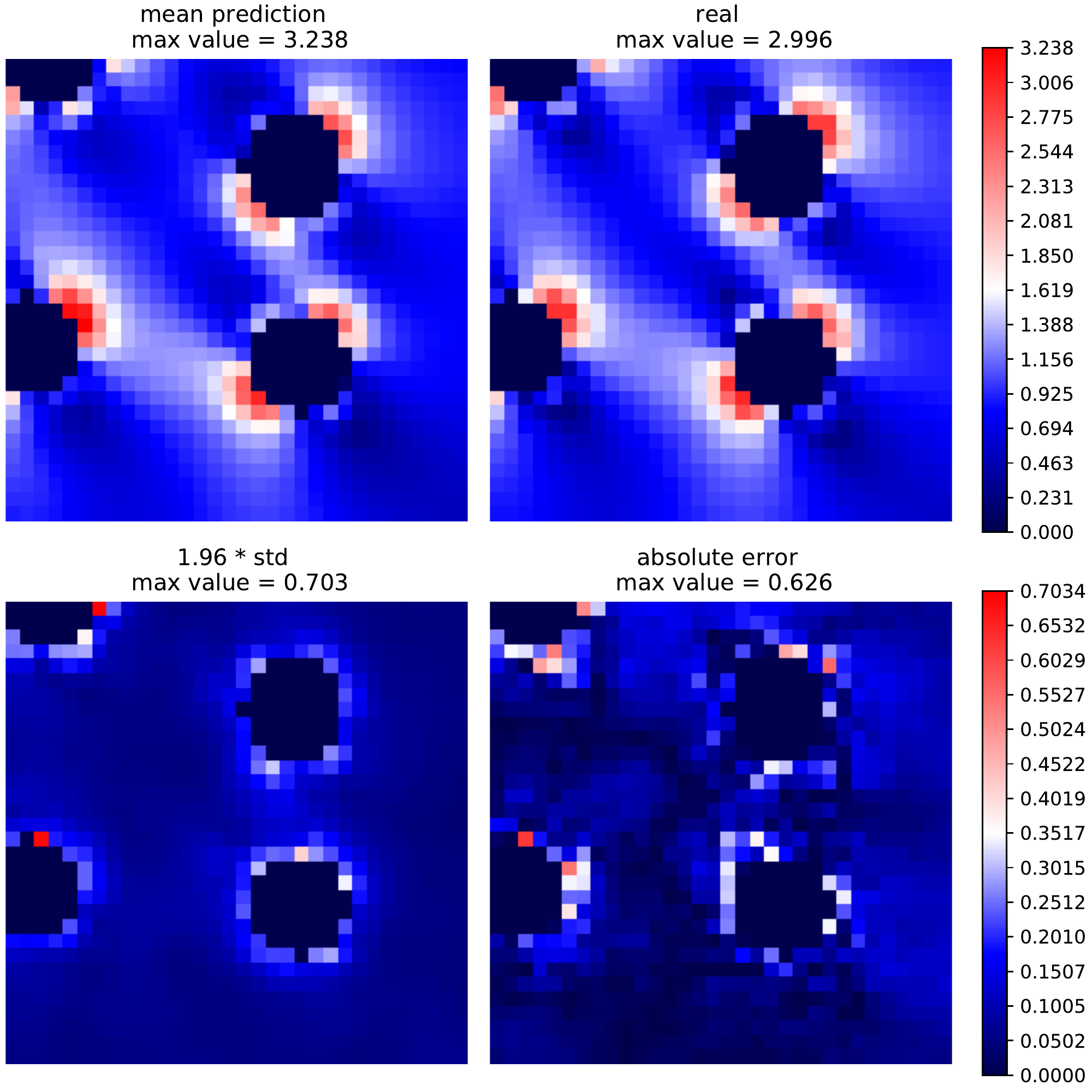} 
                \caption{} 
                \label{fig:BNN_res_images:a} 
              \end{subfigure}
              \begin{subfigure}[b]{0.5\linewidth}
                \centering
                \includegraphics[width=0.69\linewidth]{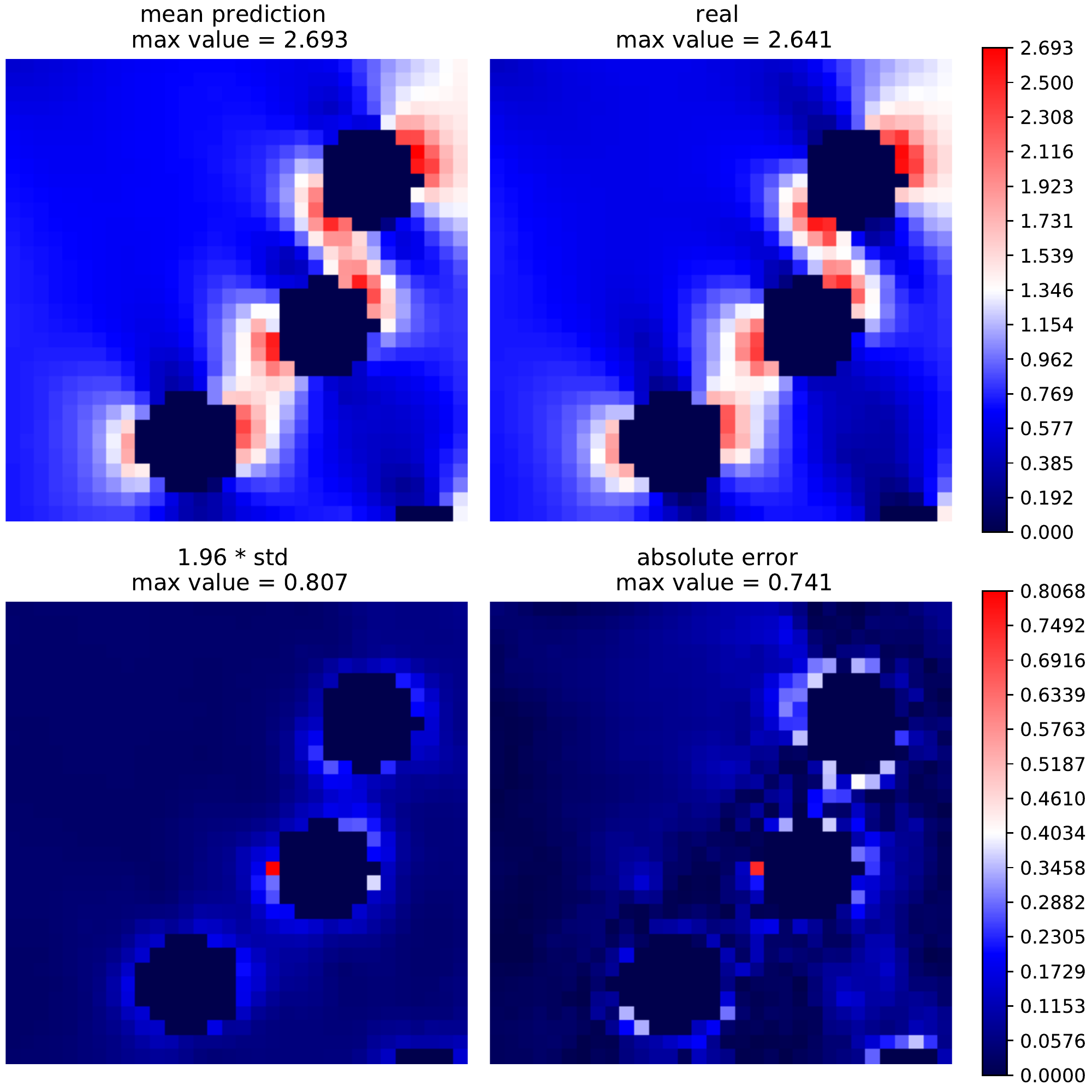} 
                \caption{} 
                \label{fig:BNN_res_images:b} 
              \end{subfigure} 
              \begin{subfigure}[b]{0.5\linewidth}
                \centering
                \includegraphics[width=0.69\linewidth]{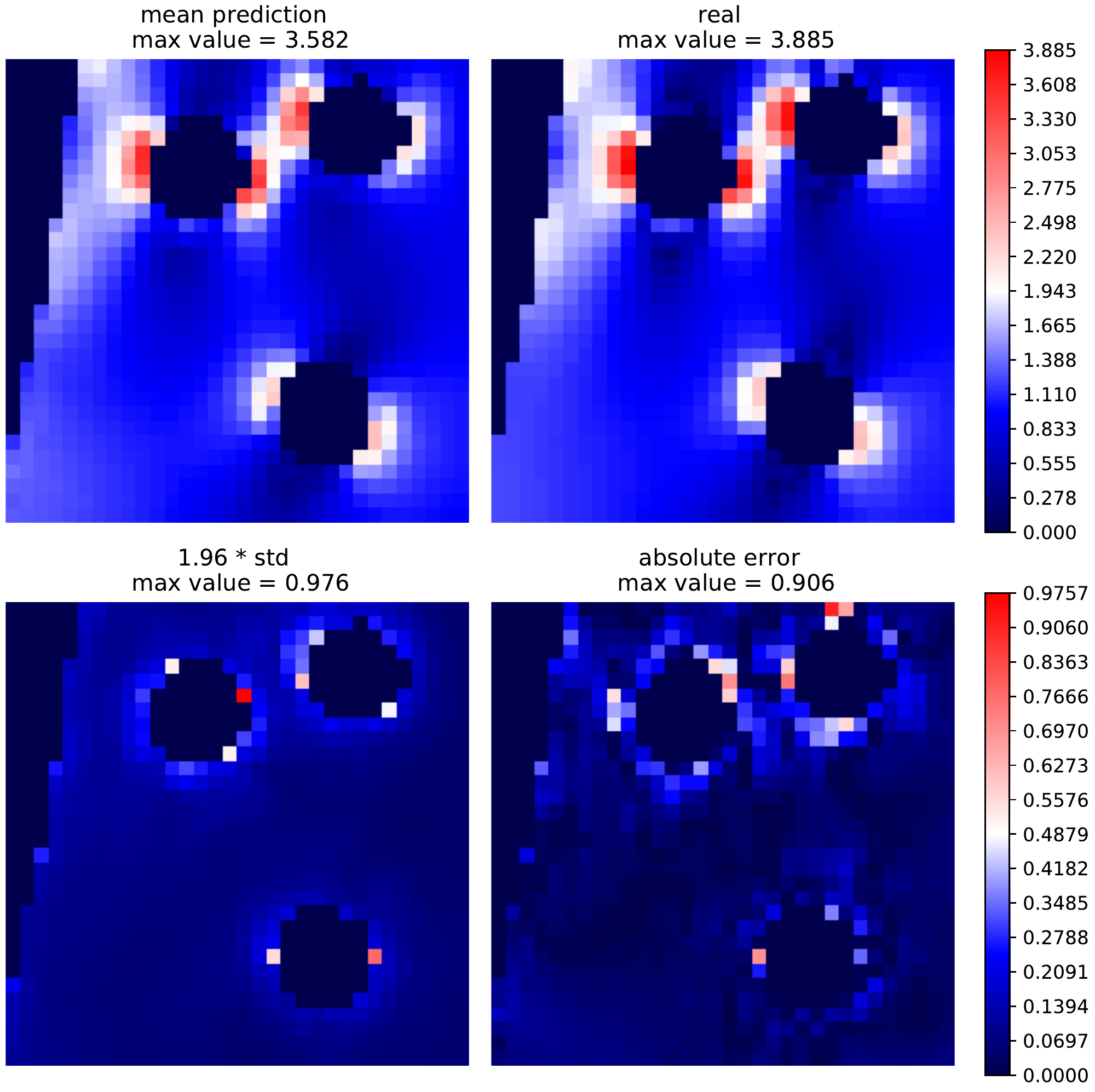}
                \caption{} 
                \label{fig:BNN_res_images:c} 
              \end{subfigure}
              \begin{subfigure}[b]{0.5\linewidth}
                \centering
                \includegraphics[width=0.69\linewidth]{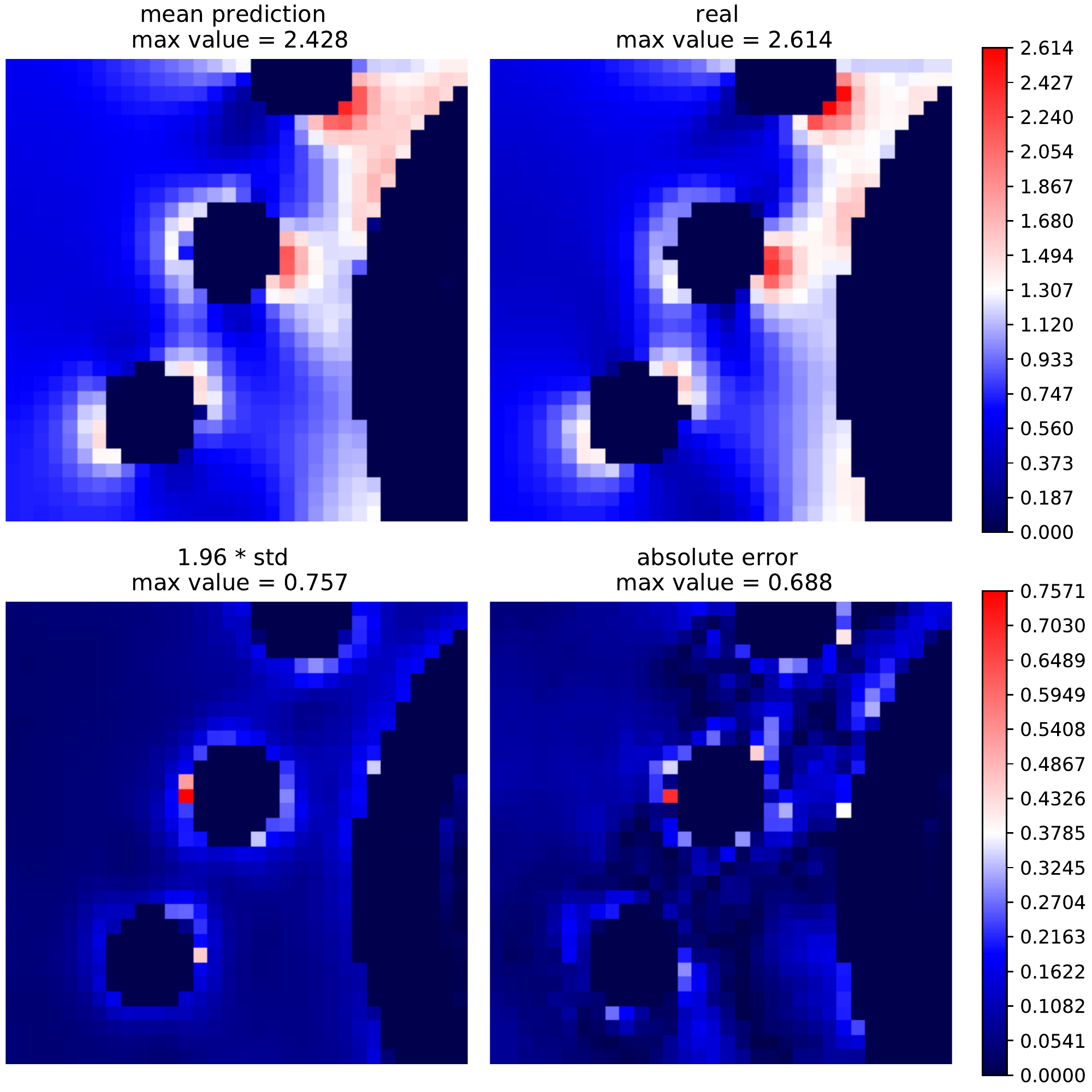}
                \caption{} 
                \label{fig:BNN_res_images:d} 
              \end{subfigure} 
              \begin{subfigure}[b]{0.5\linewidth}
                \centering
                \includegraphics[width=0.69\linewidth]{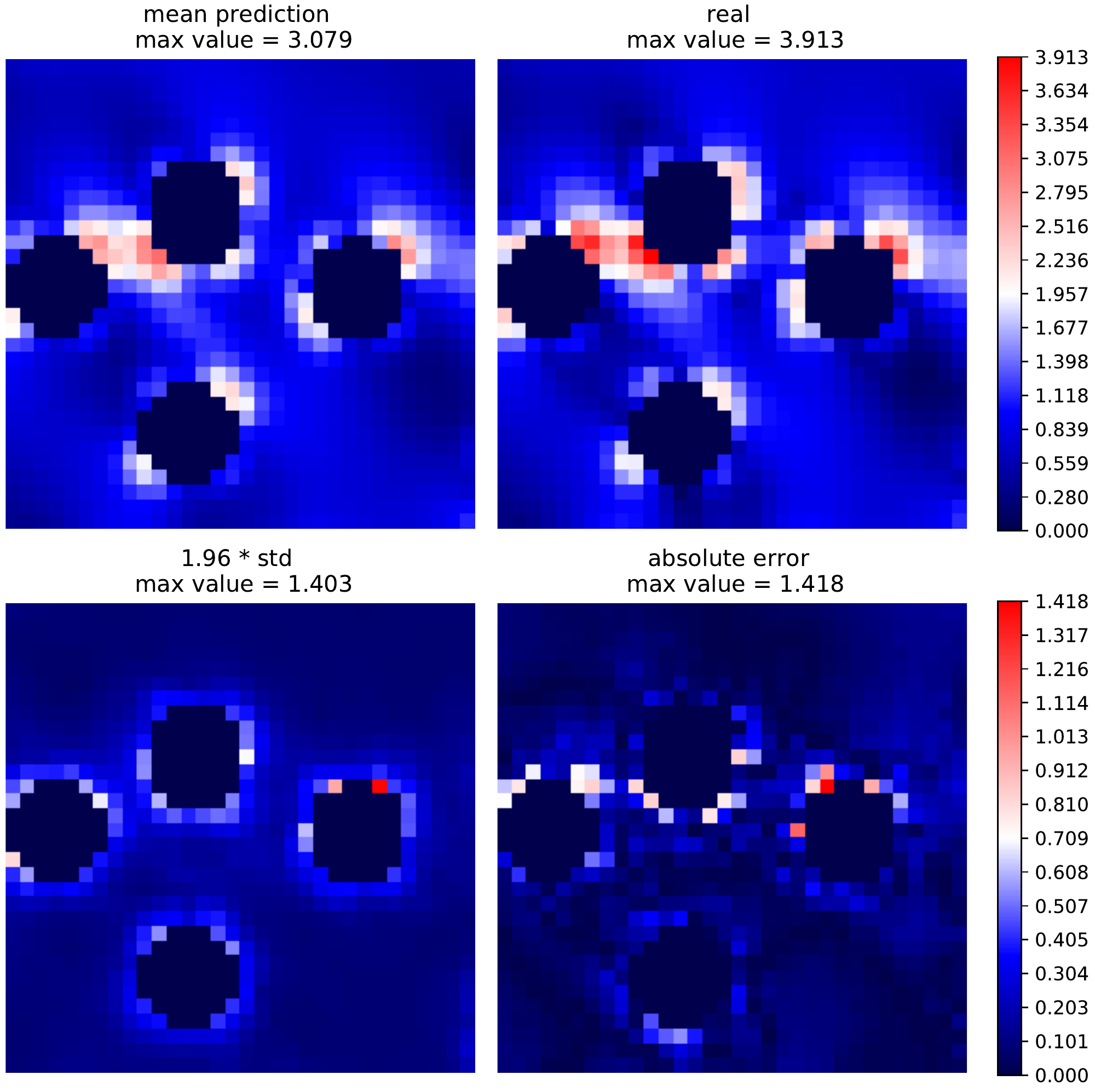}
                \caption{} 
                \label{fig:BNN_res_images:e} 
              \end{subfigure}
              \begin{subfigure}[b]{0.5\linewidth}
                \centering
                \includegraphics[width=0.69\linewidth]{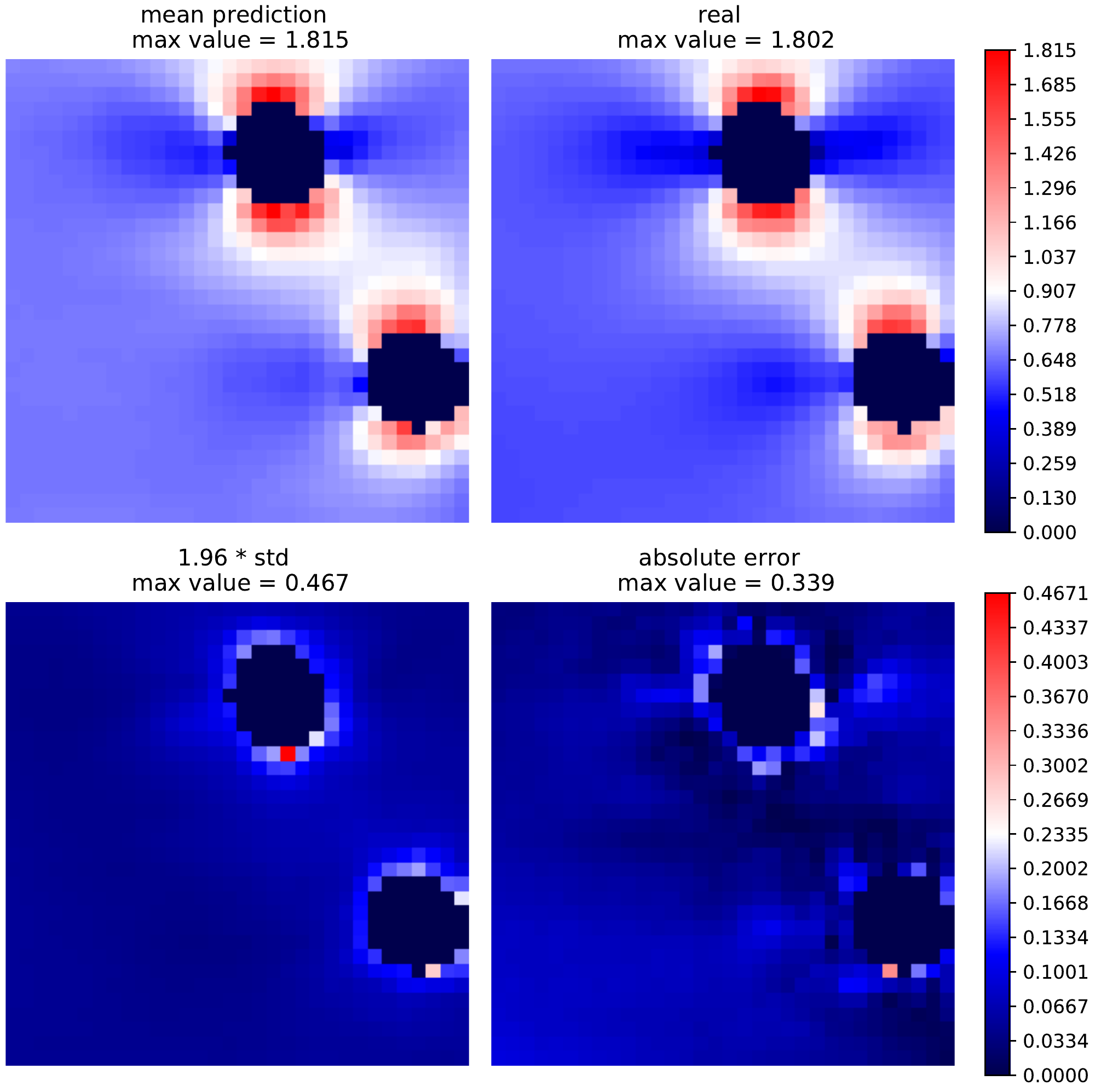}
                \caption{} 
                \label{fig:BNN_res_images:f} 
              \end{subfigure}
              \caption{Examples of BNN predictions. All images correspond to the ROI of the patches. For each of the 6 images the first row corresponds to the NN mean prediction on the left and to the scaled Tresca stress field computed by FEA and converted into an image on the right. The second row corresponds to the NN uncertainty, expressed as 1.96 $\times$ standard deviation, on the left and to the absolute error between the NN mean prediction and the FE results on the right. In images (a) and (b), we can see strong interactions between micro scale features and in images (c) and (d) we can see strong interactions between macro and micro scale features. In all these cases we can observe that the uncertainty is higher in areas where the error is higher. In image (e) we observe that the error is very high but the NN was able to identify it and provided high uncertainty for the prediction. Lastly, in image (f) we observe an image with low error and low uncertainty.}
              \label{fig:BNN_res_images} 
            \end{figure}

            Results from the uncertainty estimation on image level can be found in [Fig \ref{fig:BNN_res_images}]. On the top 2 cases [Fig \ref{fig:BNN_res_images:a}, \ref{fig:BNN_res_images:b}] we can see some examples of good mean predictions where there are clear interactions between multiple micro scale features. The middle images [Fig \ref{fig:BNN_res_images:c}, \ref{fig:BNN_res_images:d}] are examples of good mean predictions where interactions between multiple microscale features and a macro scale feature can be seen. We can observe that the uncertainty, expressed as 1.96 $\times$ standard deviation, is higher in the vicinity of the higher error pixels indicating that the BNN has successfully identified the unseen interactions (interactions that where not in the training dataset or were underrepresented). [Fig \ref{fig:BNN_res_images:e}] is an example of a case where the maximum value is miss-predicted with a large error of about 1 unit. Fortunately, we can observe that the uncertainty is also very large, specifically 1.96 $\times$ standard deviation has a value of about 1.5 unit meaning that the true maximum value is between the mean prediction and the 95\% CI. Image [Fig \ref{fig:BNN_res_images:f}] is an example of a case with low uncertainty and low error. This means that the CIs are very tight and the BNN is very confident about the prediction. That was an expectable result in the sense that this is a very simple case, 2 circular micro scale features are weakly interacting, and we would expect from the BNN to handle it without a problem because the training dataset contains a huge number of these examples.
            \par
            
\clearpage 
    
    \subsection{Nonlinear elasticity} \label{Non Linear Section}

        In this section we demonstrate the applicability of our method for multiscale problems in finite strain elasticity. Compared to previous section, the macroscale geometry of the dataset is constrained is order to showcase the ability of the framework to tackle microscale-informed stress analysis during the macroscale design of an engineering component.

        \subsubsection{Training Dataset}
            
\begin{figure}[h]
                \begin{center}
                    \includegraphics[width=\linewidth]{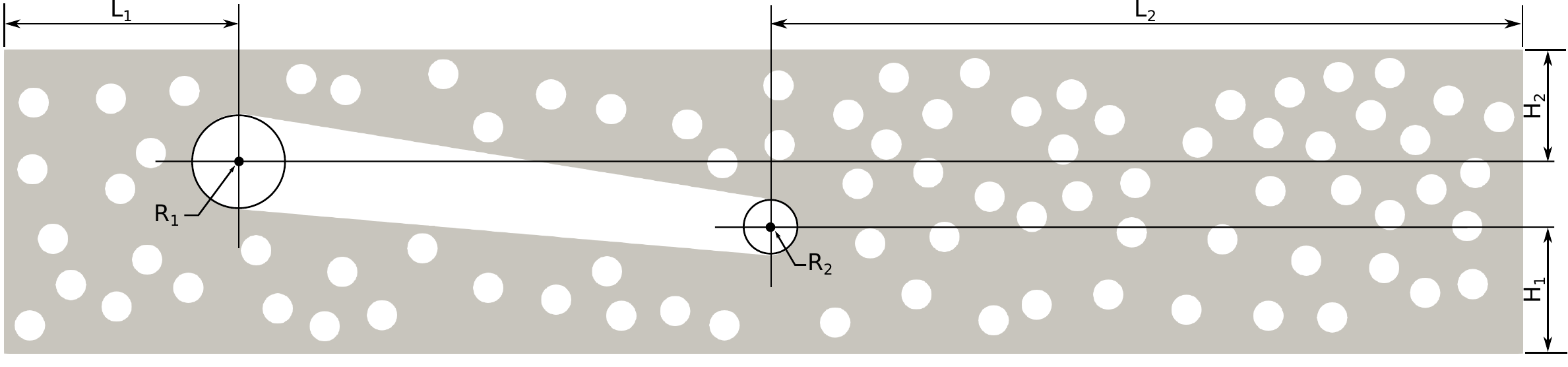}
                    \caption{Structure in the reference (undeformed) configuration. $L_1=0.76$, $R_1=0.06$, $L_2=2.45$, $R_2=0.035$, the width of the structure is 5 and the height 1}
                    \centering
                    \label{fig:structure_184_reference}
                \end{center}
            \end{figure}

            The structure that we study in this section is rectangular with one macroscale geometrical feature and a distribution of disks as microscale features. The macroscale feature whose parameters can be optimised is composed by 2 disks with random radii and centers connected by their common external tangents [Fig \ref{fig:structure_184_reference}]. Multiple instances of this structure are created by randomly choosing values for the macroscale parameters and changing the distribution of the microscale features. After training our NN will be able to evaluate the micro Cauchy stress distribution in different macroscale geometries (of the same family) under any realisation of the microscale random texture.
            
\begin{figure}[h]
                \begin{center}
                    \includegraphics[width=\linewidth, trim={10cm 5cm 10cm 5cm}, clip]{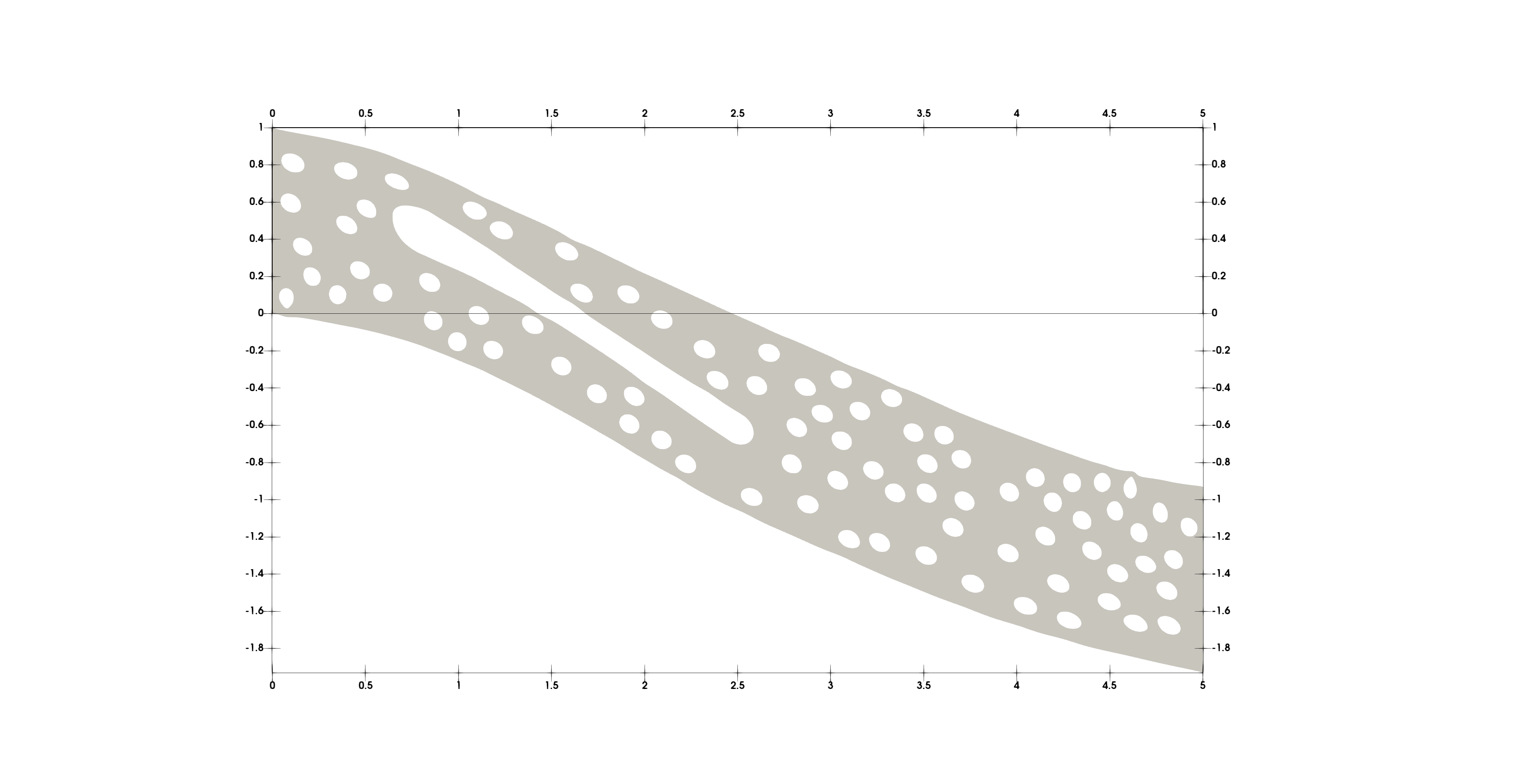}
                    \caption{Structure in the deformed configuration (Finite Strain elasticity theory)}
                    \centering
                    \label{fig:structure_184_deformed}
                \end{center}
            \end{figure}

            To create the training dataset we choose length 5 and height 1 as overall dimensions for the structure. For the boundary conditions, the structure is clamped at $x=0$ and a random displacement along the $-y$ direction ranging from 0 to 2 is applied to the other end, at $x=5$, while the displacement along the x direction is zero. An example of the deformed structure for a displacement of size -1.93 can be found in [Fig \ref{fig:structure_184_deformed}]. The radius of the disks is 0.02 units or 4 pixels as in the linear elasticity case. 
            
            The FE solution is mapped into an image of size [80 $\times$ 400]. The patch and the ROI have the same size as the linear elasticity case [72 $\times$ 72] and [32 $\times$ 23] respectively.
            
        \subsubsection{Numerical results in finite strain elasticity}
        
\begin{figure}[h]
                \begin{center}
                    \includegraphics[width=0.75\linewidth]{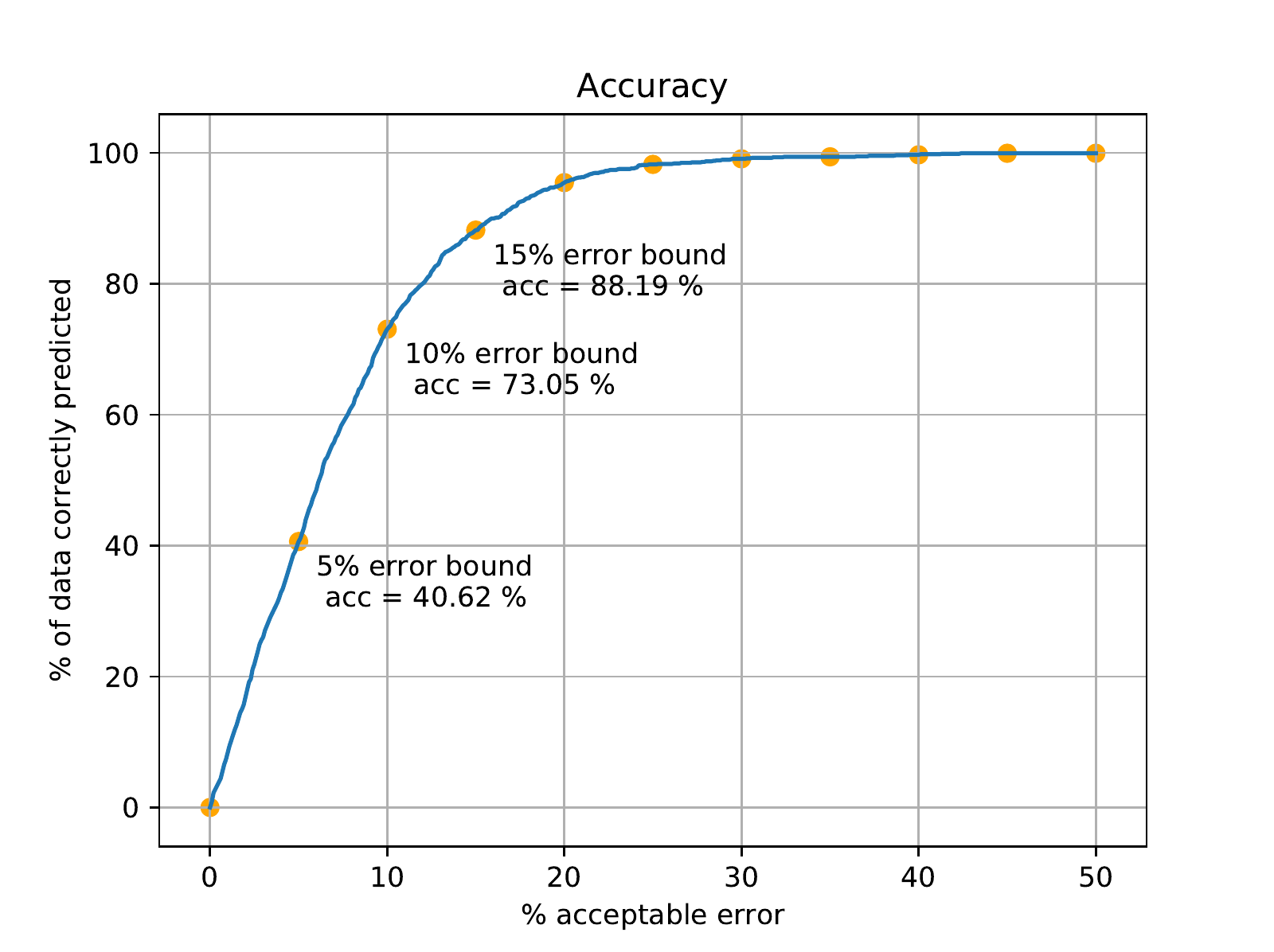}
                    \caption{Accuracy as function of the threshold. Here, accuracy is defined as the percentage of patches in the dataset for which the relative error between the max NN prediction and the max FE result in the ROI is less than a predefined threshold. The accuracy for threshold values 5\%, 10\% and 15\% is 40\%, 73\% and 88\% respectively. For threshold values higher than 20\% the accuracy is more than 95\%.}
                    \centering
                    \label{fig:nonLinear_CNN_accuracy}
                \end{center}
            \end{figure}

             We performed 200 FE simulations, which took 16 hours on an Intel\textsuperscript{\textregistered} Core\textsuperscript{\texttrademark} i7-6820HQ CPU, and we extracted 12,000 patches. Of those, 1,200 were used as a test set and 10,800 as a training set. We trained the same CNN as in the linear elastic case, section [\ref{Advanced Dataset}], without scaling the data as a preprocessing step. Training with the Adam optimizer for 300 epochs required 3 hours on an NVIDIA T4 GPU. The results can be found in [Fig \ref{fig:nonLinear_CNN_accuracy}]. The accuracy for the 10\% threshold is 73\%.
            
\begin{figure}[ht] 
              \begin{subfigure}[b]{0.5\linewidth}
                \centering
                \includegraphics[width=0.75\linewidth]{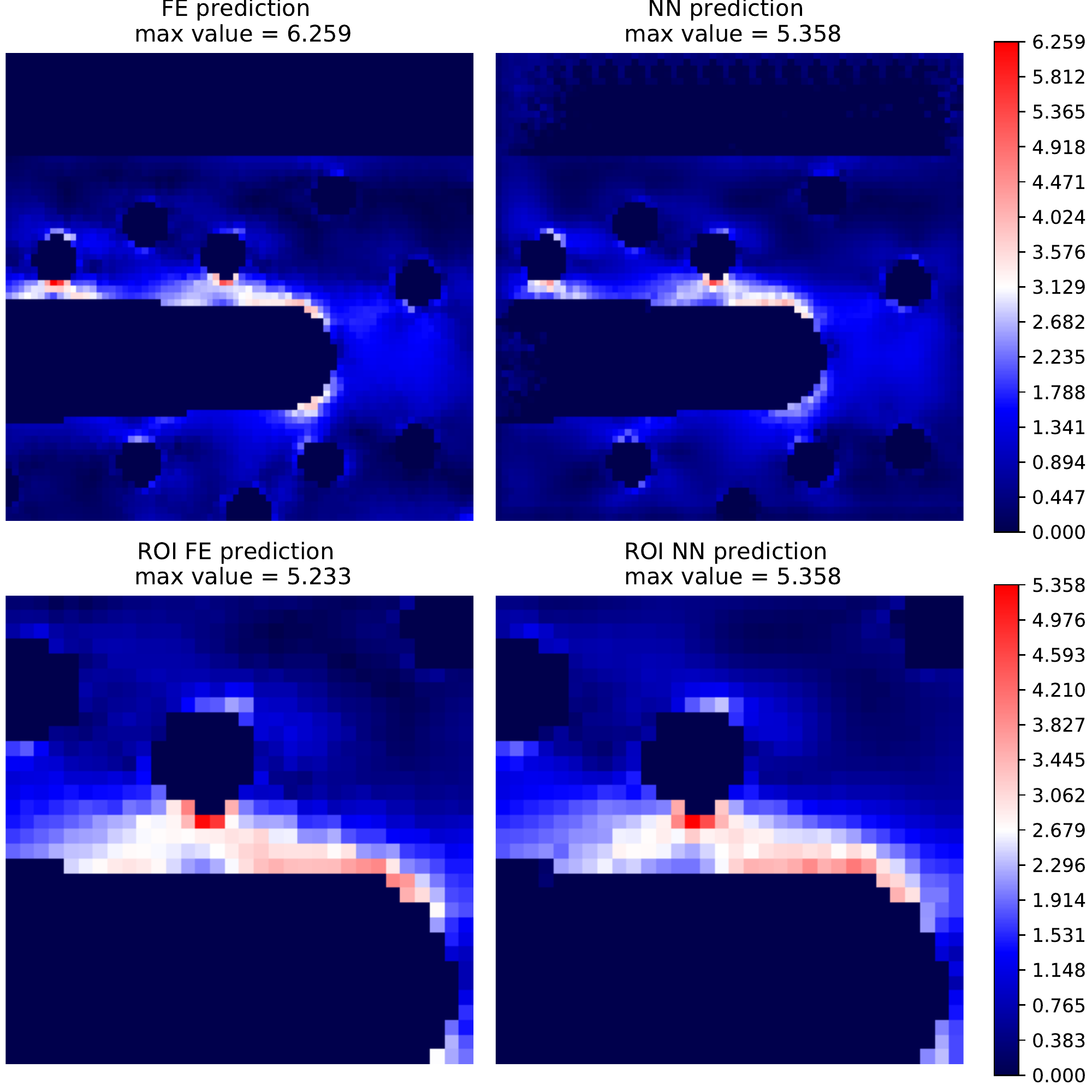} 
                \caption{} 
                \label{fig:4_examples_nonlinear:a} 
                \vspace{4ex}
              \end{subfigure}
              \begin{subfigure}[b]{0.5\linewidth}
                \centering
                \includegraphics[width=0.75\linewidth]{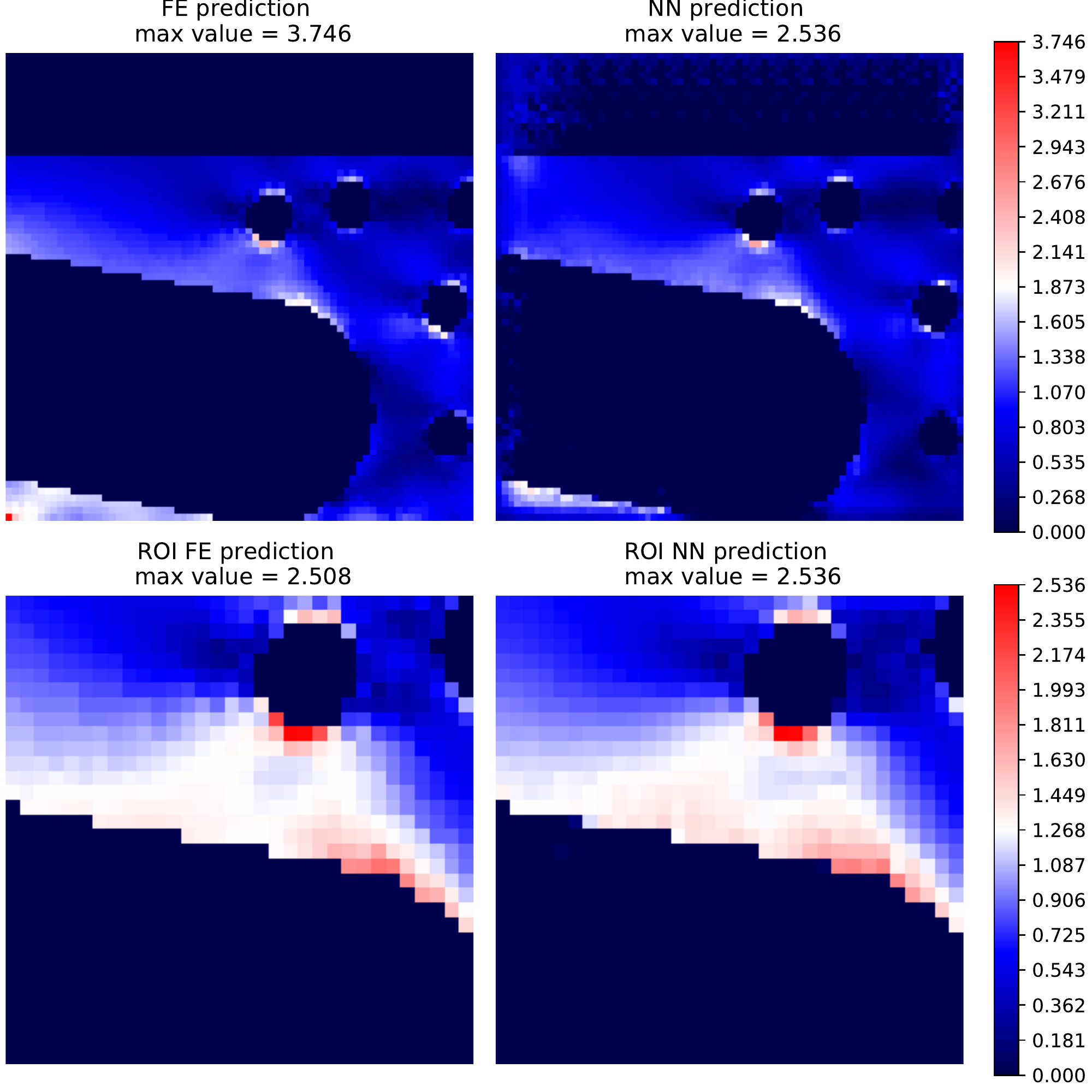} 
                \caption{} 
                \label{fig:4_examples_nonlinear:b} 
                \vspace{4ex}
              \end{subfigure} 
              \begin{subfigure}[b]{0.5\linewidth}
                \centering
                \includegraphics[width=0.75\linewidth]{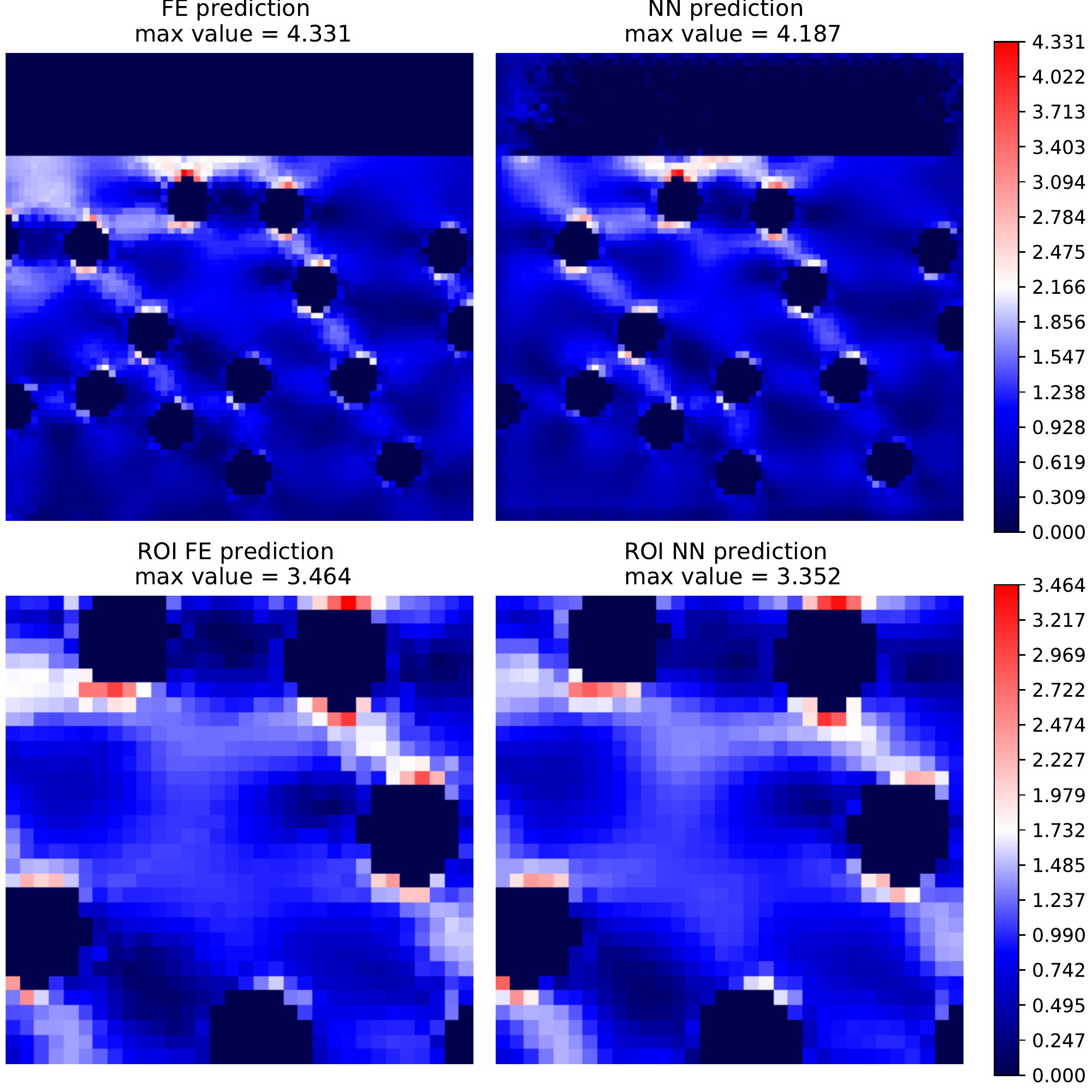}
                \caption{} 
                \label{fig:4_examples_nonlinear:c} 
              \end{subfigure}
              \begin{subfigure}[b]{0.5\linewidth}
                \centering
                \includegraphics[width=0.75\linewidth]{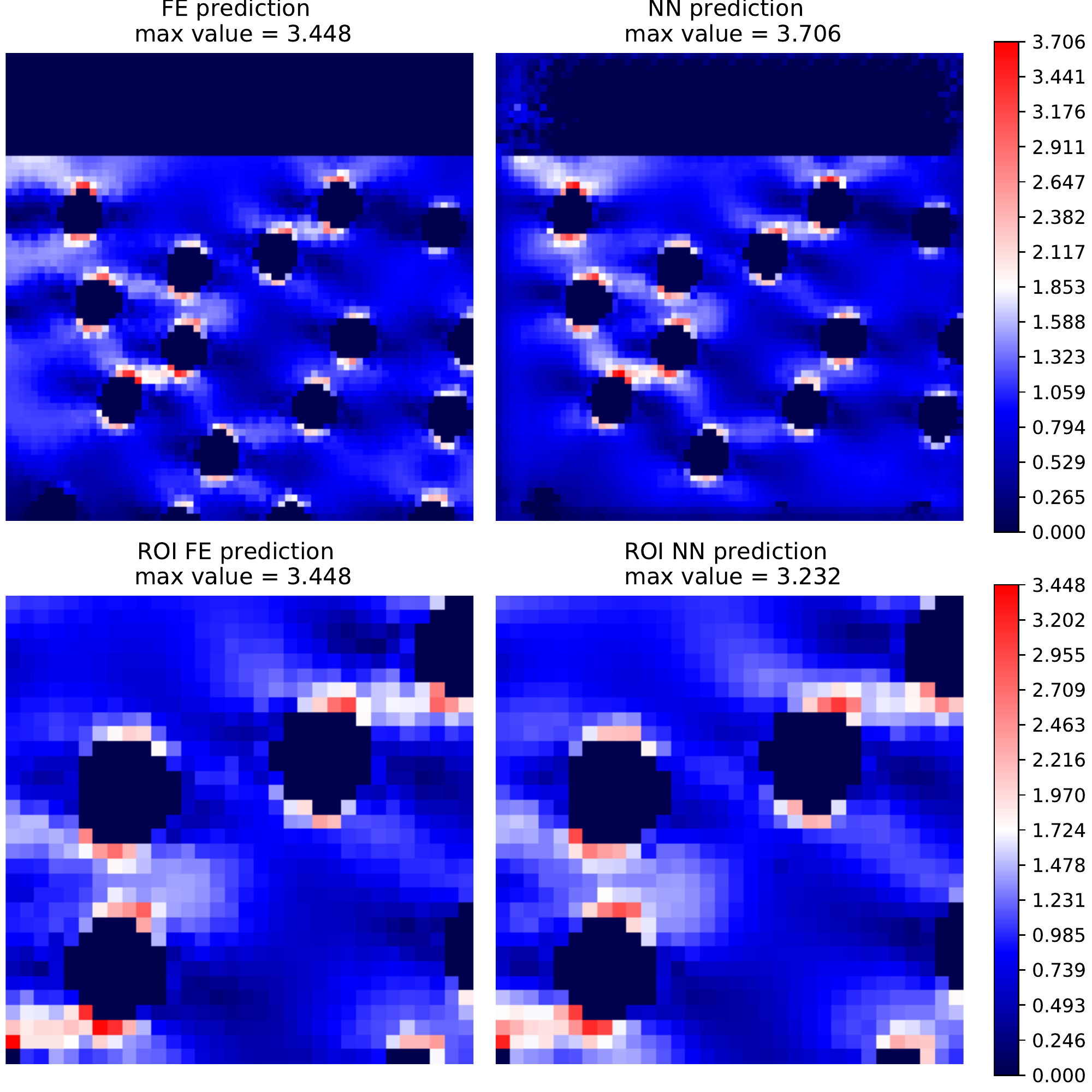}
                \caption{} 
                \label{fig:4_examples_nonlinear:d} 
              \end{subfigure} 
              
              \caption{Comparison between CNN and FE prediction in 4 patches. In each of the 4 images the first row from left to right corresponds to the scaled Tresca stress field in the patch computed by FEA and converted into an image and the CNN prediction in the patch. The second row from left to right is the scaled Tresca stress field in the ROI computed by FEA and converted into an image and the CNN prediction in the ROI. Images (a) and (b) correspond to patches where there are interactions between macro and micro scale features. Images (c) and (d) correspond to patches where there are interactions between multiple micro scale features. In all the cases the error in max values in the ROI is less than 10\% and the stress distribution is accurately predicted.}
              \label{fig:4Exmpls_non_linear} 
            \end{figure}

            Additionally, in [Fig \ref{fig:4Exmpls_non_linear}] we see the prediction of the CNN in 4 patches. The two top rows include interactions between macro and micro scale features. In both cases the max values are computed with a relative error less than 10\% and the stress distribution is also correctly predicted. The two bottom rows include interactions between micro scale features. In both cases the max values are computed with an error of less than 10\% and the stress distribution is also correctly predicted.
            
\begin{figure}[h]
                \begin{center}
                    \includegraphics[width=\linewidth]{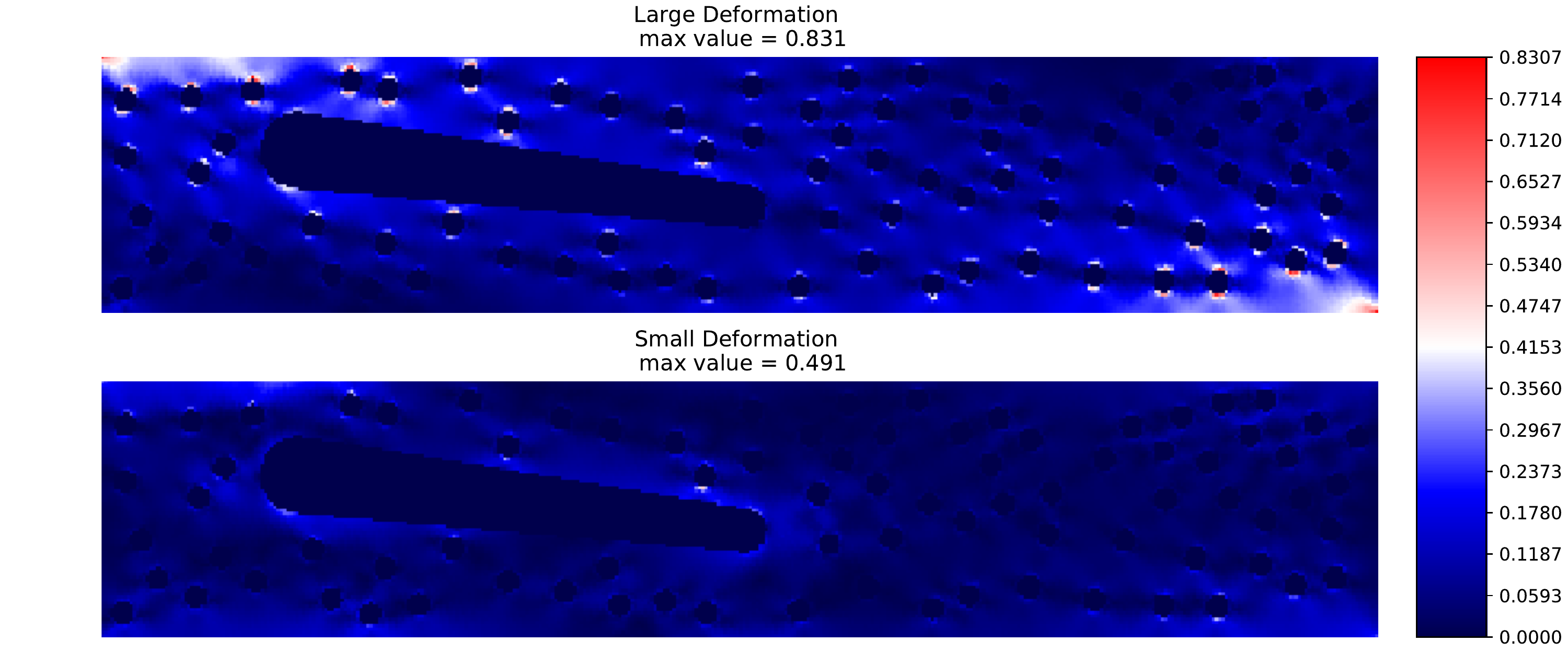}
                    \caption{Comparison between the Tresca stress field computed by FEA and converted into an image for the non linear elasticity case, on top, and the Tresca stress field computed by FEA and converted into an image for the linear elasticity case, on the bottom, for the structure described in [Fig \ref{fig:structure_184_deformed}].}
                    \centering
                    \label{fig:large_small_def}
                \end{center}
            \end{figure}

\begin{figure}[h]
                \begin{center}
                    \includegraphics[width=\linewidth]{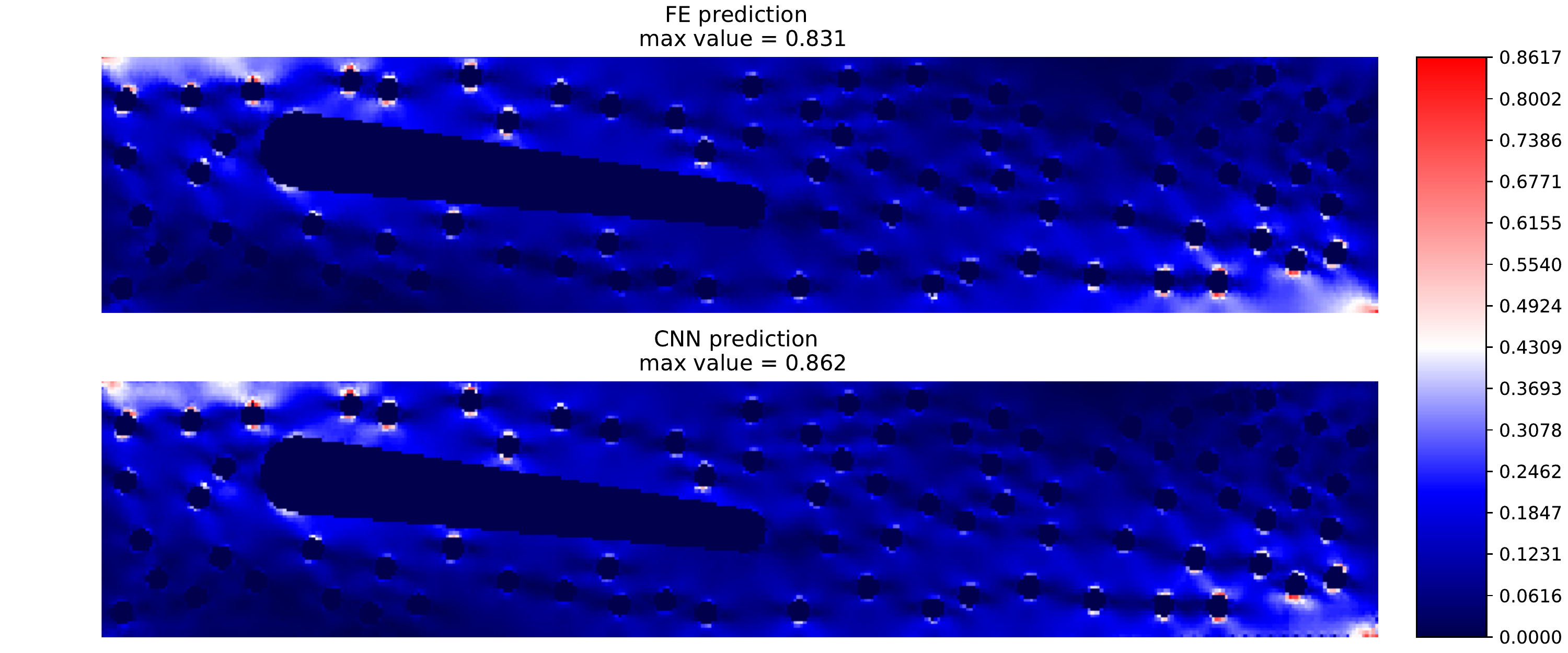}
                    \caption{Comparison between the Tresca stress field computed by FEA and converted into an image, on the top, and an image reconstructed using the
                    CNN predictions, on the bottom, for the structure described in [Fig \ref{fig:structure_184_deformed}].}
                    \centering
                    \label{fig:Example_184}
                \end{center}
            \end{figure}

\begin{figure}[h]
                \begin{center}
                    \includegraphics[width=\linewidth]{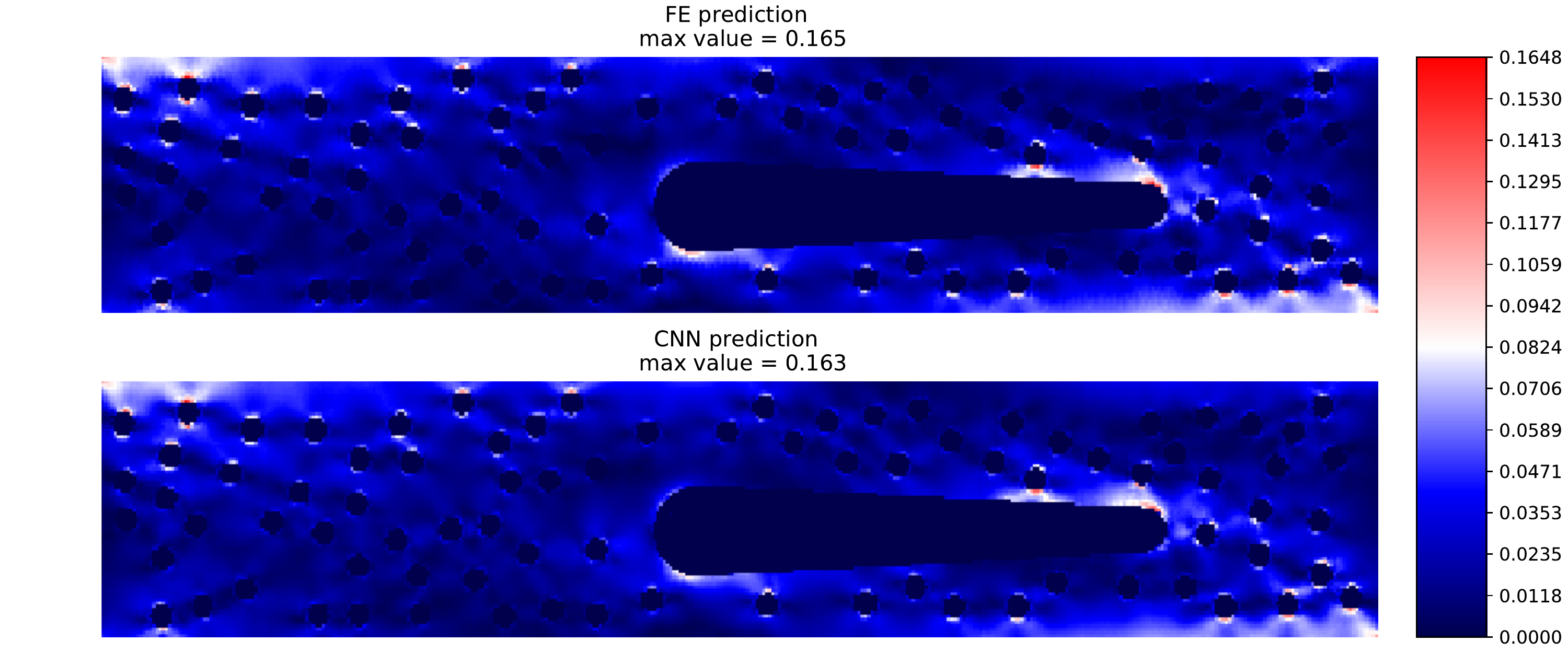}
                    \caption{Comparison between the Tresca stress field computed by FEA and converted into an image, on the top, and an image reconstructed using the CNN predictions, on the bottom.}
                    \centering
                    \label{fig:Example_174}
                \end{center}
            \end{figure}

            Lastly, we show a comparison between the CNN prediction and the FE prediction in the entire structure. We remind the reader that the CNN prediction happens at patch level and then the patches are rearranged to create the entire solution field as has already been described in [Fig \ref{fig:patch_full}]. In [Fig \ref{fig:large_small_def}] we show the difference between a structure modeled with linear elasticity and the same structure modeled with non-linear elasticity as has been calculated with FE simulations for the structure described in [Fig \ref{fig:structure_184_deformed}]. We observe that in general the 2 predictions are different and specifically that the linear elastic model underestimates the stress magnitude in regions of very large deformations. In [Fig \ref{fig:Example_184}] we show the comparison between the CNN prediction and the FE prediction for the non-linear elasticity case. We observe that the max values are very close with a relative error between the maximum values less than 4\% and also that the stress distribution is correctly predicted. We also see the same comparison for another structure in [Fig \ref{fig:Example_174}] where we can observe a similar behaviour.
            
\clearpage
    \subsection{Selective Learning} \label{Selective Learning}
    
    In this section we investigate the idea of Selective Learning to reduce the labelled data requirements for training the BNN. We need an initial dataset with labeled data so we can initially train the BNN, a bigger dataset with only unlabeled data and finally a validation set with labelled data. The principles of this framework are described below. 
    
   \begin{enumerate}
      
      \item We use an acquisition function to select small batches from the unlabeled dataset
      
      \item We label the selected data points and \say{move} them to the training set
      
      \item We train the BNN with the new training dataset
      
      \item We measure the accuracy of the BNN using the validation set
      
      \item We repeat the same process until the accuracy converges or we label the entire unlabeled set.
          
    \end{enumerate}
    
\begin{figure}[h]
        \begin{center}
            \includegraphics[width=0.5\linewidth]{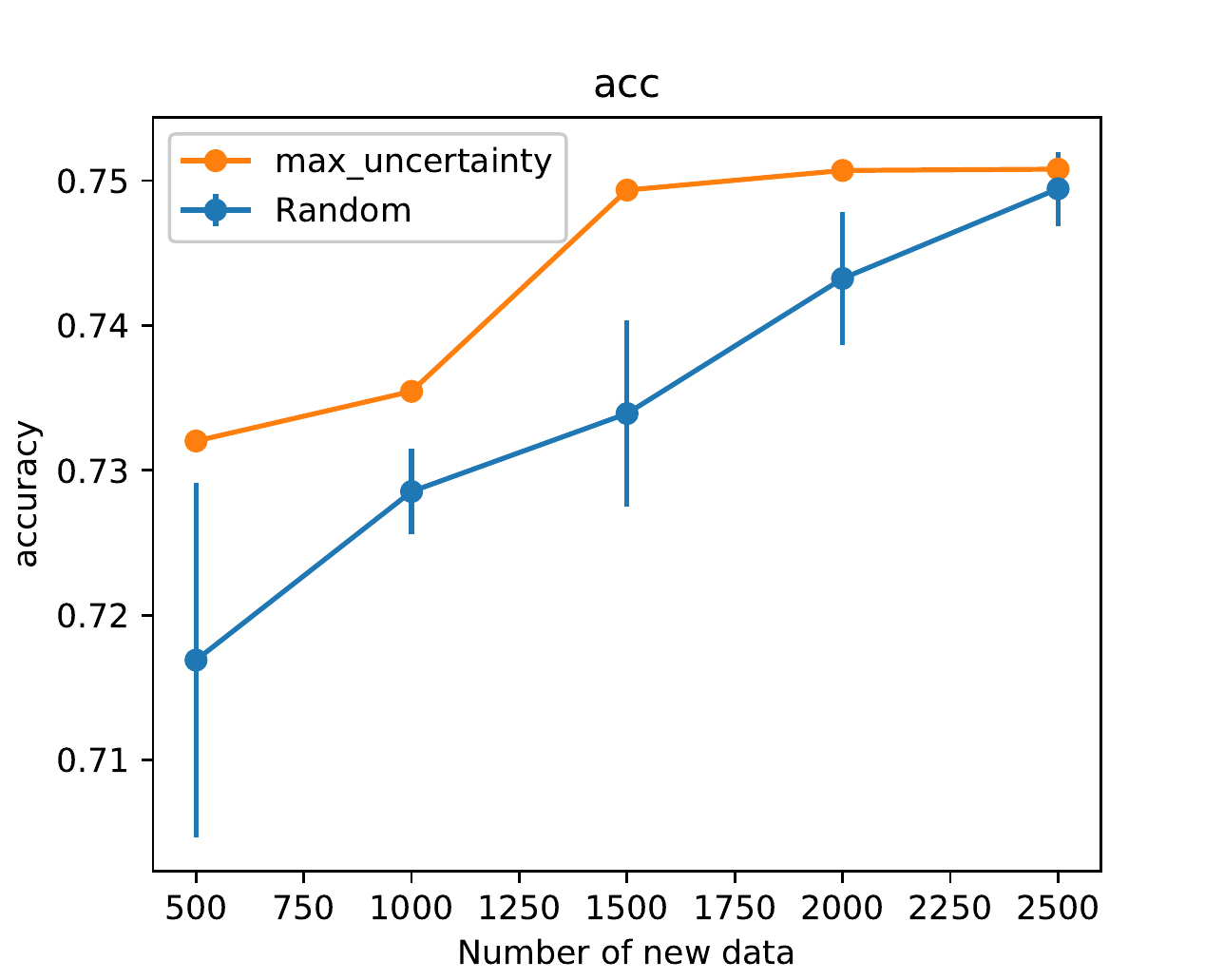}
            \caption{Results of Selective Learning for an initial set of 2,500 patches. 50 epochs per training, 50 forward passes for the uncertainty quantification and 500 added data in each iteration. The orange line corresponds to the max uncertainty acquisition function and the blue to the random acquisition function. For the random acquisition function we repeated the experiment five times and reported the mean values and the 95\% confidence interval. We can observe that the orange curve is consistently above the blue one. This means that with the maximum uncertainty acquisition function we can achieve high accuracy with less data. Specifically in this case by using only 1,500 labelled data we achieve accuracy of about 75\% with the maximum uncertainty acquisition function while we need to use all the 2,500 patches to achieve the same accuracy with the random acquisition function.}
            \centering
            \label{fig:SL_small}
        \end{center}
    \end{figure}

    The data that will be used in this section come from the linear elasticity problem [\ref{Advanced Dataset}]. We designed a small experiment to validate our approach, inspired by \citep{gal2017deep}. Here we make a comparison between a network trained using the max uncertainty acquisition function, choosing first the patches with higher uncertainty, and a second one trained using the random acquisition function, that chooses patches randomly. For the random selection approach we repeated the experiments 5 times and presented the mean and the 95\% confidence interval. We used the following setup: 2,500 patches for the initial set, 2,500 patches as the unlabeled set and 11,000 patches as validation set. We trained each network for 50 epochs, we performed 50 forward passes for the uncertainty estimation and we added 500 patches in the labeled set at each iteration, query rate = 500. The accuracy is calculated from the mean prediction of the network. The results can be found in [Fig \ref{fig:SL_small}]. We observe that the results produced by the max uncertainty acquisition function consistently present higher accuracy. More specifically, with this unlabeled data set we can reach an accuracy of about 75\%. This can be achieved using 1,500 patches with the max uncertainty acquisition function but requires all the 2,500 patches if we choose them randomly. This means that we reduced the labeled data requirement by 40\%. 
    \par
    
\begin{figure}[h]
        \begin{center}
            \includegraphics[width=\linewidth]{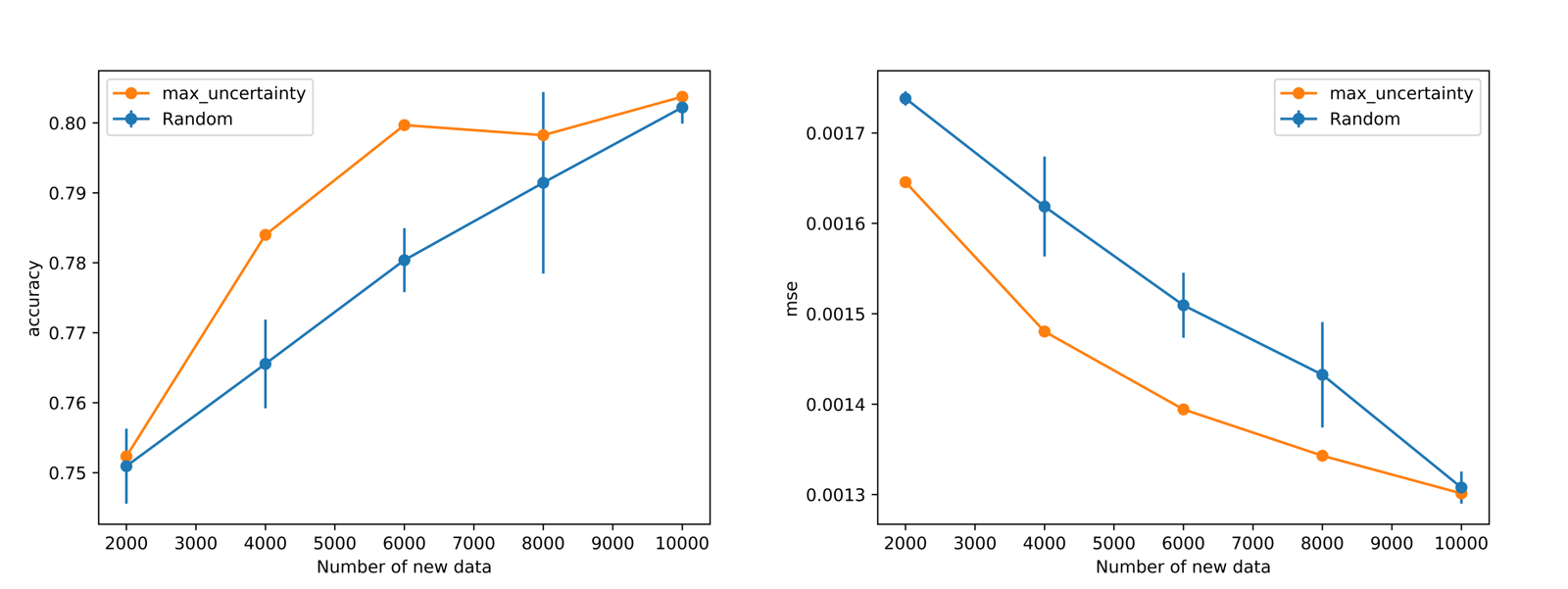}
            \caption{Demonstration of the selective learning framework for an unlabelled set of size 10,000 patches. On the left a diagram depicting the accuracy and on the right a diagram depicting the loss. The orange line corresponds to the max uncertainty acquisition function and the blue to the random acquisition function. For the random acquisition function we repeated the experiment five times and reported the mean values and the 95\% confidence interval. For the accuracy we can see that the orange line is consistently above the blue line. Specifically, we can see that we can achieve a decrease of 40\% in the labelled data requirement because the orange line reaches 80\% accuracy with only 6,000 patches while the blue one with 10,000. For the loss we can see similar results where the loss for the maximum uncertainty case is consistently below the loss for the random case.}
            \centering
            \label{fig:SL_bigg}
        \end{center}
    \end{figure}

    Now we will use a larger unlabelled dataset consisting of 10,000 patches. We compare again the max uncertainty acquisition function and a random acquisition function. The initial training set has 5,000 patches. We train for 150 epochs every network. We perform 100 forward passes for the uncertainty quantification and we label 2,000 unlabeled patches at each iteration (we calculate the micro scale stress field for them), query rate = 2,000. The results can be found in [Fig \ref{fig:SL_bigg}]. The accuracy increases faster for the max uncertainty acquisition function and also the loss function is decreasing faster until it reaches 6,000 new patches. At this point the accuracy practically stops increasing and the loss gradually approaches the same value as with the random acquisition function. Using the max uncertainty acquisition function we can reach the max accuracy using 6,000 patches while we need all the 10,000 patches when randomly choosing new data. Again we have a decrease of 40\% in the labelled data requirement.
    \par
    
\begin{figure}[h]
        \begin{center}
            \includegraphics[width=0.5\linewidth]{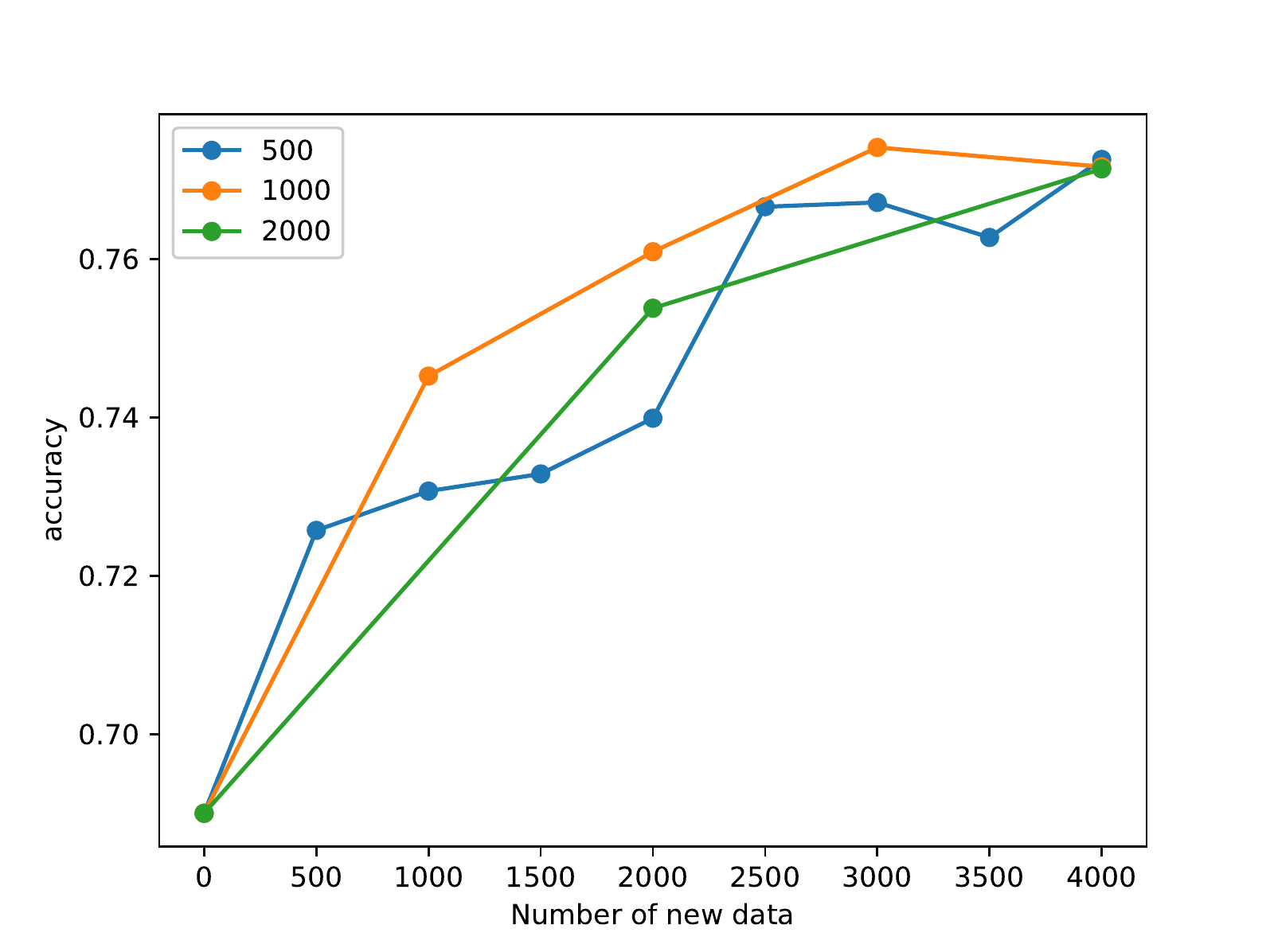}
            \caption{Selective learning with query rates of different sizes. The green line corresponds to a large query rate (of size 2,000 patches), the orange line to a medium query rate (of size 1,000 patches) and the blue line to a small query rate (of size 500 patches). The small and larger query rates result in sub-optimal behaviour and specifically small query rates result in noisy results.} 
            \centering
            \label{fig:SL_steps}
        \end{center}
    \end{figure}

    This time we want to perform a similar experiment but we are interested in examining the effect of query rate on the results. Specifically, we will use an initial set of 5,000 patches and we will perform Selective Learning on an unlabeled dataset of 4,000 patches. We will repeat the experiment 3 times, with query rates 500, 1,000 and 2,000. A similar experiment was conducted by \citep{Islam2016ActiveLF}, where he concluded that using very small query rates results in sub-optimal performance, higher simulation times and noisy behaviour. There are two reasons why the results are worse in this case. Firstly, adding only a few patches compared to the size of the initial dataset might result in overfitting and secondly, these patches might get smoothed out in the loss function. The simulation time increases because the network needs to be retrained a considerable number of times. On the other hand using too large query rates also results in worse results because the weights of the network are not updated frequently enough so new information is rarely incorporated in the network and we end up again labeling and training on patches that do not contain new information. The results of our experiment can be found at [Fig \ref{fig:SL_steps}]. We have reached the same conclusions. When query rate is 1,000 we have the optimal behaviour, when we double it we observe slower convergence and when we use a small query rate we observe noisy sub-optimal behaviour.
    \par
    
\begin{figure}[h]
        \begin{center}
            \includegraphics[width=0.5\linewidth]{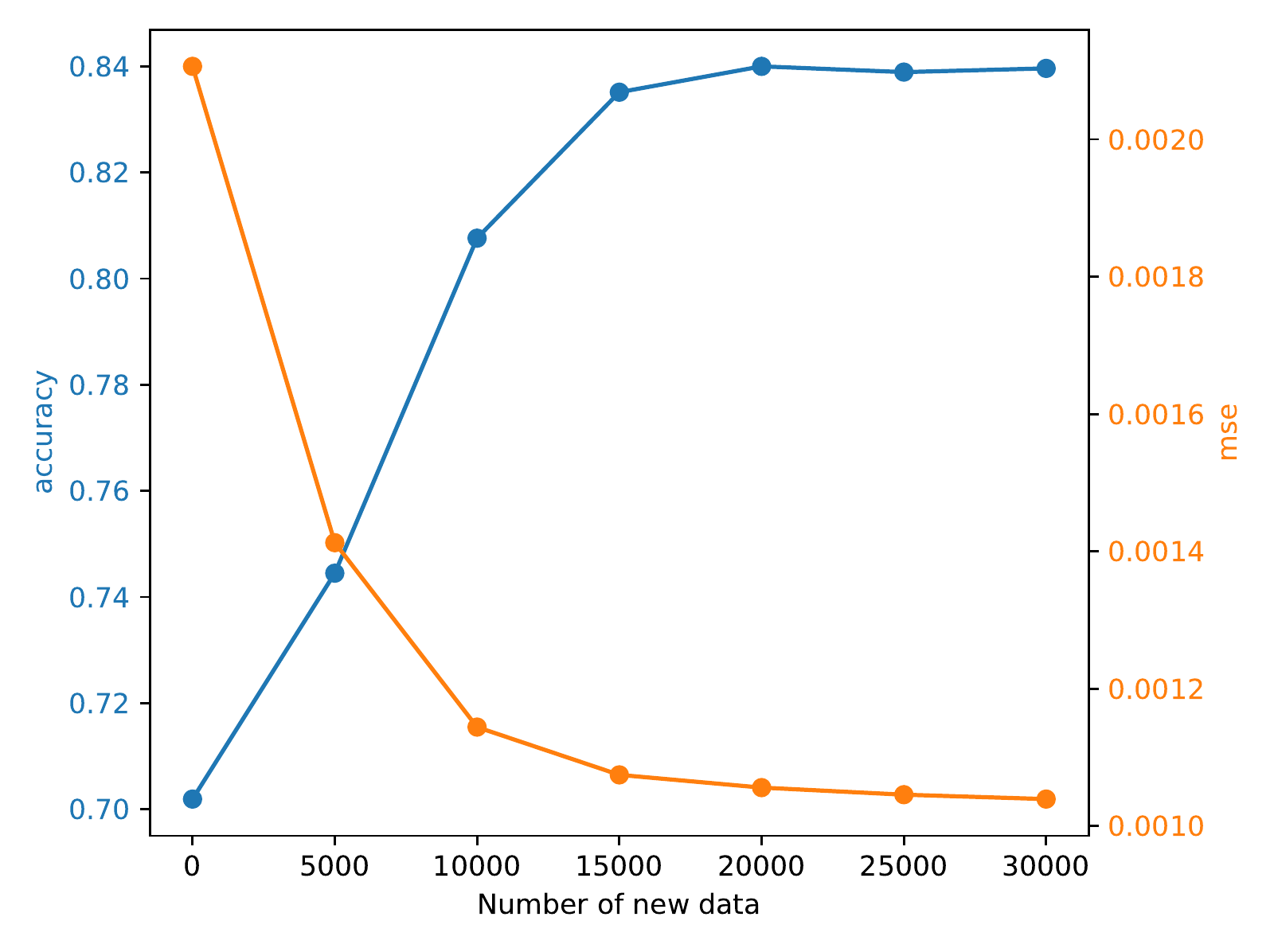}
            \caption{Accuracy and mean squared error plots with respect to train data chosen by a maximum uncertainty acquisition function. The blue line corresponds to the accuracy and the orange one to the mse. The accuracy initially sharply increases until reaching 15,000 patches and after that no further increase is observed. The same can be seen for the loss that sharply decreases until reaching 15,000 patches and then no further decrease is observed. In this case we can reduce the labelled data requirements by 50\% if we only choose the first 15,000 patches indicated by the maximum uncertainty acquisition function and not the entire dataset of size 30,000 patches.}
            \centering
            \label{fig:BigBNN}
        \end{center}
    \end{figure}

    After validating the Selective Learning framework we will now use it without the random acquisition function as baseline. We will use all the 30,000 available data to train the network. As initial set we will use again 5,000 patches. We will query 5,000 unlabeled patches at each iteration chosen by the max uncertainty acquisition function. We will train for 300 epochs and perform 100 forward passes for the uncertainty quantification. The results can be found in [Fig \ref{fig:BigBNN}]. It is clear that the accuracy is not improving after the third iteration, 15,000 patches, but we continued labelling points only for demonstration reasons. The mean squared error decreases for the first 3 iterations and then stops decreasing as well. In this example we could reach the maximum accuracy using 15,000 out of the 30,000 patches, so we managed to reduce the labelled data requirements by 50\%.
    \par
    
    \clearpage
    \subsection{Out of distribution study}
        
        Lastly, we want to test the BNN in data outside of the training set. One way to realise this study is to keep the same micro scale features and create new micro distributions. That could be done using the \say{1 Ellipse Dataset} from section [\ref{Initial Dataset}] and drawing patches from the \say{3 Ellipses Dataset} from section [\ref{Advanced Dataset}], to obtain out-of-distribution samples. Instead, we choose to completely change the micro scale features as we believe that this will be more challenging for the network to predict.
        \par
        
        In this study we will use ellipses as micro features to get out of distribution samples. 
        Neural Networks extrapolate when they make predictions outside of the data set and they are notoriously bad at extrapolating. What we are hoping for is that the BNN will understand that the ellipses are not in the dataset and will assign high variance to most of the patches. 
        \par
        
\begin{figure}[h]
            \centering
            \begin{subfigure}{.5\textwidth}
              \centering
                \includegraphics[width=\linewidth]{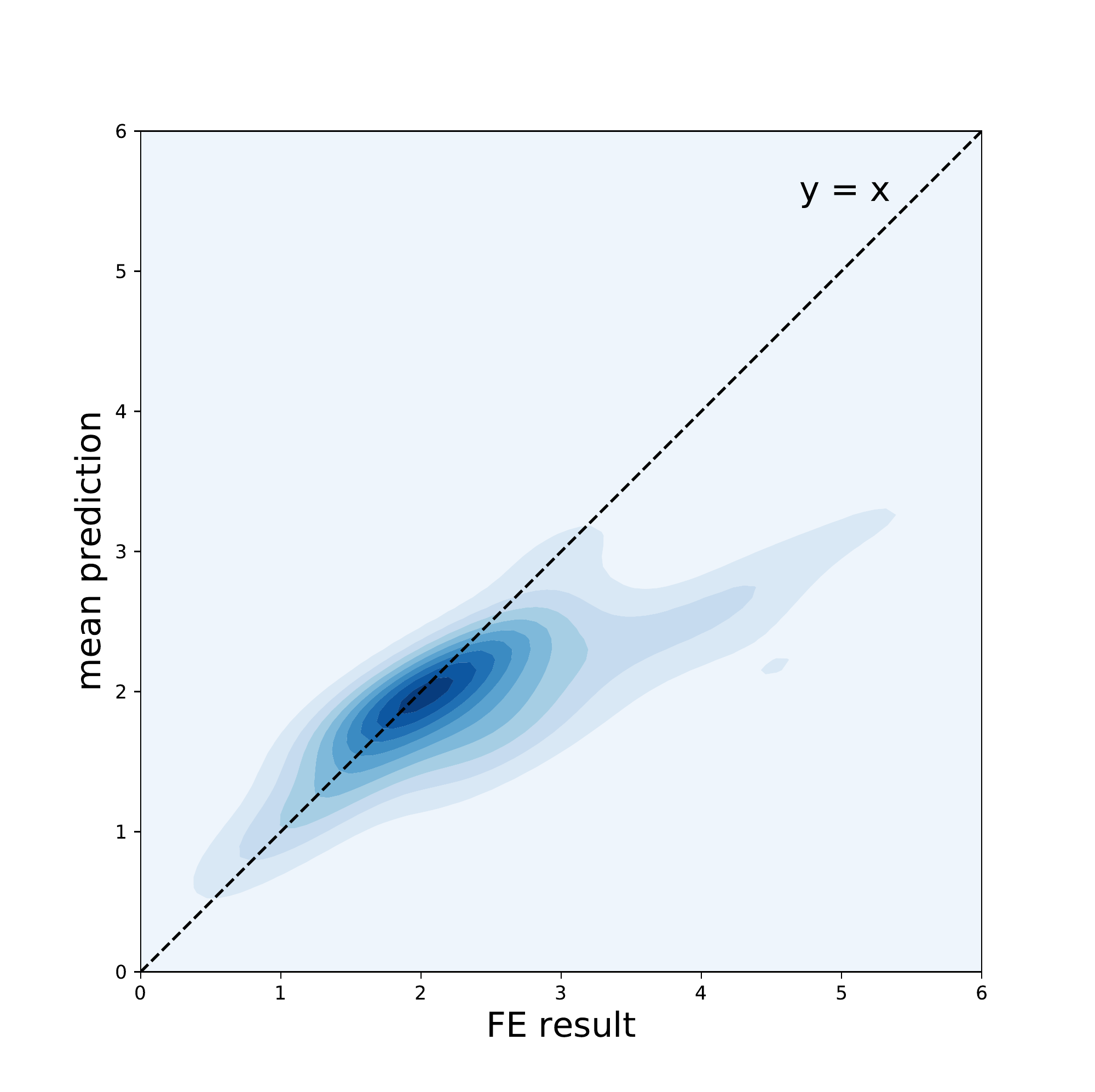}
              \caption{}
              \label{fig:BNN_res_Ellipse:a}
            \end{subfigure}%
            \begin{subfigure}{.5\textwidth}
              \centering
                \includegraphics[width=\linewidth]{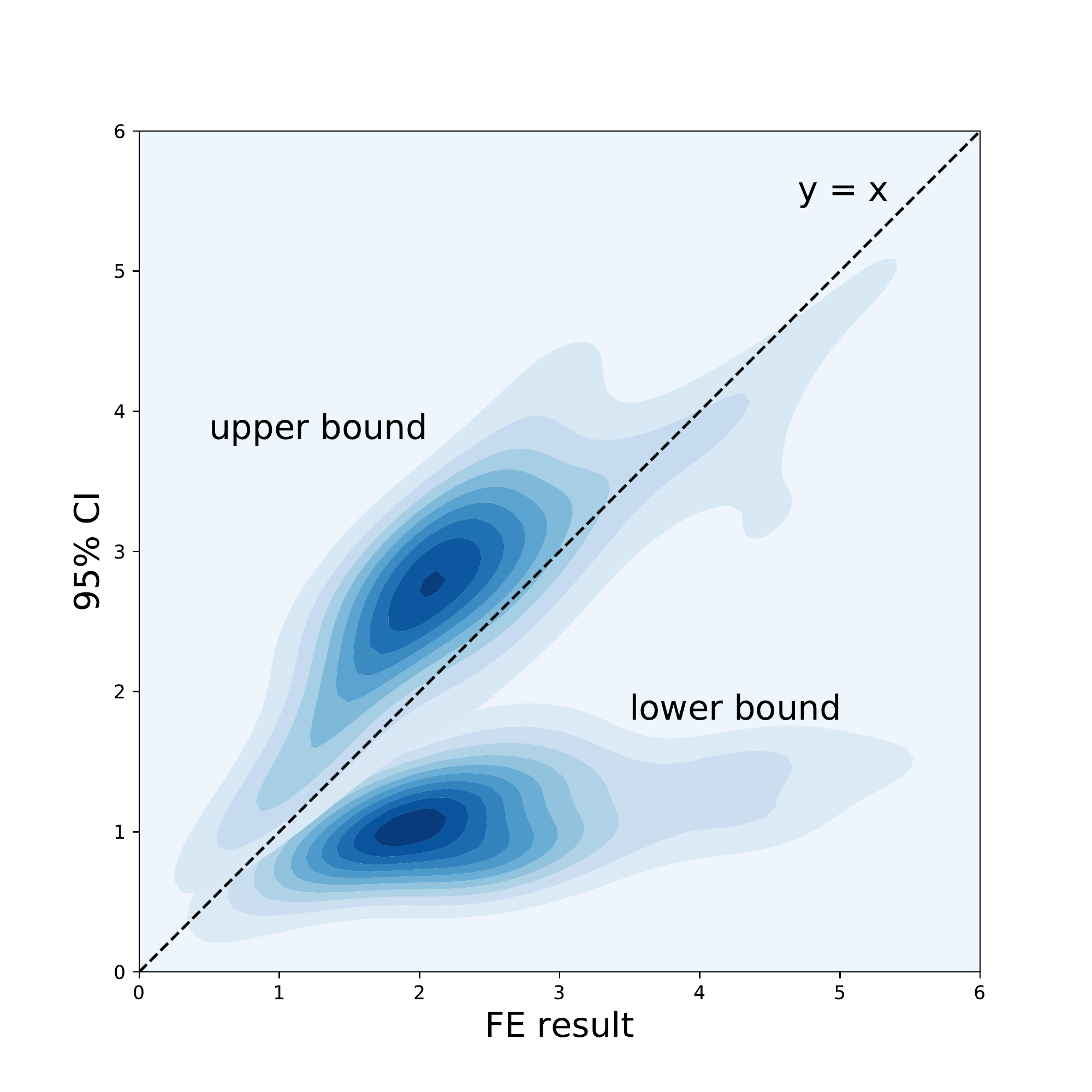}
              \caption{}
              \label{fig:BNN_res_Ellipse:b}
            \end{subfigure}
            \caption{In these 2 figures we see point densities where darker colors correspond to higher point density. On the left a diagram (a) showing the relationship between NNs' mean prediction and FE results for the maximum value in the ROI. We can observe that the maximum value is underestimated in a lot of patches from the BNN. On the right (b) a diagram showing the upper and lower 95\% CIs. We can observe that in most cases the real maximum value is inside the 95\% CIs of the prediction.}
            \label{fig:BNN_res_Ellipse}
        \end{figure}

\begin{figure}[h]
            \begin{center}
                \includegraphics[width=0.5\linewidth]{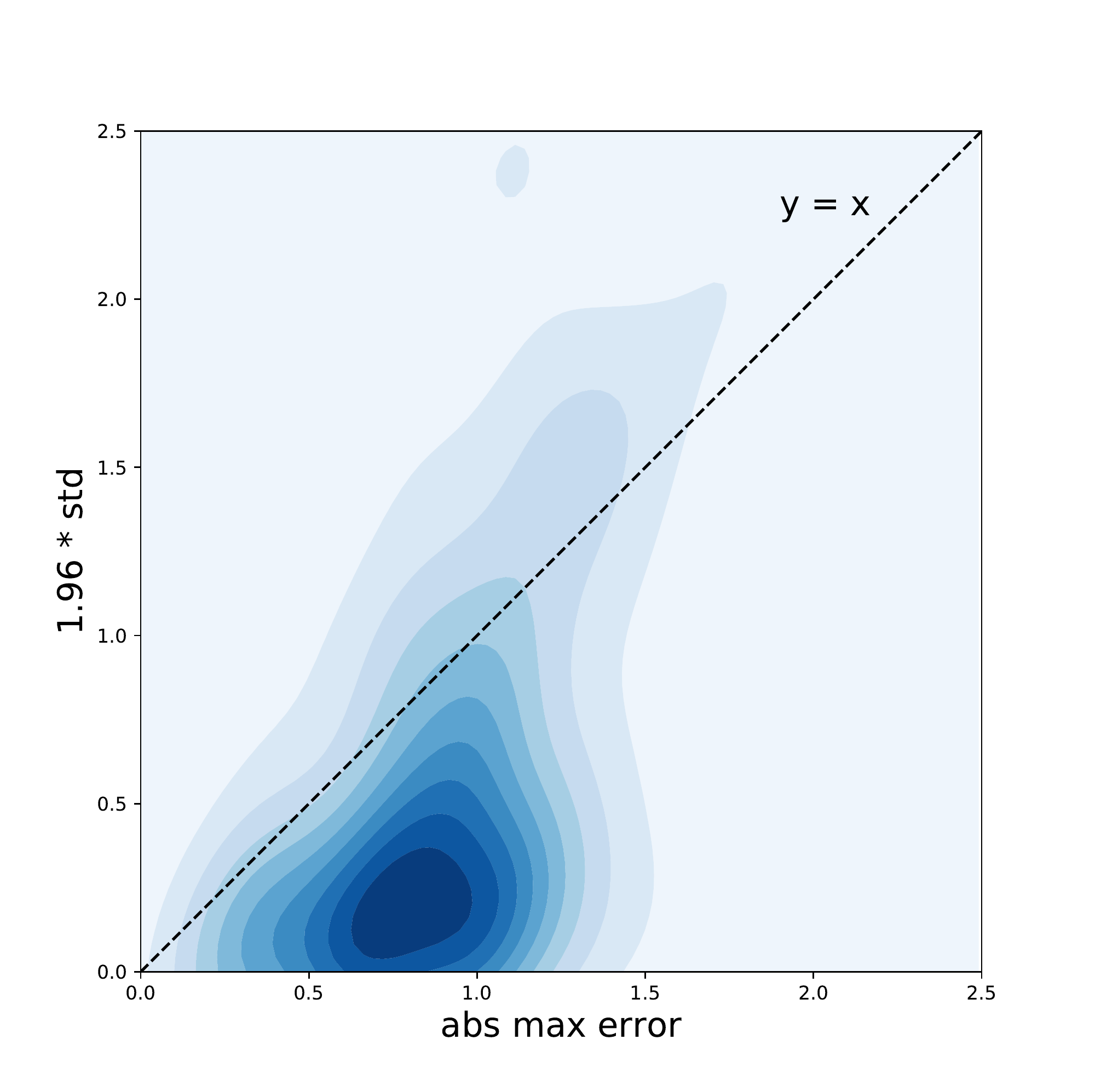}
                \caption{A diagram where the $x$ axis is the absolute error between the real maximum value in the ROI and the predicted one and the $y$ axis is 1.96 $\times$ the standard deviation. We can observe a correlation between high uncertainty and high error.}
                \centering
                \label{fig:Error_Ellipse_BNN}
            \end{center}
        \end{figure}

        We solve the same linear elastic problem we solved in the linear elasticity section [\ref{linear elastic}] with the same BCs. We created 500 patches and made a prediction with the BNN from [\ref{Selective Learning}]. The results can be found in [Fig \ref{fig:BNN_res_Ellipse}]. From the first plot [Fig \ref{fig:BNN_res_Ellipse:a}]  we can see that the mean prediction from the BNN for the max values in the patch is not close to the real max value for a big percentage of the data, accuracy $\approx$ 50\%, but is not unreasonable. Nevertheless, in a lot of cases the network successfully identified the interactions produced by the ellipses even if it was never trained on these. On the other hand, the second plot [Fig \ref{fig:BNN_res_Ellipse:b}] shows that in most cases, $\approx$ 80\%, the true max value is indeed inside the 95\% CI. Even more encouraging is the fact that higher uncertainty corresponds to higher error as can be seen from [Fig \ref{fig:Error_Ellipse_BNN}]. This also implies that selective learning is very promising in this case.

\begin{figure}[h] 
          \begin{subfigure}[b]{0.5\linewidth}
            \centering
            \includegraphics[width=0.70\linewidth]{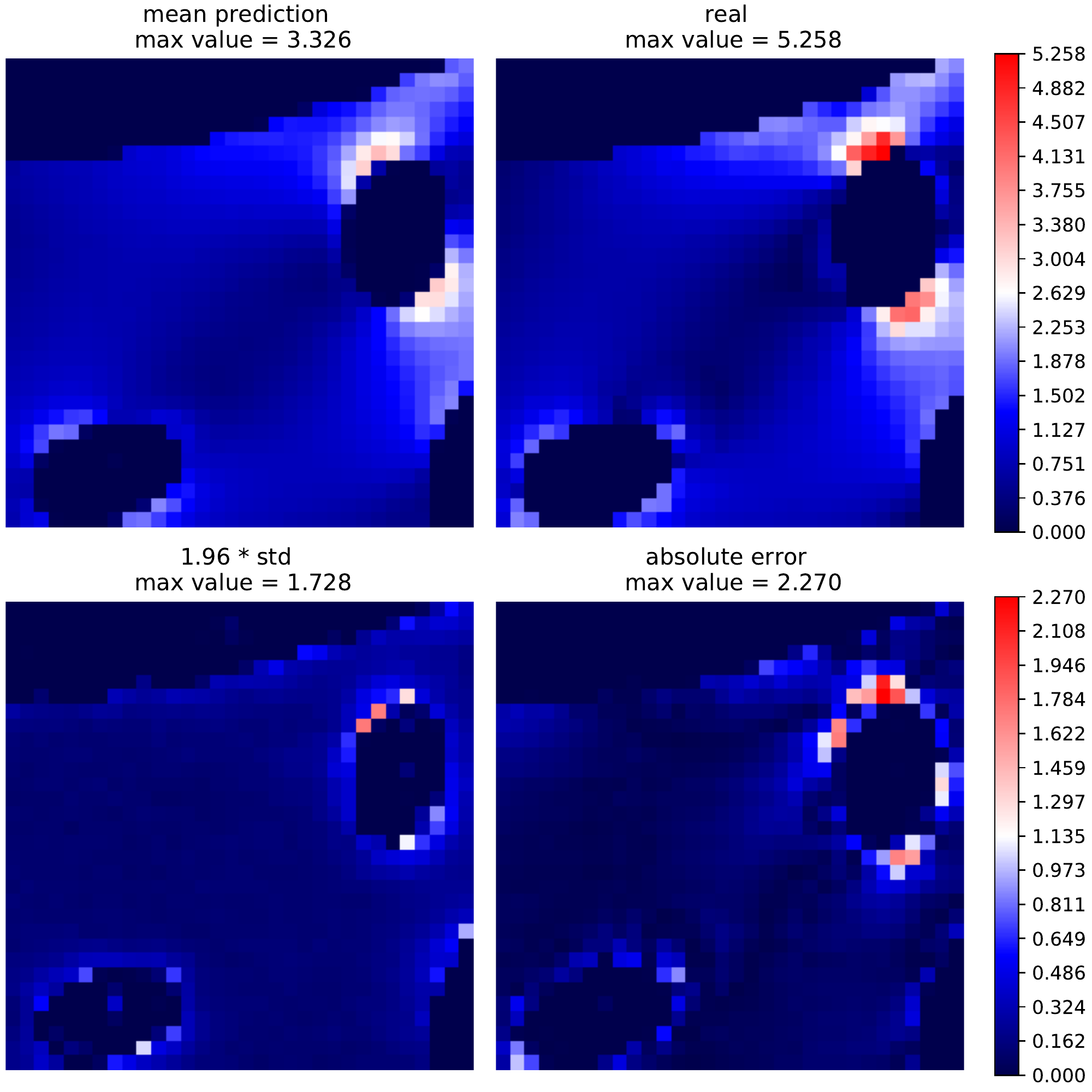}
            \caption{} 
            \label{fig:BNN_OOD:a} 
          \end{subfigure}
          \begin{subfigure}[b]{0.5\linewidth}
            \centering
            \includegraphics[width=0.70\linewidth]{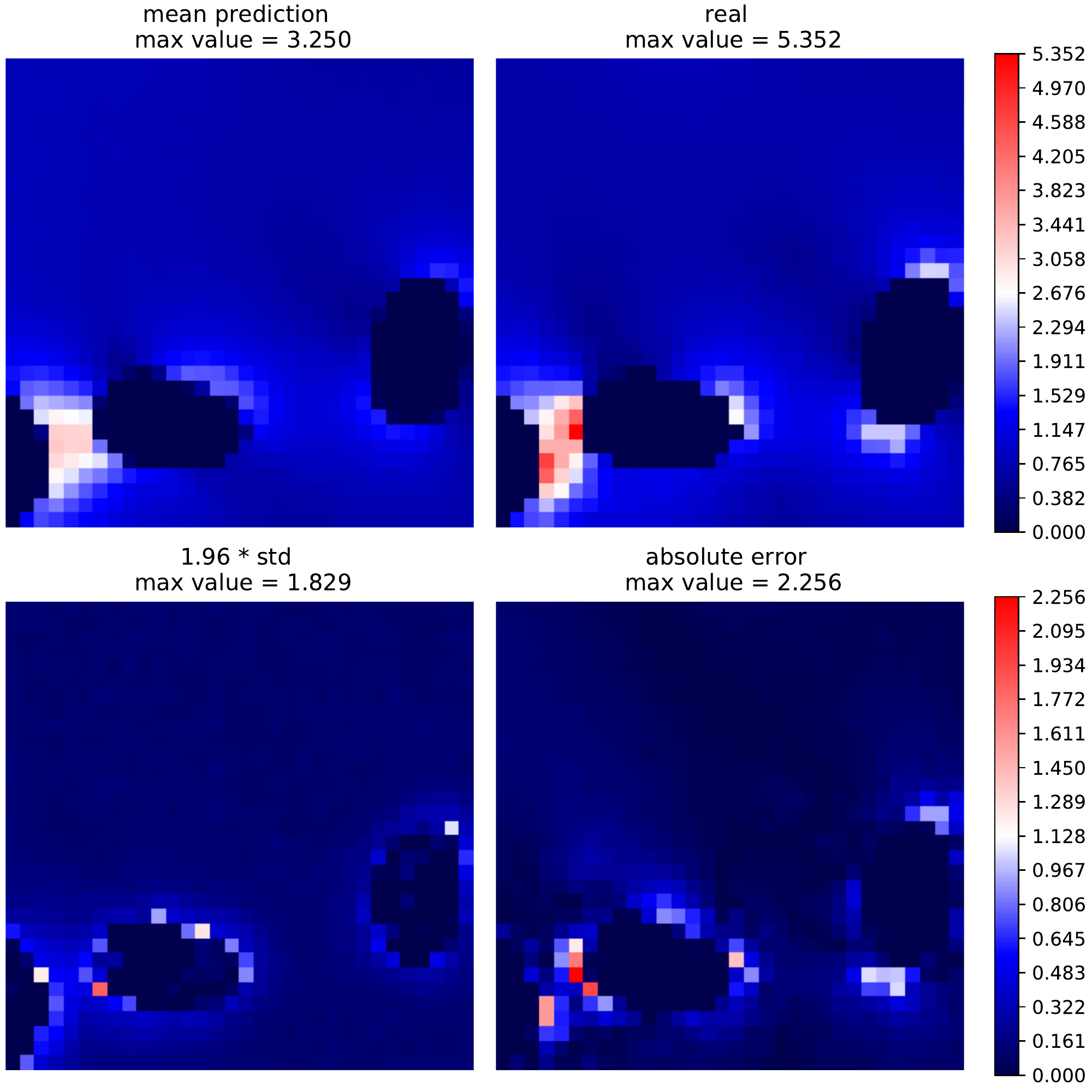}
            \caption{} 
            \label{fig:BNN_OOD:b} 
          \end{subfigure} 
          \begin{subfigure}[b]{0.5\linewidth}
            \centering
            \includegraphics[width=0.70\linewidth]{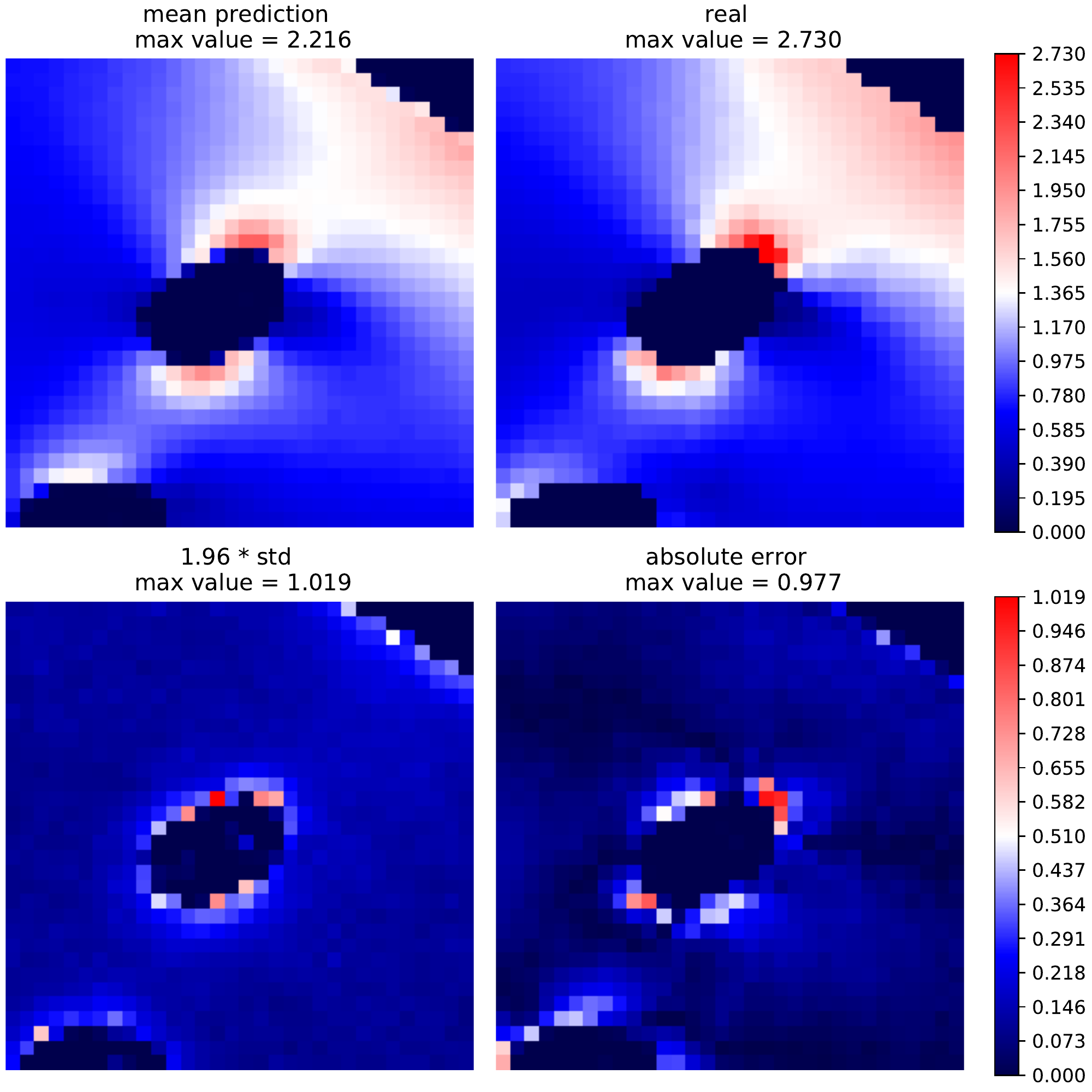} 
            \caption{} 
            \label{fig:BNN_OOD:c} 
          \end{subfigure}
          \begin{subfigure}[b]{0.5\linewidth}
            \centering
            \includegraphics[width=0.70\linewidth]{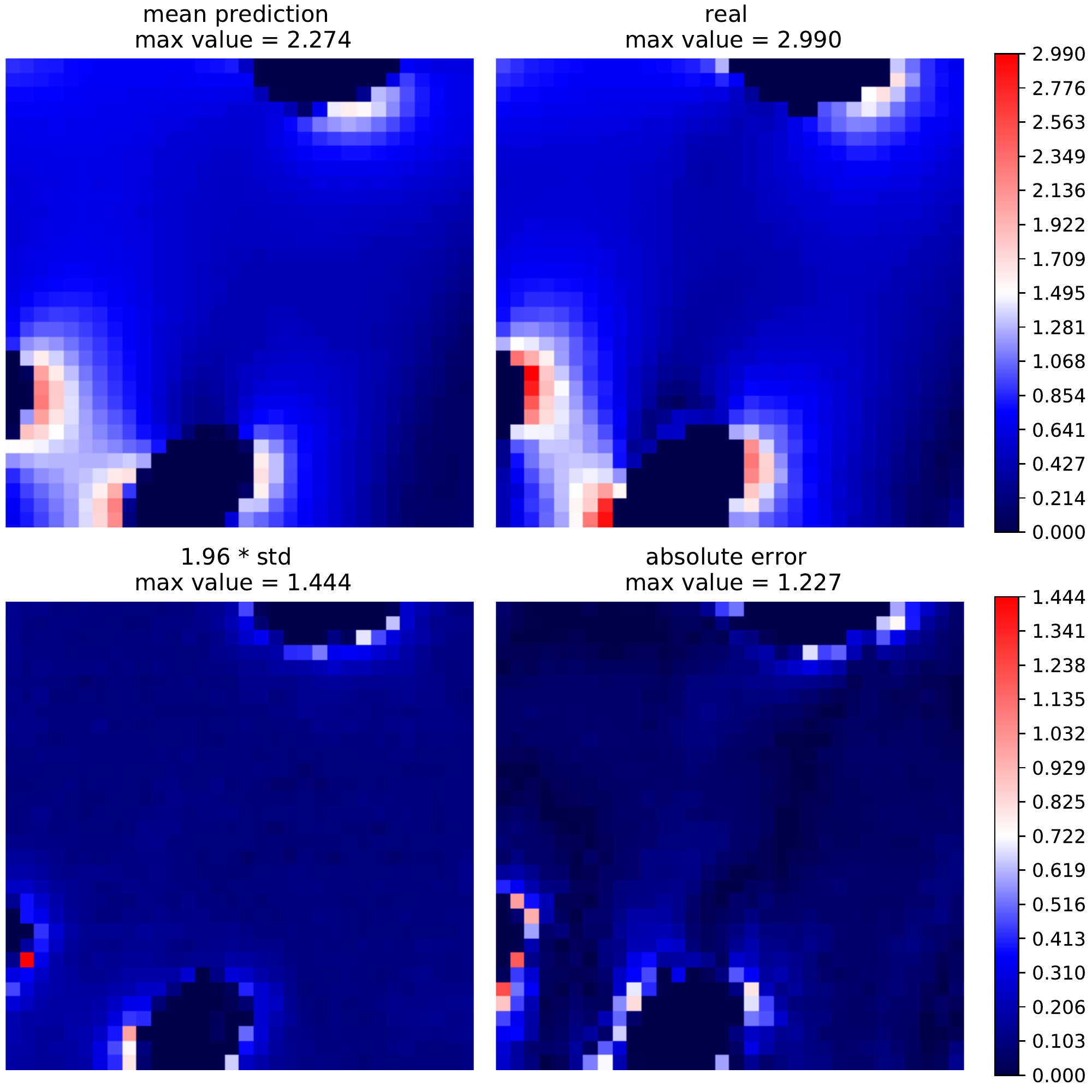}
            \caption{} 
            \label{fig:BNN_OOD:d} 
          \end{subfigure} 
          \begin{subfigure}[b]{0.5\linewidth}
            \centering
            \includegraphics[width=0.70\linewidth]{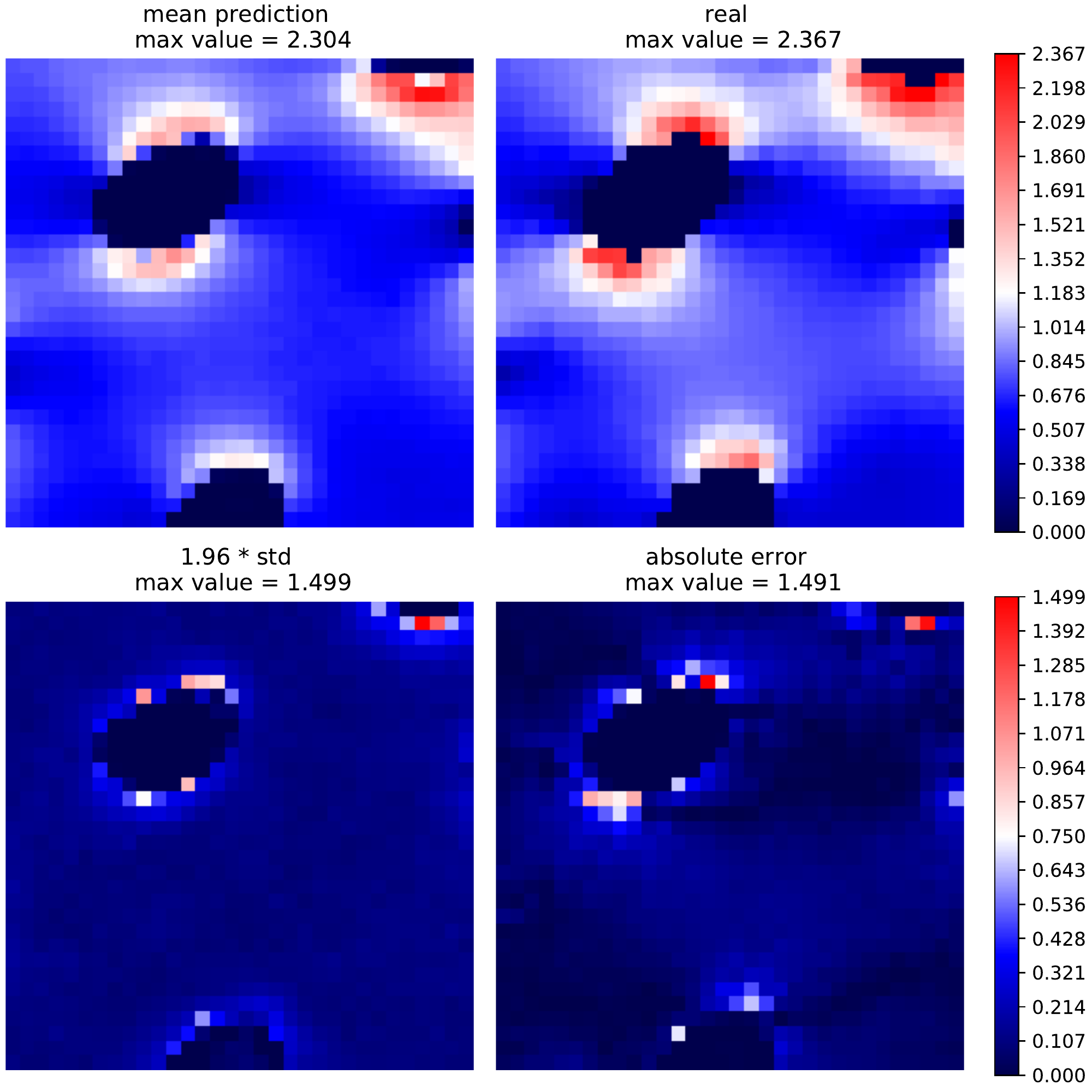}
            \caption{} 
            \label{fig:BNN_OOD:e} 
          \end{subfigure}
          \begin{subfigure}[b]{0.5\linewidth}
            \centering
            \includegraphics[width=0.70\linewidth]{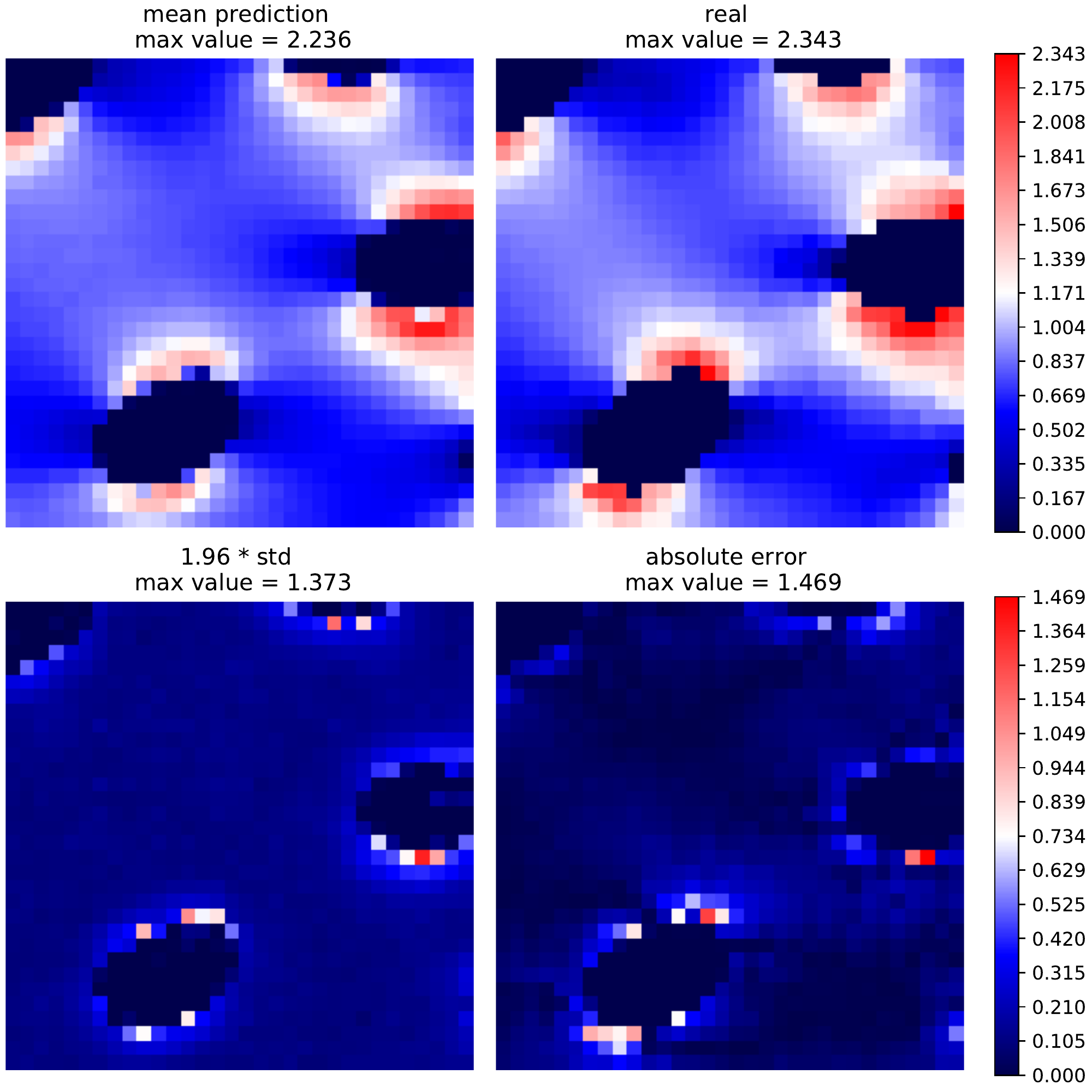}
            \caption{} 
            \label{fig:BNN_OOD:f} 
          \end{subfigure}
          \caption{BNN predictions on out of distribution examples. All images correspond to the ROI of the patches. In all of the 6 images the first row corresponds to the NN mean prediction on the left and to the scaled Tresca stress field computed by FEA and converted into an image on the right. The second row corresponds to the NN uncertainty, expressed as 1.96 $\times$ standard deviation, on the left and to the absolute error between the NN mean prediction and the FE results on the right. In figures (a) and (b) we can see cases with high error in the maximum values that is outside the 95\% CIs while in figures (c) and (d) there is high error in the maximum values but this error is contained inside the 95\% CIs. Lastly in figures (e) and (f) we can see that the maximum values are predicted with high accuracy by the BNN and the error that exists in other areas is captured by the uncertainty of the BNN.}
          \label{fig:BNN_OOD} 
        \end{figure}

        We can also see examples of predictions in 6 patches of this new dataset. In [Fig \ref{fig:BNN_OOD:a}, \ref{fig:BNN_OOD:b}] we can see 2 examples of cases where the error in max values is relatively high and even though the 95\% CIs are very broad they fail to contain the real value. In [Fig \ref{fig:BNN_OOD:c}, \ref{fig:BNN_OOD:d}] we can see 2 examples of cases where the error is high but inside the 95\% CI. Lastly in [Fig \ref{fig:BNN_OOD:e}, \ref{fig:BNN_OOD:f}] we can see 2 examples where the mean prediction of the BNN is very close to the real value. Some error is present in other areas of the patch but this error is captured by the uncertainty of the BNN. 
        
\section{Discussion} \label{Discussion}

The aim of this section is to provide clarifications regarding the applicability of the methodology developed in this paper. In terms of multiscale modelling, the approach is concurrent -which differs from homogenisation- thereby circumventing the need for scale separability but with drawbacks that are discussed below. Building upon these remarks, we identify classes of applications for which our methods may be of practical interest.

    \subsection{Homogenisation}
    
        In methodologies based on homogenisation, scale separation is postulated, and RVE is defined to describe the heterogeneous material. Spatial and material coordinates are independent, and the material only interacts with the boundary of the macro domain through the local values of the macroscale mechanical fields. Subsequently, macroscale properties are computed numerically by testing the RVE, and relocalisation may be performed to compute \textit{statistics} of the microscale fields.
        
        
        Recent work in this field is described in \citep{FFT_homo} where a Lippmann–Schwinger formulation for the unit cell problem of periodic homogenization of elasticity at finite strains is introduced and more recently the work published in \citep{LIU_HiDeNN}, where a multilevel Neural Network approach to solve a multiscale mechanics problem is developed. The macroscale is solved online, using a Physics Informed Neural Network (PINN) while at the micro scale a separate Neural Network is used, trained offline, to recover the average micro stress in the RVE given the macro strain and material parameters. To reduce the offline computational cost, a reduced order modelling approach is used \citep{LIU2016319}. The same reduced order modeling approach is also used in \citep{Liu2018} for multiscale elastoplasticity problems.
        
        In elasticity, homogenisation works well if the size of the micro scale features is much smaller than the size of macroscale stress concentrators. This is clearly not the case in the examples proposed in this paper. Conversely, our method is only useful in cases where scales are not well separated, as it will be bettered by homogenisation-based approaches when they are applicable. Indeed, homogenisation-based approaches "only" need to learn material responses in spaces of coarse strain trajectories \citep{goury_amsallem_bordas_liu_kerfriden_2016, RochaKerfriden2021}
        
\begin{figure}[h]
            \begin{center}
                \includegraphics[width=.75\linewidth]{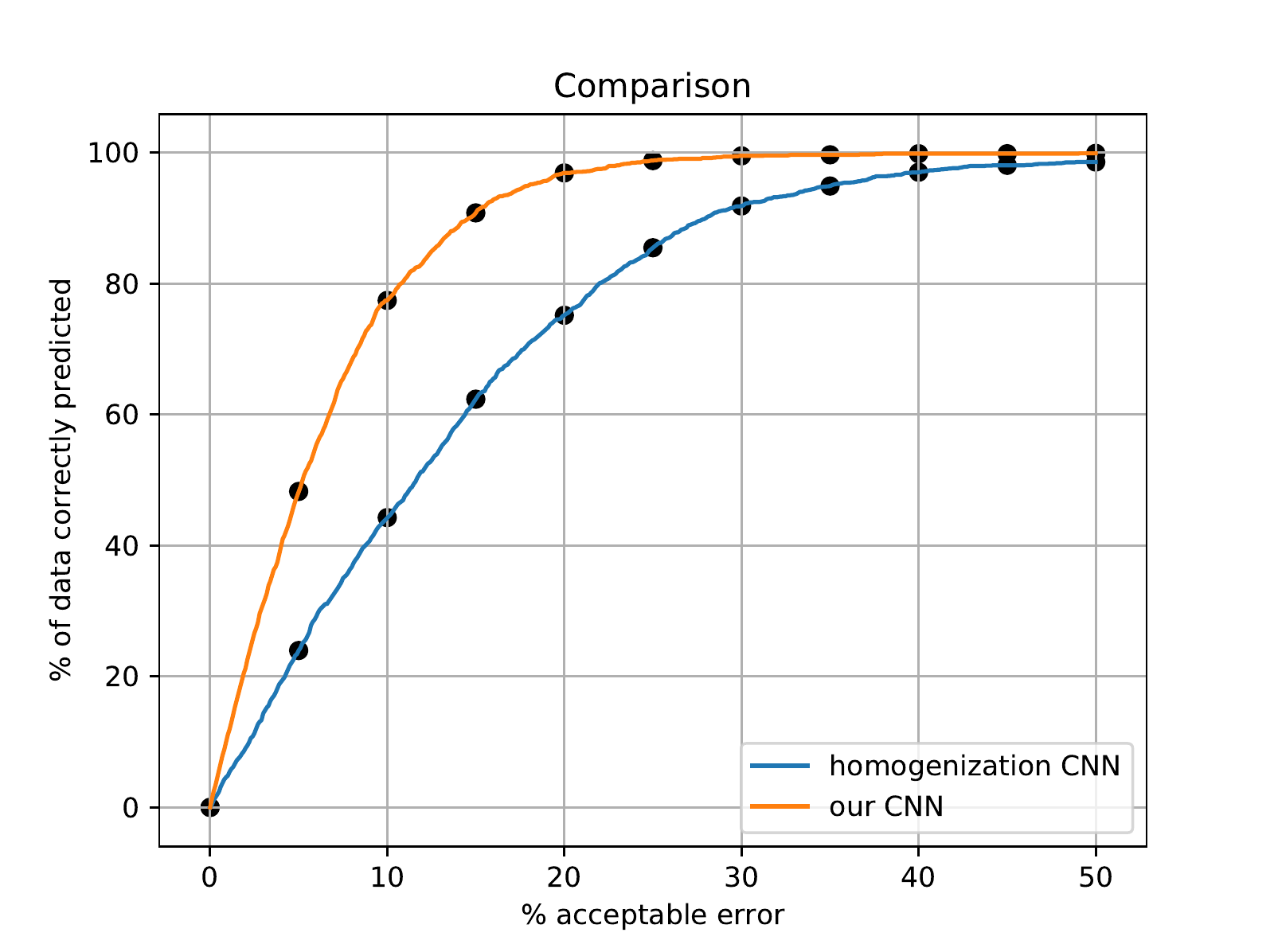}
                \caption{Comparison between 2 CNNs trained with the same architecture and dataset but one takes as input the average stress in the patch and the other the entire stress field. Blue line corresponds to the CNN with the average stress as input and the orange line to the CNN trained with the full stress as input. The x-axis of the diagram corresponds to the threshold level used to define the accuracy and the y-axis to the accuracy.}
                \centering
                \label{fig:Homogenization_vs_ours_acc}
            \end{center}
        \end{figure}

\begin{figure}[h]
            \begin{center}
                \includegraphics[width=\linewidth]{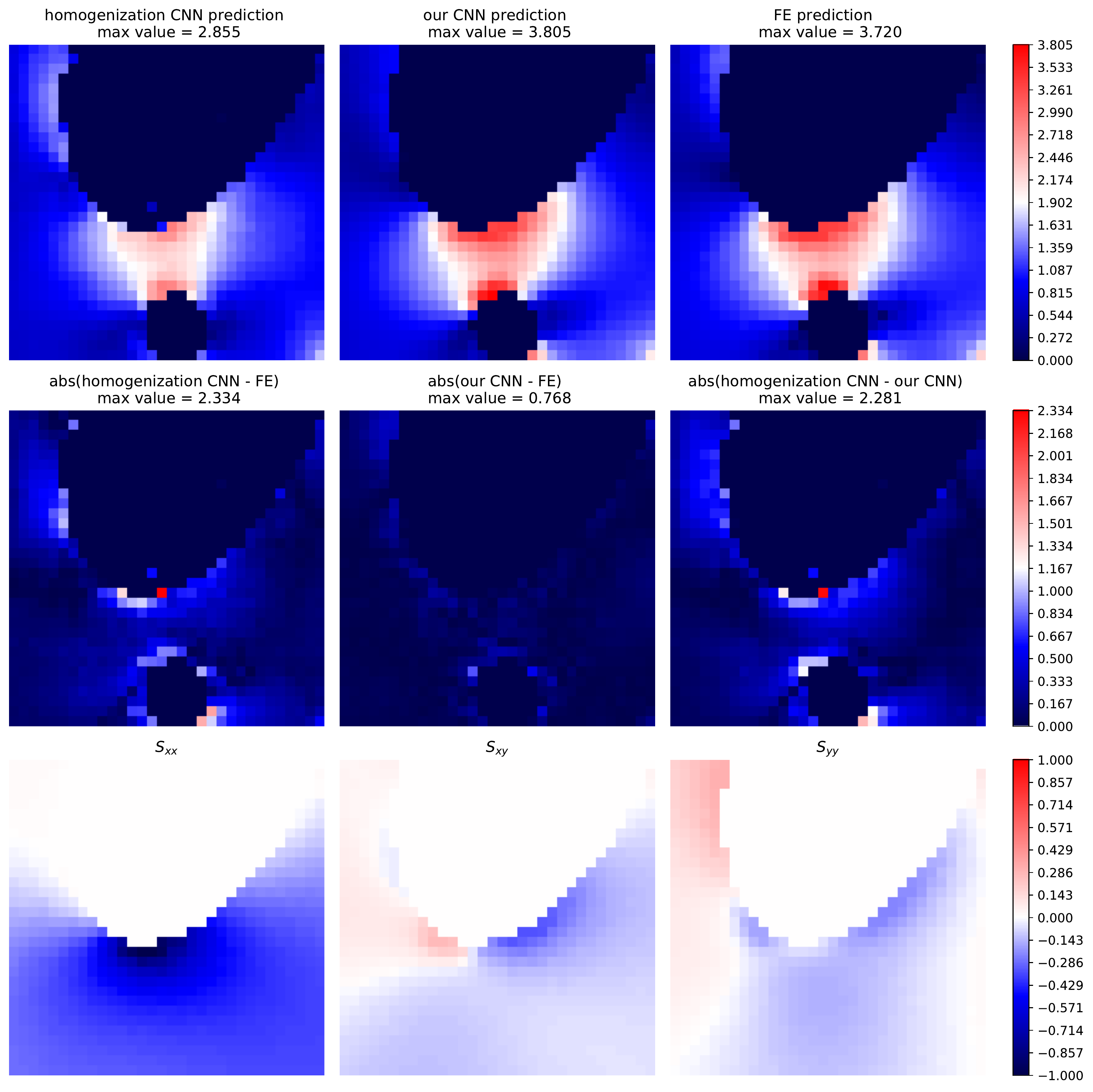}
                \caption{A figure where we compare the homogenisation CNN to our CNN for a patch where the macro stress field is not constant. In the first row we see from left to right, the homogenisation CNN prediction, our CNN prediction and the scaled Tresca stress field computed by FEA and converted into an image. We observe that the homogenisation CNN fails to predict the correct stress distribution in contrast to our CNN. In the second row from left to right we see the absolute error between the homogenisation CNN and the FE results, our CNN and the FE results, and the homogenisation CNN and our CNN. In the last row from left to right we see the xx, xy and yy components of the macro stress tensor.}
                \centering
                \label{fig:Homogenisation_vs_noHmog_difference}
            \end{center}
        \end{figure}

\begin{figure}[h]
            \begin{center}
                \includegraphics[width=\linewidth]{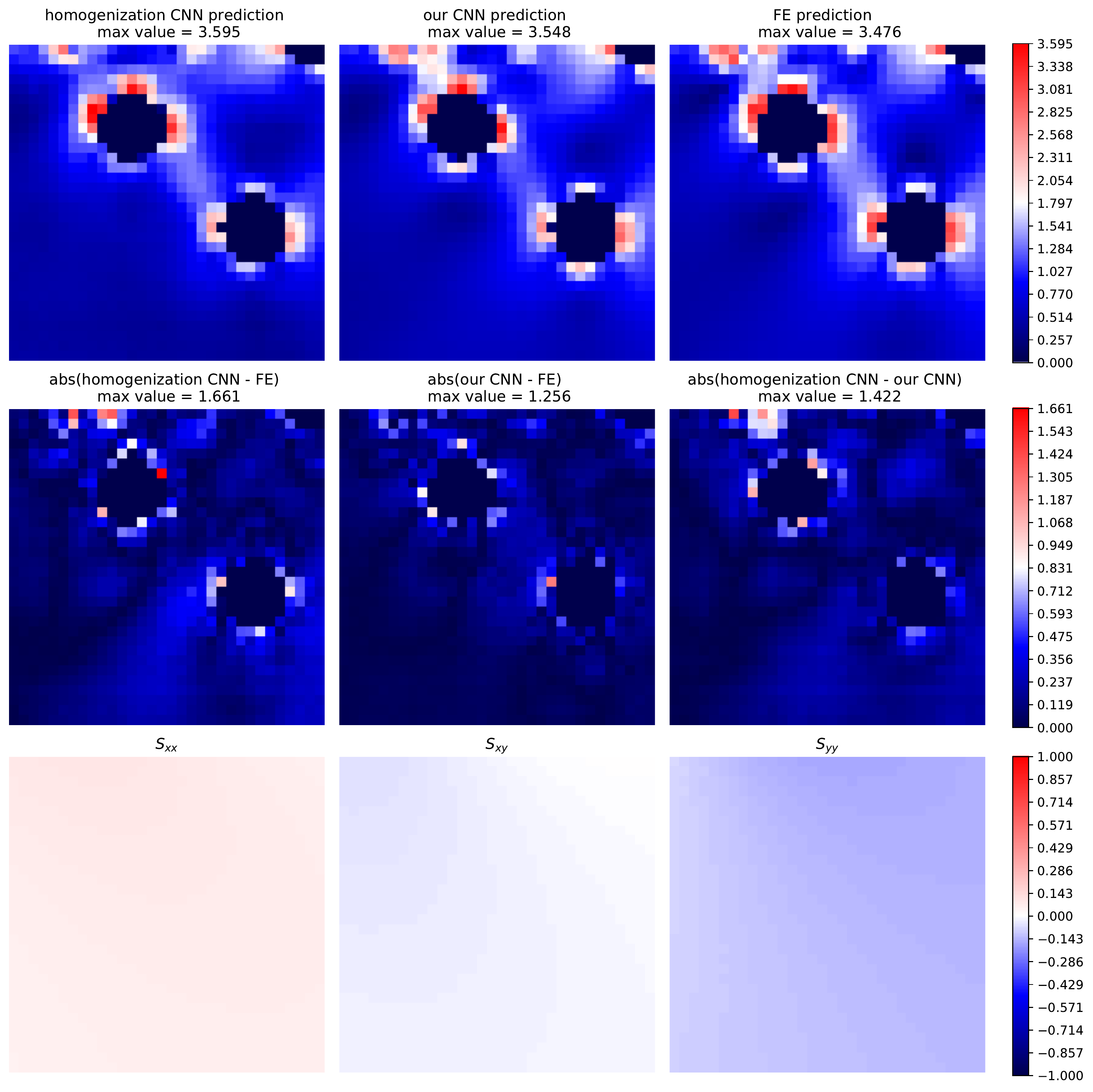}
                \caption{A figure where we compare the homogenisation CNN to our CNN for a patch where the macro stress field is constant. In the first row we see from left to right, the homogenisation CNN prediction, our CNN prediction and the scaled Tresca stress field computed by FEA and converted into an image. We observe that both the homogenisation CNN and our CNN manage to predict the correct stress distribution. In the second row from left to right we see the absolute error between the homogenisation CNN and the FE results, our CNN and the FE results, and the homogenisation CNN and our CNN. In the last row from left to right we see the xx, xy and yy components of the macro stress tensor.}
                \centering
                \label{fig:Homogenisation_vs_noHmog_same}
            \end{center}
        \end{figure}

        We propose to illustrate the qualitative point made above. To this end, we train a CNN that takes as input the average stress field over the patch and not the entire stress field, hence mimicking the effect of spatial averaging due to first-order homogenisation schemes. We call this network an homogenisation CNN. We compare this CNN with the CNN we proposed earlier in this paper. Our goal is to show that our CNN outperforms the homogenisation CNN and specifically that this happens in patches where there are interactions between micro and macro scale features. We trained the 2 CNNs with the same architecture and training dataset, 27,000 training data and 3,000 validation data. In [Fig \ref{fig:Homogenization_vs_ours_acc}] we see that the first CNN achieves higher accuracy, specifically for the 10\% threshold our CNN achieves 79\% accuracy while the homogenization CNN 42\%. Additionally, in Fig [Fig \ref{fig:Homogenisation_vs_noHmog_difference}] we compare the prediction of the 2 CNNs in a patch where the macro stress field is not constant. We see that in contrast to our CNN the homogenization CNN fails to predict the correct stress field. We can also see that the higher absolute error between the homogenisation CNN and both our CNN and the FE solution is close to the ellipse where the macro stress field varies faster. Lastly, in Fig [Fig \ref{fig:Homogenisation_vs_noHmog_same}] we compare the prediction of the 2 CNNs in a patch where the macro stress field is constant. We see that both our CNN and the homogenization CNN manage to predict the correct stress field.

    \subsection{Use cases}
    
        Although our deep learning strategy works incredibly well when generating new realisations of the random material model used to generate training examples, generalisability to other classes of random material distributions is rather poor. Similarly, predictions cannot be expected to be accurate for macroscale geometries that are not included in the training set. Both are limiting points, which points towards specialised applications of the investigated framework.
        
        For example, our method may be applied in a context where engineering components are produced in large series, potentially subject to small parameter changes. In this setting, a macroscale parametrised CAD model should be appropriately devised. Subsequently, realisations of the CAD model, complemented by random realisations of the microscale parameter field, may be used to train the NN. After training, the NN-based micro/macro simulator could help compute ensemble statistics of fatigue life.
        
        Alternatively, if the macroscale geometrical variations correspond to manufacturing defects that can be measured in manufacturing chains, 
        one could further condition the probabilistic micro/macro stress predictions to conform to partial observations of the macroscale geometrical inaccuracies to provide component-specific life predictions.
        
        As another example, the proposed approach method may be used to help optimise a CAD model under uncertainty stemming from the existence of a distribution of microscale defects. Similar to the previous use case, the CAD model needs to be appropriately parametrised, the parameter space being explored in advance during training.
        
        Both applications are made possible by the small cost of the NN predictions. This has been demonstrated in \citep{NVIDIA_PINN} where a parameterized structure is optimized using a PINN trained offline. The authors report a 45,000x speedup compared to a commercial solver and  135,000x compared to OpenFOAM (an open source FE solver). That work  differs from ours as the authors use a PINN and they don't solve multiscale problems but the basic principal of using a trained NN for geometry optimisation is exaclty the same.

    \subsection{History dependent problems}
    
        Many problems arising in computational mechanics are history dependent, for instance elastoplasticity or viscoelasticity. Unfortunately, our current multiscale CNN architecture is not able to tackle all history dependent problems in general, but it could be used for a limited set of problems where one-to-one mapping between the macro stress and the micro scale correction exists. For instance, in  \citep{RochaKerfriden2021}, we replaced constitutive relations in multiscale plasticity by instantaneous Gaussian Processes, the strain history being discarded; an approach that proved successful for monotonic macroscopic loading. It may also be the case when looking for stabilised plasticity solutions in fatigue but further investigations would be needed to prove this point. If we want to address a broader set of history dependent problems, we could add extra channels to the input that would correspond to snapshots of the stress distribution from the past \citep{Yan_temporal_cnn}. Another common approach for time series prediction are the Long Short-term Memory Networks. Recently \citep{Cracks2021} developed a convolutional-aided bidirectional Long Short-Term Memory network to predict the sequence of maximum stress until material failure. This architecture combines CNNs and LSTMs and leverages the advantages of both. Other approaches to tackle history dependent problems in mechanical engineering are not purely data-driven but try to also exploit the physical insights obtained from well established plasticity theory and experiments. This approach is illustrated by the work of \citep{Tang_MAP123-EP} for elastoplastic materials and \citep{TANG2021113484} for elastoplastic materials undergoing finite strain.

\clearpage 
\section{Conclusions}
   
    The goal of this work was to develop a multiscale meta-modelling technique that can be used to perform fast stress predictions in structures exhibiting spatially random microscopic features, without prior assumption of scale separability, and without prior parametrisation of the multiscale problem.

    The meta-modelling approach that we have designed is based on a CNN technology. We have shown that it is able to predict how macroscale solution fields should be corrected by taking into account the existence of microscale pores, interacting arbitrarily with one another and with the boundary of the computational domain. The framework is designed in such a way that it does not require any knowledge of the PDE system to generate microscale corrections, and therefore is not intrusive. Moreover, the methodology is not a priori limited to \textit{adhoc} parametrisations of the multiscale boundary value problem, as it treats both geometry and morphology of microscopic patterns as arbitrarily large images. Incidentally, the method is Bayesian and provides credible intervals for the microscale field predictions. 
    
    Experiments with data from linear elastic simulations with a variety of macroscale structures, of the same family, with randomly distributed circular pores under different boundary conditions showed good performance in terms of mean values and uncertainty intervals. This suggests that the method generalises well in new realisations of known macroscale structures and microscale distributions. We also proved that the method readily extends to multiscale predictions with geometrical nonlinearities. 
    
    
    Additionally, we investigated two other features of our framework. We addressed the problem of limited labeled data using mechanically-consistent rotations as data augmentation technique. Furthermore, to reduce the large computational cost associated with the creation of labeled data (i.e. multiscale FEA simulations) we used selective learning to choose and label only the data that contains new information for the network.
    
        
    Even though our framework works well in the aforementioned cases, some limitations need to be highlighted.
    
    Firstly, we assumed that the local macro scale fields that we use as input to our CNN are sufficient to predict all the micro scale features, everywhere in space and for all the structural problems that belong to the particular class of parametrised problems over which training is performed. This is an assumption of locality that cannot in general be used in the context of non-diffusive problems (e.g. wave propagation, crack propagation, scattering, plastic localisation).
    
    A second limitation of the methodology is its rather poor extrapolation ability. Indeed, when the CNN that was trained with disks as microscale features was used to make predictions on data with elliptically-shaped pores as microscale features, the accuracy decreased substantially. However, the uncertainty interval remained reasonably accurate and could be used to indicate that the requested predictions are too far from the training set, for instance in a selective or active learning framework. Additionally, when the CNN that was trained on the one-ellipse dataset was used to make predictions on the three-ellipses dataset, the accuracy also decreased, albeit not as sharply as in the former case. To summarise, the method does not generalise well outside the set of examples (micro or macro scale) of the parametrised problem over which training was performed, which is unsurprising. A corollary to the previous statement is that it is necessary to restrict training to relatively narrow families of boundary value problems, as the space of heterogeneous structures with randomly distributed microscopic components is arbitrarily vast.

    Lastly, the proposed method is not able to tackle all history dependent problems in general, but it could be used for a limited set of problems where one-to-one mapping between the macro stress and the micro scale correction exists. Possible modifications of the architecture have been discussed in section [\ref{Discussion}] that would extend the applicability of the proposed method to a wider set of history dependent problems.

\clearpage 

\section*{Acknowledgments} 
\begin{minipage}{0.5\textwidth}
    \includegraphics[width=\textwidth]{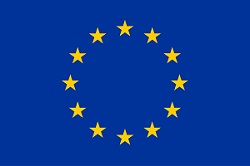}
\end{minipage}\hspace{10pt}
\begin{minipage}{0.5\textwidth}
    This project has received funding from the European Union’s Horizon 2020 research and innovation programme under the Marie Sklodowska-Curie grant agreement No. 764644.\\
    
    This paper only contains the authors' views and the Research Executive Agency and the Commission are not responsible for any use that may be made of the information it contains.\\
\end{minipage}
\hfill \break
\hfill \break
%
\hfill \break
Stéphane P.A. BORDAS received funding from the European Union’s Horizon 2020 research and innovation programme under grant agreement No 811099 TWINNING Project DRIVEN for the University of Luxembourg and from the Fonds National de la Recherche Luxembourg FNR under grant O17-QCCAAS-11758809.\\


\bibliography{BCNN}


\end{document}